%% file: fivePoints.tex
\documentclass[11pt,a4paper]{article}
\usepackage{jheppub}
\usepackage{amsmath,amssymb,amsfonts}
\usepackage{booktabs}
\usepackage{epsfig}
\usepackage{graphicx}
\usepackage{color}
\usepackage{shuffle}
\usepackage{hyperref}
\usepackage[dvipsnames]{xcolor}
\usepackage{colordvi}
\usepackage{colortbl}
\usepackage{float}
\usepackage{floatflt}
\usepackage{graphicx}
\usepackage{hhline}
\usepackage{contour}
\usepackage[tableposition=above]{caption}
\usepackage[normalem]{ulem}
\usepackage{tikz-feynman} 

\tikzfeynmanset{compat=1.1.0}
\tikzfeynmanset{/tikzfeynman/warn luatex = false}

\usetikzlibrary{backgrounds,calc}

\usepackage{wrapfig}
\usepackage{cases}
\usepackage[utf8]{inputenc}
\usepackage{mathtools,tabu}
\usepackage{slashed}

\usepackage{amsmath}
\DeclareMathOperator{\Tr}{Tr}

\usepackage{stackengine}
\stackMath
\usepackage{graphicx}

\newcommand{\rOne}{\mathfrak{r}^{(1)}}
\newcommand{\cHD}{c^{\text{HD}}}

\newcommand{\ap}{{\alpha'}}

\newcommand{\Aym}{A^{\rm YM}}

\def\be{\begin{equation}}
\def\ee{\end{equation}}
\newcommand\bea{\begin{align}}
\newcommand\eea{\end{align}}

\def\eqn#1{eq.~(\ref{#1})}

\def\eqns#1#2{eqs.~(\ref{#1}) and~(\ref{#2})}

\def\EqnsTh#1#2#3{eqs.~(\ref{#1}), (\ref{#2}), and~(\ref{#3})}
 
\def\sect#1{Section~\ref{#1}}
\def\app#1{Appendix~\ref{#1}}

\newcommand{\compAdj}[2]{\left(#1\,\textcircled{$\mathfrak{a}$}\,#2\right)}
\newcommand{\compRel}[2]{\left(#1\,\textcircled{$\mathfrak{r}$}\,#2\right)}
\newcommand{\compHybrid}[2]{\left(#1\,\textcircled{$\mathfrak{h}$}\,#2\right)}
\newcommand{\compPerm}[2]{\left({#1}\,\textcircled{\small$\mathfrak{p}$}\,{#2}\right)}
\newcommand{\compArb}[3]{\left({#1}\,\textcircled{\small$#3$}\,{#2}\right)}

\newcommand{\compSandwich}[2]{\left(#1\,\textcircled{$\mathfrak{s}$}\,#2\right)}

\newcommand{\fancyColor}{\mathfrak{c}}

\newcommand{\adjNum}{\mathfrak{a}}
\newcommand{\permNum}{\mathfrak{p}}
\newcommand{\hybridNum}{\mathfrak{h}}
\newcommand{\relaxedNum}{\mathfrak{r}}
\newcommand{\sandwichNum}{\mathfrak{s}}

\newcommand{\fancyFunc}[1]{\mathfrak{f}_{#1}}
\newcommand{\hatFancyFunc}[1]{\hat{\mathfrak{f}}_{#1}}
\newcommand{\cFunc}{\mathfrak{f}_{\fancyColor}}
\newcommand{\tcFunc}{\tilde{\mathfrak{f}}_{\fancyColor}}
\newcommand{\hatcFunc}{\hat{\mathfrak{f}}_{\fancyColor}}

\newcommand{\compNest}[2]{\left({\mathfrak{a}_{\relaxedNum}}^{({#1})}({#2})\right)}

\usepackage[normalem]{ulem}

\definecolor{hgreen}{rgb}{0,0.545,0}
\definecolor{hblue}{rgb}{0,0,0.545}
\definecolor{hred}{rgb}{0.475,0.0,0.15}

\newcommand{\fourgraph}[4]{ {
\begin{tikzpicture}[baseline=(current  bounding  box.center)]
\begin{feynman}
\vertex (a1) at (-1, -0.6) {\(#1\)};
\vertex (a2) at (-1, 0.6) {\(#2\)};
\vertex (mid1) at (-1,0);
\vertex (mid2) at (0.3,0);
\vertex (a3) at (0.3, 0.6) {\(#3\)};
\vertex (a4) at (0.3,- 0.6) {\(#4\)};
\diagram{
(mid2) --[thick](a3),
(mid2) --[thick](a4),
(mid2) -- [thick](mid1),
(a1)--[thick](mid1),
(a2) -- [thick](mid1)
};
\end{feynman}
\end{tikzpicture}
}
}

\newcommand{\fourgraphE}[4]{ 
{
\begin{tikzpicture}[baseline=(current  bounding  box.center)]
\begin{feynman}
\vertex (a1) at (-1, -0.6) {\(#1\)};
\vertex (a2) at (-1, 0.6) {\(#2\)};
\vertex (mid1) at (-1,0);
\vertex (mid2) at (0.3,0);
\vertex (a3) at (0.3, 0.6) {\(#3\)};
\vertex (a4) at (0.3,- 0.6) {\(#4\)};
\diagram {
(mid2) --[thick](a3),
(mid2) --[thick](a4),
(mid2) -- [line width=4pt,hgreen](mid1),
(a3)--[thick](mid2)-- [thick] (a4),
(a1)--[thick](mid1) -- [thick] (a2)
};
\end{feynman}
\end{tikzpicture}
}
}

\newcommand{\contactFour}[4]{ {
\begin{tikzpicture}[baseline=(current  bounding  box.center)]
\begin{feynman}
\vertex (a1) at (-1, -0.6) {\(#1\)};
\vertex (a2) at (-1, 0.6) {\(#2\)};
\vertex (mid1) at (-.35,0);
\vertex (a3) at (0.3, 0.6) {\(#3\)};
\vertex (a4) at (0.3,- 0.6) {\(#4\)};
\diagram{
(mid1)--[thick](a1),
(mid1)--[thick](a2),
(mid1)--[thick](a3),
(mid1)--[thick](a4)
};
\end{feynman}
\end{tikzpicture}
}
}

\newcommand{\fivegraph}[5]{ {
\begin{tikzpicture}[baseline=(current  bounding  box.center)]
\begin{feynman}
\vertex (a1) at (-1, -0.9) {\(#1\)};
\vertex (a2) at (-1, 0.9) {\(#2\)};
\vertex (mid1) at (-1,0);
\vertex (mid2) at (0.3,0);
\vertex (mid3) at (1.6, 0);
\vertex (a5) at (0.3, 0.9) {\(#3\)};
\vertex (a3) at (1.6, 0.9) {\(#4\)};
\vertex (a4) at (1.6, -0.9) {\(#5\)};
\diagram{
(mid2) --[thick](a5),
(mid2) -- [thick](mid3),
(a1)--[thick](mid1),
(a2) -- [thick](mid1), 
(mid1)--[thick](mid2), 
(mid3)--[thick](a3), 
(a4) -- [thick](mid3)
};
\end{feynman}
\end{tikzpicture}
}
}

\newcommand{\quadCubeFive}[5]{ {
\begin{tikzpicture}[baseline=(current  bounding  box.center)]
\begin{feynman}
\vertex (a1) at (-.1, -0.9) {\(#1\)};
\vertex (a2) at (-1, 0) {\(#2\)};
\vertex (mid2) at (-0.1,0);
\vertex (mid3) at (1.2, 0);
\vertex (a5) at (-.1, 0.9) {\(#3\)};
\vertex (a3) at (1.2, 0.9) {\(#4\)};
\vertex (a4) at (1.2, -0.9) {\(#5\)};
\diagram{
(mid2) --[thick](a5),
(mid2) -- [thick](mid3),
(a1)--[thick](mid2),
(a2) -- [thick](mid2),
(mid3)--[thick](a3), 
(a4) -- [thick](mid3)
};
\end{feynman}
\end{tikzpicture}
}
}

\newcommand{\quadCubeFiveL}[5]{ {
\begin{tikzpicture}[baseline=(current  bounding  box.center)]
\begin{feynman}
\vertex (a1) at (-.1, -0.9) {\(#1\)};
\vertex (a2) at (-1, 0) {\(#2\)};
\vertex [empty dot, ultra thick, hblue] (mid2) at (-0.1,0) {};
\vertex (mid3) at (1.2, 0);
\vertex (a5) at (-.1, 0.9) {\(#3\)};
\vertex (a3) at (1.2, 0.9) {\(#4\)};
\vertex (a4) at (1.2, -0.9) {\(#5\)};
\diagram{
(mid2) --[thick](a5),
(mid2) -- [thick](mid3),
(a1)--[thick](mid2),
(a2) -- [thick](mid2),
(mid3)--[thick](a3), 
(a4) -- [thick](mid3)
};
\end{feynman}
\end{tikzpicture}
}
}

\newcommand{\quadCubeFiveR}[5]{ {
\begin{tikzpicture}[baseline=(current  bounding  box.center)]
\begin{feynman}
\vertex (a1) at (-.1, -0.9) {\(#1\)};
\vertex (a2) at (-1, 0) {\(#2\)};
\vertex (mid2) at (-0.1,0);
\vertex [crossed dot, ultra thick, hgreen] (mid3) at (1.2, 0) {};
\vertex (a5) at (-.1, 0.9) {\(#3\)};
\vertex (a3) at (1.2, 0.9) {\(#4\)};
\vertex (a4) at (1.2, -0.9) {\(#5\)};
\diagram{
(mid2) --[thick](a5),
(mid2) -- [thick](mid3),
(a1)--[thick](mid2),
(a2) -- [thick](mid2),
(mid3)--[thick](a3), 
(a4) -- [thick](mid3)
};
\end{feynman}
\end{tikzpicture}
}
}

\newcommand{\contactFive}[5]{ {
\begin{tikzpicture}[baseline=(current  bounding  box.center)]
\begin{feynman}
\vertex [empty dot, ultra thick, hblue] (mid2) at (0,0) {};
\vertex (a1) at (-0.58778525229, -0.80901699437) {\(#1\)};
\vertex (a2) at (-0.9510565,0.30901699437) {\(#2\)};
\vertex (a3) at (0,1) {\(#3\)};
\vertex (a4) at (0.9510565,0.30901699437) {\(#4\)};
\vertex (a5) at (0.58778525229, -0.80901699437) {\(#5\)};
\diagram{
(mid2)--[thick](a1),
(mid2)--[thick](a2),
(mid2)--[thick](a3),
(mid2)--[thick](a4), 
(mid2)--[thick](a5) 
};
\end{feynman}
\end{tikzpicture}
}
}

\newcommand{\contactFiveNoBlue}[5]{ {
\begin{tikzpicture}[baseline=(current  bounding  box.center)]
\begin{feynman}
\vertex (mid2) at (0,0);
\vertex (a1) at (-0.58778525229, -0.80901699437) {\(#1\)};
\vertex (a2) at (-0.9510565,0.30901699437) {\(#2\)};
\vertex (a3) at (0,1) {\(#3\)};
\vertex (a4) at (0.9510565,0.30901699437) {\(#4\)};
\vertex (a5) at (0.58778525229, -0.80901699437) {\(#5\)};
\diagram{
(mid2)--[thick](a1),
(mid2)--[thick](a2),
(mid2)--[thick](a3),
(mid2)--[thick](a4), 
(mid2)--[thick](a5) 
};
\end{feynman}
\end{tikzpicture}
}
}

\newcommand{\fourgraphL}[4]{ 
{
\begin{tikzpicture}[baseline=(current  bounding  box.center)]
\begin{feynman}
\vertex (a1) at (-1, -0.6) {\(#1\)};
\vertex (a2) at (-1, 0.6) {\(#2\)};
\vertex[crossed dot,ultra thick, hgreen](mid1)at (-1,0){};
\vertex (mid2) at (0.3,0);
\vertex (a3) at (0.3, 0.6) {\(#3\)};
\vertex (a4) at (0.3,- 0.6) {\(#4\)};
\diagram {
(mid2) --[thick](a3),
(mid2) --[thick](a4),
(mid2) -- [thick](mid1),
(a1)--[thick](mid1)[crossed dot,green] -- [thick] (a2),
(a2) -- [thick](mid1)
};
\end{feynman}
\end{tikzpicture}
}
}

\newcommand{\fourgraphR}[4]{ 
{
\begin{tikzpicture}[baseline=(current  bounding  box.center)]
\begin{feynman}
\vertex (a1) at (-1, -0.6) {\(#1\)};
\vertex (a2) at (-1, 0.6) {\(#2\)};
\vertex (mid1) at (-1,0);
\vertex[crossed dot,ultra thick, hgreen](mid2) at (0.3,0){};
\vertex (a3) at (0.3, 0.6) {\(#3\)};
\vertex (a4) at (0.3,- 0.6) {\(#4\)};
\diagram {
(mid2) --[thick](a3),
(mid2) --[thick](a4),
(mid2) -- [thick](mid1),
(a3)--[thick](mid2)[crossed dot,hgreen] -- [thick] (a4),
(a1)--[thick](mid1) -- [thick] (a2)
};
\end{feynman}
\end{tikzpicture}
}
}

\newcommand{\fivegraphL}[5]{ 
{
\begin{tikzpicture}[baseline=(current  bounding  box.center)]
\begin{feynman}
	\vertex (a1) at (-1, -0.9) {\(#1\)};	
	\vertex (a2) at (-1, 0.9) {\(#2\)};
	\vertex[crossed dot,ultra thick, hgreen](mid1) at (-1,0) {};
	\vertex (mid2) at (0.3,0);
	\vertex (mid3) at (1.6, 0);
	\vertex (a5) at (0.3, 0.9) {\(#3\)};
	\vertex (a3) at (1.6, 0.9) {\(#4\)};
	\vertex (a4) at (1.6, -0.9) {\(#5\)};
\diagram {
(mid2) --[thick](a5),
(mid2) -- [thick](mid3),
(a1)--[thick](mid1)[crossed dot,green] -- [thick] (a2),
(mid1)[crossed dot,green]--[thick](mid2), 
(mid3)--[thick](a3), 
(a4) -- [thick](mid3)
};
\end{feynman}
\end{tikzpicture}
}
}

\newcommand{\fivegraphR}[5]{ 
{
\begin{tikzpicture}[baseline=(current  bounding  box.center)]
\begin{feynman}
	\vertex (a1) at (-1, -0.9) {\(#1\)};	
	\vertex (a2) at (-1, 0.9) {\(#2\)};
	\vertex (mid1) at (-1,0);
	\vertex (mid2) at (0.3,0);
	\vertex[crossed dot,ultra thick, hgreen] (mid3) at (1.6, 0) {};
	\vertex (a5) at (0.3, 0.9) {\(#3\)};
	\vertex (a3) at (1.6, 0.9) {\(#4\)};
	\vertex (a4) at (1.6, -0.9) {\(#5\)};
\diagram {
(mid2) --[thick](a5),
(mid2) -- [thick](mid3),
(a1)--[thick](mid1) -- [thick] (a2),
(mid1)--[thick](mid2), 
(mid3)--[thick](a3), 
(a4) -- [thick](mid3)
};
\end{feynman}
\end{tikzpicture}
}
}

\newcommand{\fivegraphM}[5]{ 
{
\begin{tikzpicture}[baseline=(current  bounding  box.center)]
\begin{feynman}
	\vertex (a1) at (-1, -0.9) {\(#1\)};	
	\vertex (a2) at (-1, 0.9) {\(#2\)};
	\vertex (mid1) at (-1,0);
	\vertex [crossed dot,ultra thick, hgreen] (mid2) at (0.3,0) {};
	\vertex (mid3) at (1.6, 0);
	\vertex (a5) at (0.3, 0.9) {\(#3\)};
	\vertex (a3) at (1.6, 0.9) {\(#4\)};
	\vertex (a4) at (1.6, -0.9) {\(#5\)};
\diagram {
(mid2) --[thick](a5),
(mid2) --[thick](mid3),
(a1)--[thick](mid1) -- [thick] (a2),
(mid1)--[thick](mid2), 
(mid3)--[thick](a3), 
(a4) -- [thick](mid3)
};
\end{feynman}
\end{tikzpicture}
}
}

\newcommand{\fivegraphMS}[5]{ 
{
\begin{tikzpicture}[baseline=(current  bounding  box.center)]
\begin{feynman}
	\vertex (a1) at (-1, -0.9) {\(#1\)};	
	\vertex (a2) at (-1, 0.9) {\(#2\)};
	\vertex (mid1) at (-1,0);
	\vertex [empty dot, ultra thick, hblue] (mid2) at (0.3,0) {};
	\vertex (mid3) at (1.6, 0);
	\vertex (a5) at (0.3, 0.9) {\(#3\)};
	\vertex (a3) at (1.6, 0.9) {\(#4\)};
	\vertex (a4) at (1.6, -0.9) {\(#5\)};
\diagram {
(mid2) --[thick](a5),
(mid2) --[thick](mid3),
(a1)--[thick](mid1) -- [thick] (a2),
(mid1)--[thick](mid2), 
(mid3)--[thick](a3), 
(a4) -- [thick](mid3)
};
\end{feynman}
\end{tikzpicture}
}
}

\newcommand{\fivegraphEL}[5]{ 
{
\begin{tikzpicture}[baseline=(current  bounding  box.center)]
\begin{feynman}
	\vertex (a1) at (-1, -0.9) {\(#1\)};	
	\vertex (a2) at (-1, 0.9) {\(#2\)};
	\vertex (mid1) at (-1,0);
	\vertex (mid2) at (0.3,0);
	\vertex (mid3) at (1.6, 0);
	\vertex (a5) at (0.3, 0.9) {\(#3\)};
	\vertex (a3) at (1.6, 0.9) {\(#4\)};
	\vertex (a4) at (1.6, -0.9) {\(#5\)};
\diagram {
(mid2) --[thick](a5),
(mid2) --[thick](mid3),
(a1)--[thick](mid1) -- [thick] (a2),
(mid1)--[line width=4pt,hgreen](mid2), 
(mid3)--[thick](a3), 
(a4) -- [thick](mid3)
};
\end{feynman}
\end{tikzpicture}
}
}

\newcommand{\fivegraphER}[5]{ 
{
\begin{tikzpicture}[baseline=(current  bounding  box.center)]
\begin{feynman}
	\vertex (a1) at (-1, -0.9) {\(#1\)};	
	\vertex (a2) at (-1, 0.9) {\(#2\)};
	\vertex (mid1) at (-1,0);
	\vertex (mid2) at (0.3,0);
	\vertex (mid3) at (1.6, 0);
	\vertex (a5) at (0.3, 0.9) {\(#3\)};
	\vertex (a3) at (1.6, 0.9) {\(#4\)};
	\vertex (a4) at (1.6, -0.9) {\(#5\)};
\diagram {
(mid2) --[thick](a5),
(mid2) --[line width=4pt,hgreen](mid3),
(a1)--[thick](mid1) -- [thick] (a2),
(mid1)--[thick](mid2), 
(mid3)--[thick](a3), 
(a4) -- [thick](mid3)
};
\end{feynman}
\end{tikzpicture}
}
}

\newcommand{\contactFiveNew}[5]{ {
\begin{tikzpicture}[baseline=(current  bounding  box.center)]
\begin{feynman}
\vertex [dot, ultra thick, teal] (mid2) at (0,0) {};
\vertex (a1) at (-0.58778525229, -0.80901699437) {\(#1\)};
\vertex (a2) at (-0.9510565,0.30901699437) {\(#2\)};
\vertex (a3) at (0,1) {\(#3\)};
\vertex (a4) at (0.9510565,0.30901699437) {\(#4\)};
\vertex (a5) at (0.58778525229, -0.80901699437) {\(#5\)};
\diagram{
(mid2)--[thick](a1),
(mid2)--[thick](a2),
(mid2)--[thick](a3),
(mid2)--[thick](a4), 
(mid2)--[thick](a5) 
};
\end{feynman}
\end{tikzpicture}
}
}

\newcommand{\quadCubeFiveNew}[5]{ {
\begin{tikzpicture}[baseline=(current  bounding  box.center)]
\begin{feynman}
\vertex (a1) at (-.1, -0.9) {\(#1\)};
\vertex (a2) at (-1, 0) {\(#2\)};
\vertex [dot, ultra thick, teal] (mid2) at (-0.1,0) {};
\vertex [empty dot, ultra thick, magenta] (mid3) at (1.2, 0) {};
\vertex (a5) at (-.1, 0.9) {\(#3\)};
\vertex (a3) at (1.2, 0.9) {\(#4\)};
\vertex (a4) at (1.2, -0.9) {\(#5\)};
\diagram{
(mid2) --[thick](a5),
(mid2) -- [thick](mid3),
(a1)--[thick](mid2),
(a2) -- [thick](mid2),
(mid3)--[thick](a3), 
(a4) -- [thick](mid3)
};
\end{feynman}
\end{tikzpicture}
}
}

\AtBeginDocument{
\heavyrulewidth=.08em
\lightrulewidth=.05em
\cmidrulewidth=.03em
\belowrulesep=.65ex
\belowbottomsep=0pt
\aboverulesep=.4ex
\abovetopsep=0pt
\cmidrulesep=\doublerulesep
\cmidrulekern=.5em
\defaultaddspace=.5em
}

\title{Composing Effective Prediction at Five Points}
\author[a,b]{John Joseph M. Carrasco}
\author[b,c]{Laurentiu Rodina}
\author[a]{Suna Zekio\u{g}lu}

\affiliation[a]{Department of Physics \& Astronomy, Northwestern University, Evanston, Illinois 60208, USA}
\affiliation[b]{Institut de Physique Theorique, Universite Paris Saclay, CEA, CNRS, F-91191 Gif-sur-Yvette, France}
\affiliation[c]{Department of Physics \& Astronomy, National Taiwan University, Taipei 10617, Taiwan}

\date{\today}
\abstract{Color-kinematics duality in the adjoint has proven key to the relationship between gauge and gravity theory scattering amplitude predictions. In recent work, we demonstrated that at four-point tree-level, a small number of color-dual  EFT building blocks could encode all higher-derivative single-trace massless corrections to gauge and gravity theories compatible with adjoint double-copy. One critical aspect was the trivialization of building higher-derivative color-weights --- indeed, it is the mixing of kinematics with non-adjoint-type color-weights (like the permutation-invariant $d^4$) which permits description via adjoint double-copy. Here we find that such ideas clarify the predictions of local five-point higher-dimensional operators as well.   We demonstrate how a single scalar building block can be combined with color structures to build higher-derivative color factors that generate, through double copy, the predictions of higher-derivative gauge-theory operators. These may then be suitably mapped, through another double-copy, to higher-derivative corrections in gravity.}

\preprint{NUHEP-TH/21-02}

\begin{document}
\maketitle
   \tableofcontents
\input intro.tex

\input review.tex

\input hdFourPoint.tex

\input algebraicStructure.tex

\input colorAndScalar.tex

\input hdCorr.tex

\input otherDualities.tex

\input gaugeGravityCorr.tex

\input fixingToStrings.tex

\input conclusion.tex

\acknowledgments
We are grateful to Yu-tin Huang, Henrik Johansson, Ian Low, Nic Pavao, Radu Roiban, Oliver Schlotterer,  Aslan Seifi, Bogdan Stoica, and Ingrid Vazquez-Holm for interesting discussion of related subjects. We are additionally grateful for especially helpful comments   by Oliver Schlotterer on an earlier draft.  This work is supported by the Department of Energy under Award Number DE-SC0021485. LR is supported by Taiwan Ministry of Science and Technology Grant No. 109-2811-M-002-523. SZ would like to acknowledge the Northwestern University Amplitudes and Insight group, Department of Physics and Astronomy, and Weinberg College for support. 

\appendix

\input buildingCompAppendix.tex

\input colorStructureAppendix.tex

\input operatorAppendix.tex

\input explicitCompAppendix.tex

\bibliographystyle{JHEP}
\bibliography{fivePoints}
\end{document}

%% file: intro.tex
\section{Introduction}

  Identifying independent higher-derivative operators and calculating their predictions can be technically challenging, especially in gauge and gravity theories. Such a challenge naturally invites the application of novel techniques used to reveal hidden simplicity and structure in scattering amplitudes (see, e.g.~\cite{Elvang:2016qvq,Shadmi:2018xan,Cheung:2018oki,Kampf:2019mcd,Elvang:2019twd,Ma:2019gtx,Durieux:2019eor,Bern:2019wie,Bern:2020ikv,Durieux:2020gip,Rodina:2021isd,Arkani-Hamed:2020blm,Bern:2021ppb} and references therein).  Here we look to color-dual building blocks, discovering new structure at five points that we expect to generalize to all multiplicity.

The task of calculating gravitational amplitudes has been remarkably simplified by the discovery of color-dual double copy structure \cite{BCJ, BCJLoop}, exploiting the fact that gravity predictions are entirely encoded by the kinematic information of gauge theory amplitudes~\cite{KLT,BjerrumBohr:2009rd,Stieberger:2009hq}. Gauge theory amplitudes themselves are heavily constrained by the duality between color and kinematics, the statement that the color weights and kinematic numerators contributing to gauge theory amplitudes obey the same algebraic relations. This duality allows for the double copy procedure of replacing color weights with additional kinematic weights, generating a gravitational amplitude.

Could the calculation of higher-derivative gauge and gravity counterterms be vastly simplified by exploiting double-copy? Notably,  at four points \cite{Carrasco:2019yyn}  the answer is a definitive yes. Indeed, outside of a small number of vector building blocks, we reduce the higher-derivative adjoint-compatible gauge theory problem to one of manipulating scalar weights.   Along the way, we clarify how operators whose predictions involve the permutation-invariant color structure $d^{abcd}$ can be completely compatible with adjoint-double copy, despite the non-adjoint nature of their color weights. One must simply compensate for the color's different algebraic properties by including kinematic functions in the same graph dressing; we identify these adjoint combinations of color with kinematics as higher-derivative color weights.   This resolves a potentially subtle point --- adjoint- color-kinematics duality can be made manifest in these amplitudes, even if the coefficients of independent color traces within the full amplitude do not satisfy the so-called Bern-Carrasco-Johansson (BCJ), or $(m-3)!$-basis, field-theory ordered amplitude relations. Such amplitudes still admit an adjoint-double copy description, as they may be expressed as a sum over cubic graphs, each dressed with two adjoint-dual weights --- it is just that one of these weights is not pure color, but rather color conspiring with kinematics to satisfy adjoint relations.  In this current manuscript, we show that such simplification is very much present for the predictions of five-field operators as well.   

The color-dual structure of the open superstring serves as a proof of concept for pulling the complexity of higher-derivative gauge theory corrections into much simpler scalar corrections. Tree-level open superstring amplitudes may be understood as a field-theory double copy between super-Yang-Mills theory and a bicolored scalar theory known as Z-theory, which encodes all string higher-derivative corrections~\cite{Broedel2013tta,Carrasco2016ldy,Carrasco2016ygv,Mafra2016mcc}.  In ref.~\cite{Carrasco:2019yyn}, we computed the scalar corrections needed to span the predictions in the low-energy expansion of the open superstring (and, through double copy, the closed superstring as well) at four-point tree level via a constructive approach that focused on the principles of color-kinematics duality, gauge invariance, locality, and unitarity.   

It is worth taking the time to emphasize what may seem like a peculiar feature of our approach. Tree-level scattering admits a type of simplification that we choose to avoid.  At tree-level, functional algebraic constraints like the kinematic Jacobi-identity for adjoint structures can relax to linear constraints on a wider set of elements, if we allow weights of the same topology but with different labels to have a different functional form.  This has advantages, and indeed allows for the closed-form all-multiplicity momentum-kernel, or KLT representation of double-copy relationships.   The penalty for this approach is a factorial growth in the number of functions one must specify at each multiplicity --- which, in each case, takes a product with the infinite tower of potential higher-derivative corrections!  Instead, here we employ an approach that is indeed required for multiloop calculations --- allowing each topology to get its own functional weight.  The cost is functional algebraic constraints to satisfy color-kinematics duality, but the gain is the factorial decrease in the number of topologies needed to form a basis to describe every contribution.  At four loops in the maximally supersymmetric case at four-points, where over eighty topologies may contribute, each with relabelings, only one non-planar topology is required to span the entire integrand through functional Jacobi relations~\cite{Bern:2012uf}.  For tree-level, provably to all multiplicity,  one requires only the single functional weight of the half-ladder (multiperipheral) topology for each multiplicity to span all topologies through Jacobi relations.   

The cornerstone of our strategy is to target the construction of  higher-derivative color weights, which mix both color structure and scalar kinematics into single numerator factors that obey adjoint relations. Their adjoint properties render them appropriate building blocks for adjoint double copy with vector kinematic weights (such as from Yang-Mills) to generate gauge and gravity corrections. The procedure of building these modified color factors was guided at four points by the relevant algebraic structures: adjoint (antisymmetry about vertices and Jacobi-relations about edges of cubic graphs) and permutation invariance. Functional graph weights satisfying these algebraic constraints, either in terms of pure color information or scalar kinematics, proved to be natural building blocks for the construction of adjoint higher-derivative color weights. 

We emphasize that the key innovation of \cite{Carrasco:2019yyn} at four-point tree-level was the recognition that these types of functional weights can be composed to generate new functions that satisfy the same functional algebraic constraints.   With a unit-step  in Mandelstam invariants, one can then climb the infinite ladder of higher-derivative corrections rung by rung without needing to resort to an ansatz at each level to satisfy the functional relations. Because kinematic permutation invariants close to a basis at finite order, there are only a finite number of distinct building-blocks one must identify through composition before allowing permutation invariants to span all higher-dimension predictions.  We demonstrate here that such a kinematic unit building-block exists at five-points with a wide variety of consequences which we explore in detail in this paper.  Indeed, with the pattern at four and five-points, it becomes clear that the functional kinematic weight at the heart of Jacobi-satisfying relationships for the unit-step half-ladder for arbitrary multiplicity $m>3$ should take the following simple form:
\begin{equation}
\label{mPointConjecture}
n_{\rm unit}(1|\cdots|m) \propto (k_1 -k_2) \cdot (k_m-k_{m-1})\,.
\end{equation} 
Jacobi relations are trivially satisfied on all edges, antisymmetry is manifest algebraically at the two terminal vertices, and the only question is whether the kinematic weight is odd or even under the reflection between these terminal vertices.  At odd multiplicity, we are forced to relax the adjoint condition to allow symmetry about the middle-vertex -- yielding a different algebraic structure than adjoint (one, e.g., with fewer basis elements), but one that can be used to construct the color-dual required adjoint-type structures through composition.  At even multiplicity, one satisfies all adjoint-type algebraic relations with this simple building block.   All order construction via relaxed adjoint for odd-multiplicity in the specific case of five points is treated in detail in this paper, but we expect this to generalize to all odd multiplicity.

In this manuscript, we demonstrate that such a simple scalar structure allows the constructive building of  higher-derivative adjoint-type color-weight corrections at five-point tree level. We see an abundance of algebraic structures give rise to many distinct types of color and composition rules for generating adjoint modified color factors. These structures, along with their associated color and scalar kinematic building blocks, prove to be sufficient for writing down appropriately factorizing five-point tree-level supersymmetry-compatible gauge theory corrections; we conjecture that only a finite number of building blocks is necessary to span all higher-derivative corrections compatible with color-kinematics duality and maximal supersymmetry. Additionally, we uncover a novel form of color-kinematics duality for local amplitudes, which may be striated into alternatively an adjoint double copy structure or double copy forms defined by the new algebraic structures appearing at five points. We may construct such local solutions using the same handful of building blocks needed to write down factorizing corrections.  These constructed factorizing and local amplitudes are expected to e.g. span the entirety of the $\ap$ expansion of the open superstring, which we explicitly verify through the ninth order in $\ap$ and present in associated ancillary computer readable files~\cite{ancFiles}.  

The paper is organized as follows. We begin with a review of amplitude properties and color-kinematics duality in \sect{section:Review}, and then summarize results from our previous studies of higher-derivative counterterms at four points in \sect{section:fourPoints}. Algebraic structures and compositions at five points are enumerated in \sect{section:algebra5pt}, and then used to generate color and scalar kinematic solutions in \sect{section:buildingBlocks}. We combine these results straightforwardly to find factorizing  and local higher-derivative color factors in \sect{section:hdCorr}. New methods for constructing color-dual local solutions are introduced in \sect{section:doublyDualLocal}.  We  discuss how these higher-derivative color-weights can be exploited to build higher-derivative gauge and gravity corrections in \sect{gaugeGravCorr}.  We demonstrate that our constructive predictions can be used to span the low-energy expansion of open superstring theory amplitudes in \sect{section:Strings}.   We discuss these results and look to future avenues of exploration in \sect{section:Conclusion}.

We provide ancillary Mathematica files hosted, with versioning, on github~\cite{ancFiles}.   We discuss our approach for discovering functional algebraic composition rules in \app{appendix:BuildingComposition}, tabulate our explicit color-basis for five points in \app{appendix:colorStructure}, offer a pedagogic example of operator matching in \app{appendix:operatorComparison},  as well as the explicit composition formulae relevant at five points in \app{appendix:ExplicitComposition}.

%% file: review.tex
\section{Review of the Duality Between Color and Kinematics}
\label{section:Review}

In this section, we briefly review features of the adjoint color-kinematics duality and its associated double-copy construction~\cite{BCJ,BCJLoop}.  For more details on double-copy in general and its applications, we refer the interested reader to the recent review~\cite{BCJreview}.
 
\subsection{Color-dressed Yang-Mills amplitudes}
We can express each full (or color-dressed) $m$-point tree-level Yang-Mills theory amplitude as a sum over the contributions of $(2m-5)!!$ individual cubic (or trivalent) graphs: 
\begin{equation}
\mathcal{A}_{m}^{\text{YM}} = \sum_{g \in \Gamma_3^{(m)}} \frac{n^{\text{YM}}_g c_g}{d_g}
\end{equation}
where the sum runs over all graphs containing $m$ external edges and only cubic vertices, $\Gamma^{(m)}_3$. In Yang-Mills, the \textit{color factor} $c_g$ associated with each graph is given simply by dressing each vertex with an appropriate $f^{abc}$ structure constant, and each internal edge with a Kronecker delta in color indices.   We use $n_g$ to represent the {\em kinematic numerator weight} corresponding to a given graph, encoding external state information and the tensor structure of the theory in terms of Lorentz dot products $\left(k_i \cdot k_j \right)$, $\left(k_i \cdot \varepsilon_j \right)$, and $\left(\varepsilon_i \cdot \varepsilon_j \right)$. Finally, $d_g$ are the (massless) cubic propagators of each graph $g$. 

For example, the color-dressed four point amplitude can be written as: 
\begin{equation}
\mathcal{A}_{4}^{\text{YM}} = \frac{n^{\text{YM}}_s c_s}{s} +  \frac{n^{\text{YM}}_t c_t}{t} +  \frac{n^{\text{YM}}_u c_u}{u}
\end{equation}
where $s$, $t$, and $u$ are the four point momentum invariants $s = s_{12} = (k_1 + k_2)^2$, $t = s_{23} = (k_2 + k_3)^2$, and $u = -s -t =  s_{13} = (k_1 + k_3)^2$, with all momenta massless and outgoing. These graphs may be equivalently labeled in terms of functional numerators, as $n_s = n(1234)$, $n_t = n(4123)$, and $n_u = n(3142)$, respectively. The four point contact term of  Yang-Mills has been associated with the three cubic graphs by multiplying its contribution by appropriate factors of unity, like $s/s$. 

\subsection{Adjoint color properties}

By virtue of the Lie algebra of the gauge group, the adjoint color factors $c_g$ obey antisymmetry around all cubic vertices and Jacobi identities on all internal edges: 
\begin{equation}
f^{abc} = -f^{acb}
\end{equation} 
\begin{equation}
f^{abe}f^{ecd} = f^{dae}f^{ebc} + f^{dbe}f^{eca}
\end{equation}
These relations hold at all multiplicity (correspondingly, $a,b,c,d$ need not be external color labels). Returning to our four point example, the Jacobi identity simply relates the color factors of the three channels: 
\begin{equation}
c_s = c_t + c_u
\end{equation} 
where $c(abcd) = f^{abe}f^{ecd}$ and $c_s = c(1234)$, $c_t = c(4123)$, and $c_u = c(3142)$. 

\subsection{Ordered amplitudes}
The full amplitude $\mathcal{A}$ is gauge invariant and obeys the Ward identity $\mathcal{A}|_{\varepsilon_i \rightarrow k_i} = 0$. There is no such condition on the kinematic numerators for individual graphs: $n_g$ vary under gauge transformations. Gauge invariant quantities can, however, be constructed from kinematic numerators and propagators by expressing the full amplitude in a basis of color factors using the Jacobi identity.   In our four point example, we can express $c_t$ in terms of a basis of $c_u$ and $c_s$, yielding,  
\begin{equation}
\begin{split} 
\mathcal{A}_{4}^{\text{YM}} &= c_s \left( \frac{n_s}{s} + \frac{n_t}{t} \right) +  c_u \left( \frac{n_u}{u} - \frac{n_t}{t} \right) \\
&= c_s A(1234) + c_u A(1423) 
\end{split}
\end{equation} 
The coefficient of an independent color basis element must be itself gauge invariant. Such structures arises at every multiplicity once color weights are expressed in a minimal color-basis.  The gauge-invariant coefficients of the basis color weights are referred to as \textit{ordered amplitudes}, and can be parameterized by a fixed external ordering.  

In our four-point example, we may surmise that while $n_s$ and $n_t$ both vary under gauge transformations, they must do so commensurately in such a way there is no effect on the physical quantity $A(1234)$. In this sense, it is the fact that color factors $c_g$ obey the Jacobi identity that allows gauge-dependent terms to cancel between the distinct graph numerators $n_g$, thereby ensuring gauge invariance of the overall amplitude. This ability to shift kinematic numerators in a manner that does not affect the physical amplitude, known as generalized gauge freedom, means that kinematic numerators are not unique; we will shortly see that certain representations (which do not necessarily line up with the representation obtained via Feynman rules) of these numerators make particularly useful properties manifest. 

For $SU(N_c)$ charged theories, the ordered amplitudes like $A(1234)$ may be uncovered not only by aligning the full amplitude along a basis of graph color weights (like $c_s$ and $c_u$ above), but also by writing the color factors in a basis of traces over generators and reading off the coefficients:
\begin{equation}
\mathcal{A}_m = \sum_{\sigma \in S_{m-1}} \Tr\left(T^{a_{\sigma_1}}T^{a_{\sigma_2}} \cdots T^{a_m}\right) A(\sigma_1,\sigma_2, \cdots, m) \, .
\end{equation} 
By virtue of the cyclicity of the color traces, the ordered amplitudes must also obey cyclic invariance. 

In particular, we will often refer to the following five-point ordered amplitude: 
\begin{equation}
\label{orderedFivePointAmplitude}
A(12345) = \frac{n(12345)}{s_{12}s_{45}} + \frac{n(23451)}{s_{23}s_{15}} + \frac{n(34512)}{s_{12}s_{34}} + \frac{n(45123)}{s_{23}s_{45}} + \frac{n(51234)}{s_{15}s_{34}}
\end{equation}

\subsection{Color-kinematics duality and double copy}
Adjoint color-kinematics duality states that kinematic numerators obey the same algebraic properties ---  antisymmetry and Jacobi --- as the adjoint color factors: 
\begin{equation}
\label{adjointFourPioint}
\begin{split}
c(abc) = -c(acb) &\quad\Leftrightarrow\quad n(abc) = -n(acb)  \\
c_s = c_t + c_u &\quad\Leftrightarrow\quad n_s = n_t + n_u 
\end{split} 
\end{equation} 
Generic four-point cubic representations of Yang-Mills satisfy this kinematic Jacobi identity independent of gauge.  At higher multiplicity, not all representations are guaranteed to make this duality manifest, but by virtue of generalized gauge freedom, it is always~\cite{KiermaierTalk} possible to rearrange a given set of numerators into a {\em color-dual representation} that satisfies the corresponding kinematic Jacobi identities. 

At tree level, adjoint color-kinematics duality may be equivalently expressed in terms of the $(m-3)!$-basis BCJ relations between ordered amplitudes, 
\begin{equation}
\sum_{i=2}^{m-1} k_1 \cdot \left(k_2 + \cdots + k_i \right) A_m(2, \cdots, i, 1, i + 1, \cdots, m)  = 0
\end{equation}
We will use this to impose adjoint color-kinematics duality in cases where ordered amplitudes prove to be more closely related to the most natural choice of building blocks than numerators. At four points, these relations amount to demanding the permutation invariance of the quantity $A(abcd)/s_{ac}$. 

Guided by the knowledge that the algebraic properties obeyed by the color factors ensure the gauge invariance of the amplitude, \textit{double copy construction} allows us to build linearized diffeomorphism invariant gravity amplitudes $\mathcal{M}$ from gauge theory amplitudes $\mathcal{A}$ by replacing the color factors with kinematic numerators that also satisfy the same algebraic properties of antisymmetry and Jacobi identities: 
\begin{equation}
\mathcal{A} = \sum_g \frac{n_g c_g}{d_g}  \quad \xrightarrow{c \rightarrow \tilde{n}}  \quad
\mathcal{M} = \sum_g \frac{n_g \tilde{n}_g}{d_g}
\end{equation}
where $n$ and $\tilde{n}$ need not be numerators from the same gauge theory.

In general, we can label any adjoint-type theory's double-copy structure by the weight each single-copy theory contributes graph by graph, i.e.:
\begin{equation}
\label{genDblCpy}
{\cal A}^{j\otimes k}\equiv \sum_g \frac{j_g k_g}{d_g}
\end{equation}

General higher derivative corrections to any field theory can also be written on cubic graphs, with any contact terms assigned to cubic graphs by multiplying by inverse propagators.   We will primarily be interested in theories that manifest an adjoint double-copy structure ${\cal A}^{j\otimes k}$; higher derivative corrections to such theories will require $j$ and/or $k$ to take on additional powers of momentum invariants.    The next section will review the progress  made at four-points.

%% file: hdFourPoint.tex
\section{Higher-Derivative Adjoint-Type Corrections at Four-Points}
\label{section:fourPoints}
In ref.~\cite{Carrasco:2019yyn}, we showed that we can exploit double copy structure to write down higher-derivative corrections to gauge theory and gravity. Gauge theory corrections may be entirely captured by constructing a double copy between adjoint vector numerators and a small number of simple objects that mix color structures and scalar kinematics into adjoint-color-dual \textit{modified color factors}. When this vector copy is simply Yang-Mills, we generate all local four-point higher-derivative corrections that contribute to the tree-level open superstring. More general higher-derivative vector corrections can be achieved by consideration of a spanning set of adjoint-type vector numerators, to be discussed later in this section.  

Each and every higher-derivative correction can be described in terms of \textit{two double-copy structures}: one involving adjoint-type weights dressing cubic graphs, and one involving permutation-invariant weights dressing the quartic contact. One may wonder about four-point mixed adjoint-symmetric weights, following e.g. the algebra of $f^3d^3$ color-weights, but as there are no (non-trivial) four-point scalar weights that obey such algebraic relations, we expect such structures not to be compatible with adjoint-type double-copy. This simplifies our task to finding building blocks compatible with either adjoint or permutation-invariant algebraic constraints so that we may construct amplitudes through double copy. We will frame the discussion of four-point higher-derivative corrections by introducing first the relevant graphs, then the algebras associated with each of them as well as composition rules, and finally building blocks involving single-trace color and scalar kinematic weights.

\subsection{Algebraic structures at four-points}

There are multiple double-copy representations of the same amplitude corresponding to striations along different algebraic properties. These striations may be associated with graphical representations. Here, we will discuss this idea in the context of four-point tree-level amplitudes for massless particles. 

\subsubsection{Graph representations}
Let's first establish a notation for graphs at four-points.  There is only one cubic graph topology, also known as the half-ladder, defined by one edge and two (cubic) vertices, whose labels we can specify with an ordered list:
\be
(abcd) \equiv \fourgraph{a}{b}{c}{d} \,.
\ee
Additionally, we have a contact graph with a quartic vertex:
\be
(\langle abcd \rangle) \equiv \contactFour{a}{b}{c}{d} \,.
\ee

\subsubsection{Adjoint algebraic constraints}

Cubic graphs $g=(abcd)$ are dressed by adjoint-type weights $\adjNum_g = \adjNum(abcd)$. These weights obey antisymmetry about each vertex and satisfy Jacobi relations on each internal edge.  The antisymmetry weights are expressed as: 
\begin{align}
\adjNum(abcd) &= -\adjNum(bacd)\, \text{:} &\text{antisymmetric around vertex}\hskip .5cm\fourgraphL{a}{b}{c}{d}\,,\\
\adjNum(abcd) &= -\adjNum(abdc)\, \text{:} &\text{antisymmetric around vertex}\hskip .5cm\fourgraphR{a}{b}{c}{d} \, ,
\end{align}
and the Jacobi-like constraint as:
\begin{equation}
\adjNum(abcd)  =\adjNum(bcda) + \adjNum(cadb) \, \text{:} ~~~~ \text{Jacobi relation about edge} \fourgraphE{a}{b}{c}{d} \, .
\end{equation}
These combine to generate the familiar 2-cycle equivalence relation:
\begin{equation}
\adjNum(abcd) = \adjNum(cdab) \, ,
\end{equation}
and symmetry under reversal:
\begin{equation}
\adjNum(abcd) = \adjNum(dcba) \, .
\end{equation}

An ordered partial amplitude with leg ordering $(abcd)$ is given by summing over the two cubic channels that could contribute to that ordering:
\begin{equation}
\label{ordered4pt}
A(abcd) = \frac{\adjNum(abcd)}{(k_a+k_b)^2} +\frac{\adjNum(bcda)}{(k_b+k_c)^2} \,.
\end{equation}
Under adjoint type-constraints, all 24 ways of dressing a cubic four-point graph collapse to the three Mandelstam channels. As a convenient shorthand, we will invoke Mandelstam labels as shorthand for graphs
\begin{align}
s&=(1234)\\
t&=(2341)\\
u&=(3142)
\end{align}
as well as propagators,
\begin{align}
s&=(k_1+k_2)^2=(k_3+k_4)^2\\
t&=(k_2+k_3)^2=(k_1+k_4)^2\\
u&=(k_1+k_3)^2=(k_2+k_4)^2,
\end{align}
where, via conservation of momentum for massless particles, $s+t+u=0$.  Using conservation of momentum and the graph dressing adjoint constraints, we may see that the ordered amplitudes satisfy the standard field theory relations of Yang-Mills ordered amplitudes:
\begin{align}
A(abcd) &= A(bcda)\,  && \text{cyclic symmetry,}\\
A(abcd) &= (-1)^4 A(dcba)\, && \text{reversal symmetry,}\\
A(abcd) &= - A(adbc)-A(acdb)\, && \text{$(n-2)!$ basis relations,}\\
\frac{A(abcd)}{(a+c)^2} &= \frac{A(adbc)}{(a+b)^2} =\frac{A(acdb)}{(a+d)^2}\, && \text{$(n-3)!$ basis relations.}
\end{align}
From the second relation, we see that we can simply label each ordered amplitude in terms of its propagators: $A(s_{ab},s_{bc})\equiv A(abcd)$.  From the last relation, we see that $A(s,t)/u$ is permutation invariant and thus so is $s t A(s,t) = t \adjNum_s + s \adjNum_t$.

Making adjoint double-copy structure manifest, a full amplitude from a theory described by adjoint-type graph weights $\adjNum_g$ and $\tilde{\adjNum}_g$ is given by summing over all three distinct Mandelstam channels:
\begin{equation}
\label{adjDoubleCopyAmp4}
{\cal A}_4 =  \adjNum\otimes\tilde{\adjNum}= \frac{\adjNum_s \tilde{\adjNum}_s}{s} +\frac{\adjNum_t\tilde{\adjNum}_t }{t}  +\frac{\adjNum_u\tilde{\adjNum}_u }{u}\,.
\end{equation}
This expression is manifestly permutation invariant, consistent with Bose-symmetry of the overall amplitude. By expressing both copies $\adjNum_g$ and $\tilde{\adjNum}_g$  in terms of a Jacobi-basis, and exploiting conservation of momentum, we can rewrite it to manifest permutation-invariant double copy form: 
\begin{equation}
\label{piDoubleCopyAmp4}
{\cal A}_4  =- \frac{ ( t \adjNum_s + s \adjNum_t ) ( t\tilde{\adjNum}_s + s \tilde{\adjNum}_t)}{s t u} = -\frac{ s t A(s,t) \times s t \tilde{A}(s,t)}{s t u} 
\end{equation}
so that both copies are encoded in permutation-invariant quantities.

\subsubsection{Permutation-invariant algebraic constraints}

The four-point quartic contact graph $g=(\langle abcd \rangle)$ is dressed by permutation-invariant type graph weights $\permNum_g$. These weights are completely symmetric under all permutations, so we will often omit the graphical argument to permutation invariant weights.

By virtue of the permutation invariance of the quantity $s_{ab} s_{bc} A(s_{ab},s_{bc})$, we may express an adjoint-ordered partial amplitude with leg ordering $(abcd)$ as:
\begin{equation}
A(abcd) = \frac{\permNum }{s_{ab} s_{bc} }  
\end{equation}
This convention ensures that the adjoint-ordered partial amplitudes satisfy the $(n-3)!$ amplitude relations.  

A full amplitude with double-copy structure may be encoded using the permutation invariant weights $\permNum_g$ and $\tilde{\permNum}_g$ accordingly, 
\begin{equation}
{\cal A}_4= {\permNum}\otimes{\tilde\permNum}=-\frac{\permNum \tilde{\permNum}}{s t u}  
\end{equation}
Any amplitude with this permutation-invariant double copy structure may be equivalently expressed as an adjoint-type double copy, and vice-versa.

\subsubsection{Composition at four-points}
We may take two graph weights and compose them together to generate a new graph weight.  This type of composition obviates the need to employ an ansatz to achieve algebra-satisfying graph-weights.    This quadratic operation can be defined  by the following composition rules: 
\begin{align}
\label{compAdjaa4}
\compAdj{ \adjNum }{ \tilde{\adjNum} }_s &= (\adjNum_t \tilde\adjNum_t - \adjNum_u \tilde \adjNum_u)\\
\label{compAdjap4}
\compAdj{ \adjNum }{\permNum }_s &= \adjNum_s \,  \permNum  \\
\label{compPermaa4}
\compPerm{ \adjNum }{  \tilde\adjNum } &=(\adjNum_s \tilde{\adjNum}_s + \adjNum_t \tilde{\adjNum}_t+\adjNum_u \tilde{\adjNum}_u) \\
\label{compPermpp4}
\compPerm{ \permNum}{ \tilde\permNum } &=   \permNum  \,  \tilde{\permNum} \,. 
\end{align}
This notation $\compArb{\mathfrak{f}}{\mathfrak{g}}{\mathfrak{e}}$ may be read as a graph weight of algebraic type $\mathfrak{f}$ composed with a graph weight of algebraic type $\mathfrak{g}$ to generate a new graph weight of algebraic type $\mathfrak{e}$. We specifically note that the compositions rules that generate adjoint graph weights, $\compArb{\mathfrak{f}}{\mathfrak{g}}{\mathfrak{a}}_s$, as written will yield an s-channel, or $g = (1234)$, dressing; all other graph orderings may be obtained via functional relabeling. These definitions hold under scaling by any order of permutation invariants, including, e.g., overall constant factors.

With these composition rules, we will be able to use an incredibly small number building blocks to generate all adjoint-double-copy compatible four-point amplitudes at all orders in mass-dimension. Note particularly that one can simply multiply any graph weight by a permutation-invariant combination of Mandelstam scalars to trivially generate higher-mass-dimension weights of the same algebraic type as the initial graph weight. 

We will continue now with specific examples of scalar, color, and vector weights.  We will begin with scalar weights, as their two basis permutation invariants trivialize towers of higher-mass-dimension weights, and they provide useful building blocks for spanning higher-derivative color-weights. 
\subsection{Scalar adjoint-type weights}
 
 One can show via a generic ansatz that, up to constant factors, there is only one adjoint-type scalar-weight that is linear in Mandelstam invariants:
 \begin{equation}
 \label{simpleScalarFour}
     \adjNum^{s,1}_s = (t-u)=\adjNum^{\text{ss}}_s \, .
 \end{equation}
This is precisely the kinematic weight corresponding to a Yang-Mills scalar theory of massless adjoint scalars minimally coupled by gluons, sometimes referred to as the \textit{simple scalar} theory.  The relevant cubic interaction operator for this theory takes the form: 
\begin{equation}
\mathcal{L}_{\text{ss}} = f^{abc}A_{\mu}^a \partial^{\mu}\varphi^b \varphi^c\,, 
\end{equation}
and the resulting ordered amplitude is: 
\begin{equation}
A^{\text{ss}}_4(s,t) \propto \frac{s^2+t^2+u^2}{s t} \propto 1 + \frac{s}{t} + \frac{t}{s} \, . 
\end{equation}

We can consider two operations with this scalar weight: composing it with itself to a permutation invariant and composing it with itself to get an adjoint weight of higher mass dimension. The first operation results in something proportional to the lowest (positive) mass dimension scalar permutation invariant:
\begin{equation}
\permNum^{s,2}=\compPerm{ \adjNum^{\text{ss}} }{  \adjNum^{\text{ss}}}=  s^2+t^2+u^2 = \sigma_2\,.
\end{equation}
So we see that: $ s t A^{\text{ss}}_4(s,t)=\sigma_2$.

Composing $\adjNum^{\text{ss}}$ with itself to make an adjoint-type numerator results in a mass-dimension two adjoint-type weight,
\begin{equation}
\adjNum^{s,2}_s=\compAdj{ \adjNum^{\text{ss}} }{  \adjNum^{\text{ss}}}\propto s(t-u) = \adjNum^{\text{nlsm}}_s \,.
\end{equation}
This is proportional to the familiar Nonlinear Sigma Model kinematic weight which builds amplitudes for the action: 
\begin{equation}
\mathcal{S}_{\text{\text{NLSM}}} = \int d^D x \left( \frac{1}{2} \Tr \left[ \partial_{\mu}\varphi \frac{1}{1-\varphi^2}\partial^{\mu}\varphi \frac{1}{1-\varphi^2}  \right] \right) \,.
\end{equation}
Its ordered amplitude is seen to be:
\begin{equation}
A^{\text{nlsm}}_4(s,t)\propto \frac{s t u}{ s t} = u\,.
\end{equation}

Composing $\adjNum^{\text{nlsm}}$ with $\adjNum^{\text{ss}}$ to a permutation invariant results in the last unique permutation invariant building block:
\begin{equation}
\permNum^{s,3}=\compPerm{ \adjNum^{\text{nlsm}} }{  \adjNum^{\text{ss}}}\propto s t u = \sigma_3\,.
\end{equation}
So we see that $ s t A^{\text{nlsm}}_4(s,t) = \sigma_3$, and therefore at four-point the double-copy of any theory with the non-linear sigma model is simply the permutation invariant of that theory itself:
\begin{equation}
\adjNum^{\text{nlsm}} \otimes \adjNum  = s t A^{\adjNum}(s,t)\, .
\end{equation}

All higher order permutation invariants are made of sums of products of various powers of $\sigma_2$ and $\sigma_3$.  Because we may always generate new adjoint weights by multiplication of $\adjNum^{\text{ss}}$ or $\adjNum^{\text{nlsm}}$ with scalar permutation invariants, one might expect that running out of novel permutation invariant structures indicates that we have also now run out of novel adjoint-type scalar weights up to permutation invariants. This would indeed be correct.  The ladder of adjoint-type composition for scalars has already closed under permutation invariant products by weight three:
\begin{equation}
\compAdj{ \adjNum^{\text{nlsm}} }{  \adjNum^{\text{ss}}}_s \propto  \sigma_2 \adjNum^{\text{ss}} \,.
\end{equation}
All higher-mass-dimension $\adjNum$ type weights that generate unique ordered-amplitudes may be expressed as sums over  $\adjNum^{\text{ss}}$ and $\adjNum^{\text{nlsm}}$ each taken with appropriate polynomials in $\sigma_2$ and $\sigma_3$.  For a proof that this is sufficient see ref.~\cite{Carrasco:2019yyn}.
As such, there are only two distinct pure-scalar adjoint-type weights up to permutation invariants, $\adjNum^{s,1}$, and
$\adjNum^{s,2}\equiv\compAdj{ \adjNum^{s,1} }{ \adjNum^{{s,1}}}$, and only the first results in new ordered amplitudes (modulo scalar permutation invariants) when invoking composition.  We have the case at four-points where simply knowing the linear adjoint-type numerator as well as all composition rules gives us everything we need know about scalar adjoint-type weights.

\subsection{Color adjoint-type weights}

There are three independent color-weights linear in traces, i.e. that generate ordered trees unique up to scalar permutation invariants. The first is simply the most natural, dressing every vertex with $f^{abc}$ structure constants:
\begin{equation}
 \adjNum^{c,1}_s= f^{a_1 a_2 b} f^{b a_3 a_4}=c_s\,.
\end{equation}
This results in an ordered amplitude that looks like:
\begin{align}
A^{c,1}(s,t)&= \frac{c_s}{s} + \frac{c_t}{t} \\
&=A^{\text{bi-adj}}(s,t)
\end{align}
The second involves composition between the simple-scalar numerator and the first color-numerator:
\begin{equation}
 \adjNum^{c,2}_s= \compAdj{\adjNum^{c,1}}{\adjNum^{s,1}}= (s-u) c_t - (t-s) c_u \,. 
\end{equation}
This results in ordered amplitudes proportional to:
\begin{align}
A^{c,2}(s,t) 
  &=  \frac{c_s \adjNum^{\text{nlsm}}_s+c_t \adjNum^{\text{nlsm}}_t +c_u \adjNum^{\text{nlsm}}_u}{s t}
\end{align}
The third unique adjoint color-weight is as simple as composing the permutation invariant color-symbol $d^{abcd}$ with the adjoint scalar kinematic weight $\adjNum^{\text{nlsm}}$,
\begin{equation}
 \adjNum^{c,3}_s= \compAdj{\adjNum^{\text{nlsm}}}{d^{abcd}}_s =   d^{abcd} s \left( u - t \right)   \,.
\end{equation}

All other adjoint-type compatible color weights linear in color-traces are simply permutation invariants times these three color weights.   This means, for example, if we want to consider all higher-derivative corrections to maximally supersymmetric Yang-Mills at four-points, we simply need to construct the double-copy between the supersymmetric Yang-Mills and the following adjoint-type higher-derivative color numerator: 
\begin{multline} 
\label{eqn:fourPointcHD}
c^{\text{HD}} = \sum_m \left(\alpha^{\prime} \right)^m  \times \left\{ \sum_{2X+3Y = m} a_{c,1}^{(X,Y)} \sigma_2^X \sigma_3^Y \adjNum^{c,1}
+ \sum_{2X+3Y = m-1} a_{c,2}^{(X,Y)} \sigma_2^X \sigma_3^Y \adjNum^{c,2} \right. \\
\left. +  \sum_{2X+3Y = m-2} a_{c,3}^{(X,Y)} \sigma_2^X \sigma_3^Y \adjNum^{c,3} \right\}  \, ,
\end{multline}
where $X \geq 0$ and, for the first two sums $Y \geq 1$, and for the final sum  $Y \geq 0$, so that in all cases, we obtain contact corrections with no poles in  $s$, $t$, or $u$. We demand local contact corrections because there are no non-zero higher-derivative three-point scalar amplitudes for a four-point amplitude with pole structure to factorize into, as all Lorentz invariant dot products $ \left( k_i \cdot k_j \right)$ vanish for on-shell, real three point momenta. We additionally introduce a dimensionful parameter, $\ap$, to track the order in higher-derivative correction: this constant carries units $[ \ap ] = M^{-2}$, so corrections at $n^{\text{th}}$ order in $\ap$ must carry higher-derivative corrections of mass-dimension $2n$. This corresponds to carrying $n$ additional scalar-kinematic Lorentz dot products relative to the uncorrected theory. Finally, the coefficients $a^{(X,Y)}_{c,i}$ are free overall parameters accompanying each individual solution at each order, corresponding to Wilson coefficients for distinct higher-derivative operators. 

The full higher-derivative correction amplitude is given by double-copy as: 
\begin{equation}
{\cal A}_4^{\text{sYM} + \text{HD}} =  n^{\text{sYM}}\otimes c^{\text{HD}}= \frac{n^{\text{sYM}}_s c^{\text{HD}}_s}{s} +\frac{n^{\text{sYM}}_t c^{\text{HD}}_t }{t}  +\frac{n^{\text{sYM}}_u c^{\text{HD}}_u }{u} \, .
\end{equation}
Indeed this structure was exposed in the resummed in $\alpha'$ four-point tree-level open-superstring scattering amplitude in ref.~\cite{Carrasco:2019yyn}. More general corrections to pure Yang-Mills may be given by the double copy of these higher-derivative adjoint color numerators with any generic combination $n^{\rm vec}$ of the eight total distinct adjoint-type vector numerators as: 
\begin{equation}
{\cal A}_4^{\text{YM} + \text{HD}} =  n^{\rm vec}\otimes c^{\text{HD}}= \frac{n^{\rm vec}_s c^{\text{HD}}_s}{s} +\frac{n^{\rm vec}_t c^{\text{HD}}_t }{t}  +\frac{n^{\rm vec}_u c^{\text{HD}}_u }{u} \, .
\end{equation}
There are seven gauge-invariant tensor structures~\cite{Bern:2017tuc} (see also, e.g.~\cite{Barreiro:2013dpa, Boels:2016xhc} and references therein) that adjoint-type numerators can be built from (up to products with rational functions of permutation invariant color-weights), and this yields eight distinct (under polynomials of $\sigma_2$ and $\sigma_3$) vector building blocks, starting (with lowest mass-dimension) at Yang-Mills.  Since Yang-Mills numerators at four points have three Lorentz dot-products per term, we denote $n^{\text{YM}}_{\adjNum}$ as $\adjNum_{\text{vec}}^{3}$; then, $\adjNum_{\text{vec}}^{4}$ will generically refer to vectors with one more dot product in each numerator than Yang-Mills numerators.  There are three such $\adjNum_{\text{vec}}^{4}$, including amplitudes that arise where one vertex gets $\text{Tr}(F^3)$ and the other vertex gets $\text{Tr}(F^2)$.  Similarly there are three such $\adjNum_{\text{vec}}^{5}$, including amplitudes that arise where each vertex gets a $\text{Tr}(F^3)$, and a final $\adjNum_{\text{vec}}^{6}$.  This final unique structure, $\adjNum_{\text{vec}}^{6}$, can be built non-linearly by rational products of scalar permutation invariants and the lower-weight vector blocks, reflecting the difference between the unique seven tensor invariants and our polynomial in permutation-invariant basis of adjoint-type vector numerators.  We have not constructed a direct proof that these vector invariant blocks are sufficient to span all vector building blocks, but rely on the fact that we span the seven gauge-invariant tensor structures of ref.~\cite{Bern:2017tuc}, and have verified up to $\alpha'^{25}$ above Yang-Mills that we span all such gauge-invariant permutation invariants.  It was noted in ref.~\cite{Carrasco:2019yyn} that only four of these are needed to build the bosonic open-string. We include all eight distinct vector blocks in associated auxiliary files~\cite{ancFiles}. 

We note now that with these eight vector building blocks, and the three color building blocks, we have the ability to span every higher-derivative gauge operator that is compatible with local adjoint-double-copy by considering simply these with polynomials in the permutation-invariant scalar weights $\sigma_2$, and $\sigma_3$.  Additionally we can span every higher-derivative gravity amplitude compatible with local adjoint-double-copy.  This means that the ansatz to verify whether a gravity amplitude is consistent with adjoint double-copy is incredibly small. 

\subsection {Striating by $\mathbf{d^{abcd}}$ structures}

At four-points, adjoint-type double-copy means permutation-invariant double-copy, as can be seen in the equivalent forms of the amplitude in Eqns~\ref{adjDoubleCopyAmp4} and~\ref{piDoubleCopyAmp4}. There is no new physical information here, but it allows a double-copy description where every building block is a full physical amplitude in some theory, which does offer an interesting perspective as we will discuss. This is not the first opportunity to see that, depending upon the algebra made manifest, the same amplitudes admit multiple different decompositions into double-copies of predictions of distinct theories.  A spectacularly notable example involves considering supergravity in three-dimension, whose scattering amplitudes admit a double-copy construction from three-algebra striated BLG amplitudes~\cite{Bargheer2012gv,Huang2012wr,Huang:2013kca} as well as the more familiar adjoint-type double-copy construction from super-Yang-Mills amplitudes.

Consider the full four-point amplitudes for Yang-Mills and Gravity. Writing them as manifestly permutation-invariant double copies leads to the following representation:
\begin{align}
- { \sigma_3}  {\cal A}^{(\rm YM)}&= \left[s t A^{\rm bi-adj}(s,t)\right] \left[ s t \Aym(s,t) \right]  \\
- { \sigma_3 } {\cal A}^{(\rm GR)}&= \left[ s t \Aym(s,t) \right] ^2  \, .
\end{align}
Note that every individual permutation-invariant element of the above expressions are recognizable as proportional to the full four-point amplitudes of known theories.  Taking the color-ordered $(s,t)$ channel of a bi-adjoint scalar, and multiplying it by $s t$ yields the full color-dressed amplitude for the Nonlinear Sigma Model.   Famously $ [ s t \Aym(s,t) ]$ yields the four-point amplitude for Born-Infeld.  Finally, the purely scalar permutation invariant,  $[\sigma_3\equiv(s t u)]$, is recognizable as the four-point amplitude of the Special Galileon.  So a perhaps surprising novelty of permutation-invariant striations at four-points is that \textit{full amplitudes} for theories can serve as constructive building blocks.  As noted in the first version of ref.~\cite{Carrasco:2019yyn},  permutation-invariance admits the following whimsical but exact departures from the typical: ``GR$\sim$YM${}^2$'' slogan at four-points:
\begin{align}
{\cal A}^{(\rm YM)}_4&\equiv\frac{ {\cal \tilde{A}}^{\rm NLSM}  {\cal A}^{\rm Born-Infeld}} {{\cal A}^{\rm Spec.Gal} }\, &
{\cal A}^{(\rm GR)}_4&\equiv\frac{ ({\cal A}^{\rm Born-Infeld})^2 } {{\cal A}^{\rm Spec.Gal}}  \, .
\end{align}

\subsection{An example at four-points}

In this example, we take a closer look at the amplitudes of the maximally supersymmetric $F^4$ and $R^4$ operators \cite{Chandia:2003sh}, 
\begin{equation}
F^4_{\text{SUSY}} \equiv \Tr \left[ F_{\mu}^{\nu} F_{\nu}^{\rho} F_{\rho}^{\sigma} F_{\sigma}^{\mu}  + 2 F_{\mu}^{\nu} F_{\rho}^{\sigma} F_{\nu}^{\rho} F_{\sigma}^{\mu} - \tfrac{1}{4}F_{\mu\nu}F_{\rho\sigma}F^{\mu\nu}F^{\rho\sigma} - \tfrac{1}{2}F_{\mu\nu}F^{\mu\nu}F_{\rho\sigma}F^{\rho\sigma}    \right]  
\end{equation} 
\begin{equation}
R^4_{\text{SUSY}} \equiv t_{(8)}^{\mu_1 \nu_1 \mu_2 \nu_2 \mu_3 \nu_3 \mu_4 \nu_4}t_{(8)}^{\lambda_1 \rho_1 \lambda_2 \rho_2 \lambda_3 \rho_3 \lambda_4 \rho_4} R_{\mu_1 \nu_1 \lambda_1 \rho_1} R_{\mu_2 \nu_2 \lambda_2 \rho_2} R_{\mu_3 \nu_3 \lambda_3 \rho_3} R_{\mu_4 \nu_4 \lambda_4 \rho_4} \,  ,
\end{equation}
both through the lens of adjoint double copy construction via higher derivative numerator factors and using the permutation invariant striation explained above. 

The $F^4$ amplitude is quite naturally striated by permutation invariant color-kinematics duality, as its color structure is the totally symmetric $d^4$,
\begin{equation}
\mathcal{A}^{F^4_{\text{SUSY}}} = d^{a_1 a_2 a_3 a_4} \left( s \, t  \,A^{\text{sYM}}(1234) \right)
\end{equation} 
correspondingly accompanied by the permutation invariant quantity $s \, t \, A^{\text{sYM}}(1234)$. This form suggests an associated permutation invariant double copy via the replacement $d^{a_1 a_2 a_3 a_4} \rightarrow s \, t  \,A^{\text{sYM}}(1234)$, which yields precisely the $R^4$ amplitude: 
\begin{equation}
\mathcal{M}^{R^4_{\text{SUSY}}} = \left( s \, t  \, A^{\text{sYM}}(1234) \right)^2 = s \, t \, u \left( \mathcal{M}^{\text{SUGRA}} \right)
\end{equation} 

Both amplitudes also manifest adjoint double copy structure by virtue of higher derivative factors $\cHD$ and $n^{\text{HD}}$ at $\mathcal{O}\left( \ap^2 \right)$ and $\mathcal{O}\left( \ap^3 \right)$, respectively: 
\begin{equation}
\mathcal{A}^{F^4_{\text{SUSY}}} = \sum_g \frac{ \left(d^4\, n^{\text{NLSM}}_g \right) \left(n^{\text{sYM}}_g \right) }{d_g}
\end{equation} 
\begin{equation}
\mathcal{M}^{R^4_{\text{SUSY}}} = \sum_g \frac{ \left((stu) \, n^{\text{sYM}}_g \right) \left(n^{\text{sYM}}_g \right) }{d_g}
\end{equation} 
In particular, $\mathcal{A}^{F^4_{\text{SUSY}}}$ makes clear the power of the using modified color factors to encode higher derivative corrections: color conspires with scalar kinematics within a single adjoint $\cHD$ numerator to capture the permutation invariant $d^4$ color of $F^4_{\text{SUSY}}$ in an adjoint-dual structure. This adjoint numerator $c^{\text{HD},(2)}_g = d^4 n^{\text{NLSM}}_g$ yields the $\mathcal{O}\left( \ap^2 \right)$ term in Chan-Paton dressed $\mathcal{Z}$-theory, 
\begin{equation}
A^{\text{BA+HD},(2)}(1234) \propto d^{a_1 a_2 a_3 a_4} \, s_{13}
\end{equation} 

These adjoint forms may be lined up with the permutation invariant striations of $\mathcal{A}^{F^4_{\text{SUSY}}}$ and $\mathcal{M}^{R^4_{\text{SUSY}}}$:
\begin{equation}
-\mathcal{A}^{F^4_{\text{SUSY}}} = \frac{ \left(d^4 \, s \, t \,A^{\text{NLSM}}(1234) \right) \left( s \, t \, A^{\text{sYM}}(1234) \right)}{ s \, t \, u } = d^4 \left( s \, t \, A^{\text{sYM}}(1234) \right)
\end{equation}
\begin{equation}
-\mathcal{M}^{R^4_{\text{SUSY}}} = \frac{ \left((s\,t\,u) \, s \, t \,A^{\text{sYM}}(1234) \right) \left( s \, t \, A^{\text{sYM}}(1234) \right)}{ s \, t \, u } = \left( s \, t \, A^{\text{sYM}}(1234) \right)^2
\end{equation}

%% file: algebraicStructure.tex
\section{Algebraic Structures at Five-Points}

\label{section:algebra5pt}

Our methods for constructing higher-derivative corrections at four points were driven by the guiding principle of finding color and kinematic building blocks that fell into one of two algebraic categories: adjoint or permutation invariant. This focus on identifying the smallest elements of prediction that obeyed these two algebraic structures meant that, once such building blocks were identified, we could use universal composition rules to combine them to arbitrarily high order in mass dimension, rather than having to treat each order in mass dimension on a case-by-case basis using an ansatz. 

The success of this method hinges on the fact that our constructive composition rules simply require that their arguments obey standard algebraic relations. Once we've obtained building blocks that obey algebraic relations of interest, we can compose them in new and interesting ways to generate order after order of correction. At four points, to generate all maximally-supersymmetric-compatible corrections to gauge theory, we need only a handful of compositions rules and just a few adjoint and permutation invariant building blocks (adjoint color $f^3f^3$, adjoint scalar kinematics $\adjNum^{\text{ss}}$, and permutation invariant color $d^4$). 

We thus guide our study of five point corrections in the same manner, beginning by focusing on the relevant algebraic structures at five points and how they can be composed together. In this section, we'll first describe the algebraic structures of interest in the abstract by looking at relevant graph topologies, and then discuss how to write down general composition rules. In the following sections, we will find both specific scalar kinematic and color weights satisfying the algebraic constraints of interest, which will then serve as building blocks for generating modified color factors $\cHD$ that encode higher derivative corrections.

\subsection{Graph representations}
Let us start with graph representations of five-point graphs, as all algebraic structures discussed are naturally described in terms of these graphs.  There is only one cubic graph topology at five points, with three vertices and two internal edges, whose labels we can specify with an ordered list:
\be
(abcde) \equiv \fivegraph{a}{b}{c}{d}{e} \,.
\ee
There is one topology with one quartic vertex and one cubic vertex, connected by one internal edge between them:
\be
(\langle abc \rangle de) \equiv \quadCubeFive{a}{b}{c}{d}{e} \,.
\ee
And there is a contact graph with a quintic vertex:
\be
(\langle abcde \rangle) \equiv \contactFiveNoBlue{a}{b}{c}{d}{e} \,.
\ee
We will now introduce five algebraic structures at five-points: three that dress cubic graphs (adjoint, relaxed, and sandwich), one that dresses the quartic-cubic graph (hybrid), and one that dresses the quintic contact graph (permutation-invariant). 

\subsection{Cubic graph structures}
\subsubsection{Adjoint graph weights}
Adjoint numerators $\adjNum_g$ obey antisymmetry around each cubic vertex and satisfy Jacobi relations on each internal edge. At five points, this results in three antisymmetry constraints:
\begin{align}
\label{adjointConstraints1}
\adjNum(abcde) &= -\adjNum(bacde)\, \text{:} &\text{antisymmetric around vertex}\hskip .5cm\fivegraphL{a}{b}{c}{d}{e}\,,\\
\adjNum(abcde) &= -\adjNum(abced)\, \text{:} &\text{antisymmetric around vertex}\hskip .5cm\fivegraphR{a}{b}{c}{d}{e}\,,\\
\adjNum(abcde) &= -\adjNum(edcba)\, \text{:} &\text{antisymmetric around vertex}\hskip .5cm\fivegraphM{a}{b}{c}{d}{e}\,.
\end{align}
And two Jacobi constraints:
\begin{align}
\label{adjointConstraints2}
\adjNum(abcde) &= \adjNum(cbade) + \adjNum(acbde) \, \text{:} &\text{Jacobi relation about edge} \fivegraphEL{a}{b}{c}{d}{e} \\
\adjNum(abcde) &= \adjNum(abedc) + \adjNum(abdce) \, \text{:} & \text{Jacobi relation about edge} \fivegraphER{a}{b}{c}{d}{e}
\end{align}
Taking these constraints into consideration, any adjoint dressing of the 15 distinctly labeled cubic five point graphs $\Gamma_{3}$ may be written in a basis of six adjoint graph weights.  For an explicit example of DDM basis graphs, we refer the reader to \eqn{ddmBasis} in \app{appendix:colorStructure}.

We define the full amplitude as summing over these fifteen cubic graphs, dressing each with two adjoint weights $\adjNum$ and $\tilde{\adjNum}$ and the product $d_g$ of two associated cubic propagators, 
\begin{equation}
\mathcal{A}  = \sum_{g \in \Gamma_{3}} \frac{\adjNum_g \tilde{\adjNum}_g}{d_g} \, , 
\end{equation}
where for $g = (abcde)$, the denominator is given as $d_g = s_{ab}s_{de}$. We then define the ordered amplitude on the $\tilde{\adjNum}$ weights as the coefficient of an independent basis element $\tilde{\adjNum}$ within the full amplitude. Reading off the coefficient of $\tilde{\adjNum}(12345)$, we find the ordered amplitude:
\begin{equation}
\label{adjointOrderedAmplitude}
A_{\adjNum}(12345)  = \frac{\adjNum(12345)}{s_{12}s_{45}} + \
\frac{\adjNum(12543)}{s_{12}s_{34}} + \
\frac{\adjNum(15243)}{s_{15}s_{34}} + \
\frac{\adjNum(32415)}{s_{15}s_{23}} + \
\frac{\adjNum(45123)}{s_{23}s_{45}} \, . 
\end{equation}
This is the familiar ordered amplitude, in agreement with the definition of reading off the coefficient of an independent color trace, for Yang-Mills amplitudes, as found in~\eqn{orderedFivePointAmplitude}.

It turns out that  at five points, in stark contrast to four-points, adjoint numerators cannot be formed out of scalar monomials linear in Lorentz products, e.g. $\alpha k_1\cdot k_2 + \beta k_2\cdot k_3 \cdots$.  This motivates considering an almost adjoint structure.

\subsubsection{Almost adjoint (relaxed) graph weights}
For the almost adjoint (relaxed) numerators, $\relaxedNum_g$, we still have three vertex constraints and two edge constraints, but the middle-node antisymmetry condition is now relaxed to a symmetry condition:
\begin{align}
\label{relaxedConstraints}
\relaxedNum(abcde) &= -\relaxedNum(bacde)\, \text{:} &\text{antisymmetric around vertex}\hskip .5cm\fivegraphL{a}{b}{c}{d}{e}\,,\\
\relaxedNum(abcde) &= -\relaxedNum(abced)\, \text{:} &\text{antisymmetric around vertex}\hskip .5cm\fivegraphR{a}{b}{c}{d}{e}\,,\\
\relaxedNum(abcde) &= + \relaxedNum(edcba)\, \text{:} &\text{\textit{symmetric} around vertex}\hskip .5cm\fivegraphMS{a}{b}{c}{d}{e}\,,\\
\relaxedNum(abcde) &= \relaxedNum(cbade) + \relaxedNum(acbde) \, \text{:} &\text{Jacobi relation about edge} \fivegraphEL{a}{b}{c}{d}{e}\,, \\
\relaxedNum(abcde) &= \relaxedNum(abedc) + \relaxedNum(abdce) \, \text{:} & \text{Jacobi relation about edge} \fivegraphER{a}{b}{c}{d}{e}\, .
\end{align}
Any dressing of the 15 cubic five-point graphs with relaxed numerators is expressible in a basis of five relaxed graph weights.  It turns out that scalar kinematic relaxed graph weights exist at every mass-dimension, and there exists a composition that allows generation of all such weights from the unique solution that is linear in momentum invariants.   This algebraic structure will be the \textit{fundamental building block for five-points}.

We define the full amplitude as summing over the fifteen cubic graphs, dressing each with two relaxed weights $\relaxedNum$ and $\tilde{\relaxedNum}$ and the product $d_g$ of two associated cubic propagators, 
\begin{equation}
\mathcal{A}  = \sum_{g \in \Gamma_{3}} \frac{\relaxedNum_g \tilde{\relaxedNum}_g}{d_g} \, , 
\end{equation}
where for $g = (abcde)$, the denominator is given as $d_g = s_{ab}s_{de}$. We then define the ordered amplitude on the $\tilde{\relaxedNum}$ weights as the coefficient of an independent basis element $\tilde{\relaxedNum}$ within the full amplitude. Reading off the coefficient of $\tilde{\relaxedNum}(12345)$, we find the ordered amplitude:
\begin{equation}
A_{\relaxedNum}(12345)  = \frac{\relaxedNum(12345)}{s_{12}s_{45}} + \
\frac{\relaxedNum(12435)}{s_{12}s_{35}} + \
\frac{\relaxedNum(32415)}{s_{15}s_{23}} + \
\frac{\relaxedNum(35142)}{s_{24}s_{35}} + \
\frac{\relaxedNum(42315)}{s_{15}s_{24}} + \
\frac{\relaxedNum(45132)}{s_{23}s_{45}} \, . 
\end{equation}

Given the vertex conditions, one might imagine that this type of structure is naturally associated with combining a symmetric three-vertex color-weight $d^{abc}$ with adjoint  $f^{abc}$ structure constants:  $f^{a_1 a_2 b_1} d^{b_1 a_3 b_2} f^{b_2 a_4 a_5}$.  It turns out that having a symmetric color-weight sandwiched between antisymmetric structure constants does not obey Jacobi relations around the two edges, but instead a less constraining six-term identity --- this six-term identity can be satisfied by graph weights that obey Jacobi relations, like relaxed weights, but does not require it.

\subsubsection{Sandwich graph weights}
If we do not impose the two Jacobi requirements, but instead the six-term identities that are obeyed by $f\,d\,f$ type color-weights, we have the vertex constraints:
\begin{align}
\label{sandConstraintsSym}
\sandwichNum(abcde) &= -\sandwichNum(bacde)\, \text{:} &\text{antisymmetric around vertex}\hskip .5cm\fivegraphL{a}{b}{c}{d}{e}\,,\\
\sandwichNum(abcde) &= -\sandwichNum(abced)\, \text{:} &\text{antisymmetric around vertex}\hskip .5cm\fivegraphR{a}{b}{c}{d}{e}\,,\\
\sandwichNum(abcde) &= + \sandwichNum(edcba)\, \text{:} &\text{symmetric around vertex}\hskip .5cm\fivegraphMS{a}{b}{c}{d}{e}\,.
\end{align}
And the remaining six-term constraint:
\begin{equation}
\label{sandConstraintsSix}
\sandwichNum(abcde) + \sandwichNum(acbde) - \sandwichNum(adbce) + \sandwichNum(adceb) + \sandwichNum(aebcd) - \sandwichNum(aecdb) = 0 
\end{equation}
There is an 11-element basis of sandwich graph weights for the fifteen distinctly labeled cubic graphs at five-points. All relaxed adjoint numerators satisfy these constraints as objects that satisfy Jacobi about internal edges automatically satisfy \eqn{sandConstraintsSix}, but not all sandwich numerators satisfy the more stringent relaxed Jacobi constraints of \eqn{relaxedConstraints}.  We present one choice of such eleven spanning basis graphs in \eqn{sandwichBasis}.

We define the full amplitude as summing over the fifteen cubic graphs, dressing each with two sandwich weights $\sandwichNum$ and $\tilde{\sandwichNum}$ and the product $d_g$ of two associated cubic propagators, 
\begin{equation}
\mathcal{A}  = \sum_{g \in \Gamma_{3}} \frac{\sandwichNum_g \tilde{\sandwichNum}_g}{d_g} \, , 
\end{equation}
where for $g = (abcde)$, the denominator is given as $d_g = s_{ab}s_{de}$. We then define the ordered amplitude on the $\tilde{\sandwichNum}$ weights as the coefficient of an independent basis element $\tilde{\sandwichNum}$ within the full amplitude. Reading off the coefficient of $\tilde{\sandwichNum}(12345)$, we find the ordered amplitude:
\begin{equation}
A_{\sandwichNum}(12345)  =\frac{\sandwichNum(12345)}{s_{12}s_{45}} + \
\frac{\sandwichNum(42315)}{s_{15}s_{24}} \, . 
\end{equation}

\subsection{Quartic and quintic graph structures}

\subsubsection{Hybrid graph weights}
Five point higher-derivative corrections inherit local four-point contact term corrections, so it is worth exploring whether or not adjoint representations of five-points can live on associated hybrid quartic-cubic graphs. We introduce \textit{hybrid numerators} $\hybridNum$ that dress the 10 graphs containing one cubic and one quartic vertex, $\Gamma_{\hybridNum}$, allowing for antisymmetry around the cubic vertex and symmetry around the quartic vertex:
\begin{equation}
\label{eqn:hybridSymmetries}
\hybridNum(\langle abc \rangle de) = \hybridNum(\langle bac \rangle de) = \hybridNum(\langle acb \rangle de) \,, \hskip0.5cm \text{sym. about quartic vertex: } \, \quadCubeFiveL{a}{b}{c}{d}{e}
\end{equation}
\begin{equation}
\hybridNum(\langle abc \rangle de) = -\hybridNum(\langle abc \rangle ed)\, , \hskip1cm \text{antisym. about cubic vertex: } \, \quadCubeFiveR{a}{b}{c}{d}{e}
\end{equation}
As the order of first three external leg labels do not matter for $\hybridNum_g$ we need only provide the final two entries in order to unambiguously label the graph. Following $d^4 f^3$ color weights, hybrid graph weights are constrained to obey four-term identities: 
\begin{equation}
\begin{split}
\label{eqn:hybridJacobi}
\hybridNum(de) = \hybridNum(ec) + \hybridNum(eb) + \hybridNum(ea) \\ 
\hybridNum(de) = \hybridNum(cd) + \hybridNum(bd) + \hybridNum(ad) 
\end{split}
\end{equation}
Any of the ten distinct hybrid graph weights $\Gamma_{\hybridNum}$ are expressible in a minimal basis of six hybrid graph weights.  We present one choice for these basis weights in \eqn{hybridBasis}.

We define the full amplitude as summing over the ten hybrid graphs, dressing each with two hybrid weights $\hybridNum$ and $\tilde{\hybridNum}$ and a single cubic propagator, 
\begin{equation}
\mathcal{A}  = \sum_{g \in \Gamma_{\hybridNum}} \frac{\hybridNum_g \tilde{\hybridNum}_g}{d_g} \, , 
\end{equation}
where for $g = (\langle abc \rangle de)$, the denominator is given as $d_g = s_{de}$. We then define the ordered amplitude on the $\tilde{\hybridNum}$ weights as the coefficient of an independent basis element $\tilde{\hybridNum}$ within the full amplitude. Reading off the coefficient of $\tilde{\hybridNum}(12345)$, we find the ordered amplitude:
\begin{equation}
A_{\hybridNum}(12345)  =\frac{\hybridNum(12345)}{s_{45}} + \frac{\hybridNum(13452)}{s_{25}} \
+ \frac{\hybridNum(13524)}{s_{24}}  \, . 
\end{equation}

\subsubsection{Permutation-invariant algebraic weights}
We also discuss  purely local permutation invariant weights, $\permNum$, which naturally dress the five-point quintic contact graph:
\begin{equation}
\permNum(\langle abcde \rangle)=\permNum(\langle \sigma \rangle) ~\forall~ \sigma \in S_5(abcde) \, , \hskip1cm \text{permutation symmetry for} \, \contactFive{a}{b}{c}{d}{e} \,.
\end{equation} 

The full amplitude is given by simply dressing the single contact graph with two permutation invariants $\permNum$ and $\tilde{\permNum}$,
\begin{equation}
\mathcal{A} = \permNum \tilde{\permNum} \, ,
\end{equation}
so the ordered amplitude $A_{\permNum} = \permNum$ is simply the permutation-invariant weight itself.

\subsection{Composition rules}
Just like at four points, at five points, we are able to find \textit{composition rules} that non-trivially combine two numerators into a distinct third numerator. These constructions are useful for mixing color structures and kinematic factors into adjoint numerators $\cHD$, the foundation of our method of constructing adjoint-double copy-striations of higher-derivative predictions. At four points, we only saw two relevant algebraic structures --- adjoint and permutation invariant --- and found composition rules to combine these objects accordingly.  

At five points, there are now many more ingredients to consider. Two relaxed graph weights may be composed into a new relaxed graph weight: $\compRel{\relaxedNum}{\relaxedNum}$, or composed into an adjoint weight $\compAdj{\relaxedNum}{\relaxedNum}$, and so forth with sandwich weights and hybrid weights. In general, we can aspire to take two numerators of any given algebraic structures $\mathfrak{e}$ and $\mathfrak{f}$ and compose them into a new numerator of desired algebraic structure $\mathfrak{g}$, with a rule of the form: 
\begin{equation}
\mathfrak{g}(12345) = \compArb{\mathfrak{e}}{\mathfrak{f}}{\mathfrak{g}}  = \sum_{g \in \rho_\mathfrak{e}}\sum_{g^{\prime} \in \rho_\mathfrak{f}} b_{g,g^{\prime}}^{\mathfrak{efg}} \mathfrak{e}(g)\mathfrak{f}(g^{\prime})
\end{equation}
where the sums run over the \textit{basis graphs} $\rho$ for each given type of numerator (for instance, for relaxed weights, $\rho_{\mathfrak{e}}$ runs over the five basis graphs that arise from imposing relaxed constraints on the fifteen cubic graphs). The $b$ coefficients are fixed by imposing $\mathfrak{g}$-type algebraic constraints on this expression for $\mathfrak{g}(12345)$ while simultaneously exploiting the algebraic properties of $\mathfrak{e}$ and $\mathfrak{f}$ (these coefficients are then specific to the types $\mathfrak{e}$, $\mathfrak{f}$, and $\mathfrak{g}$ that define the particular composition rule).  We describe an approach to identifying composition rules in \app{appendix:BuildingComposition}, and present explicit formulae for relevant five-point composition rules  in \app{appendix:ExplicitComposition}.  

With these composition rules, we are prepared to generate ladders of numerators of higher and higher mass dimension from those already found: two linear weights may be composed into a quadratic one, which in turn may be composed with a linear weight to find a cubic numerator, and so on. 

We note here that, analogous to the four point case, we may generically generate permutation invariants at five points by summing over all graph orderings: 
\begin{equation}
\compPerm{\mathfrak{e}^j}{\mathfrak{e}^k} = \sum_{g \in \Gamma_\mathfrak{e}}\mathfrak{e}^j(g)\mathfrak{e}^k(g) 
\end{equation} 
where $\mathfrak{e}^j$ and $\mathfrak{e}^k$ are two weights of the same algebraic structure $\mathfrak{e}$, and $\Gamma_\mathfrak{e}$ is the list of \textit{all} graphs corresponding to that structure (the 15 cubic graphs $\Gamma_{3}$ for $\mathfrak{e} = \adjNum, \relaxedNum, \text{ or } \sandwichNum$; and the 10 graphs $\Gamma_{\hybridNum}$ for $\mathfrak{e} = \hybridNum$). As such, we will consistently find it useful to compose (relaxed) adjoint with (relaxed) adjoint to yield permutation invariants according to: 
\begin{align}
\label{eq:PIfromR}
\compPerm{\relaxedNum^j}{\relaxedNum^k}  &= \sum_{g \in \Gamma{3}}^{15} \relaxedNum^j_g \relaxedNum^k_g \,, \\
\label{eq:PIfromA}
\compPerm{\adjNum^j}{\adjNum^k}  &= \sum_{g \in \Gamma_{3}}^{15} \adjNum^j_g \adjNum^k_g \,.
\end{align} 

\subsection{Casting between hybrid and adjoint}
\label{castingDiscussion}
Inspired by the solutions we find for adjoint and hybrid color weights that will discussed in section~\ref{hybridColor}, we find a symbolic \textit{linear mapping} between hybrid and adjoint solutions that we refer to as \textit{casting}. A weight $\hybridNum$ that satisfies hybrid constraints may be cast into an adjoint weight $\adjNum[\hybridNum]$ by writing down the following linear combination: 
\begin{equation}
\begin{aligned}
\adjNum[\hybridNum](12345) =  -\frac{1}{5} \Big( 2\,\hybridNum(12345)+\hybridNum(12435)-\hybridNum(12534)\\
+2\,\hybridNum(23415)-2\,\hybridNum(23514) \Big)
\end{aligned}
\end{equation}
Similarly, an adjoint solution $\adjNum$ may be cast into a hybrid solution $\hybridNum[\adjNum]$ as follows:
\begin{equation}
\begin{aligned}
\label{adjToHybridCast}
\hybridNum[\adjNum](12345)  = \frac{1}{2} \Big( \adjNum(12345)-3\,\adjNum(12435)+\adjNum(13245)\\
-3\,\adjNum(13425)+3\,\adjNum(14235)+3\,\adjNum(14325) \Big)
\end{aligned}
\end{equation}
We also note that these casting maps are invertible, in the sense that $\adjNum[\hybridNum[a]] = a$ and $\hybridNum[\adjNum[h]] = h$. These mappings are not restricted to pure color solutions, but hold for any weights satisfying these algebraic relations. This trivializes the process of finding hybrid solutions once adjoint ones have been found, or vice versa.

%% file: colorAndScalar.tex
\section{Color and Scalar Kinematic Building Blocks} 
\label{section:buildingBlocks}
Now that the algebraic structures of interest at five points have been established, we will find both pure color and pure scalar kinematic solutions to those relevant algebraic constraints, making use of our composition rules along the way.  Our ultimate goal will be to build adjoint-structure higher-dimensional color-weights of all orders in mass-dimension so we can trivially generate gauge-invariant higher-derivative corrections to Yang-Mills.   By replacing the color in those corrections with gauge-invariant vector weights of matching algebraic structure, we will construct gauge-invariant higher derivative corrections to gravitation.  At four-points, we achieved this by considering as our primary unit-step to all orders in mass-dimension the adjoint simple-scalar numerator.   Here we will see a different story unfold, where our unit-step lives instead in the relaxed adjoint structure.   Before constructing these building blocks, we first quickly review our language for describing color and scalar kinematics and their relevant bases. 

For color, one common basis is in terms of traces over color generators, but as at four-points we will find that combining the traces into contractions of structure constants $f$ and $d$ will allow us to easily identify the solutions to our algebraic constraints. Prioritizing the color structure basis representation will allow us to uncover novel corresponding color-kinematics dualities at the level of the graph dressings within our amplitudes, a concept that may hopefully be generalizable to loop level. In contrast, relying on the trace basis of color factors will only inform us that the ordered amplitudes dressing each trace within the full amplitude are cyclically symmetric, a known concept that is not amenable to loop-level generalizations, as our favorite loop calculation methods rely on writing down and exploiting the algebraic properties of graph dressings rather than ordered amplitudes. 

For scalar kinematics, we will make repeated use of composition rules to generate ladders of solutions to algebraic constraints at each order in mass dimension. In all cases, we will find simple patterns of repeated composition will generate all desired solutions, and that each algebraic structure eventually closes at some order under repeated multiplication with scalar permutation invariants. Finally, we will see that, at all orders, all scalar kinematics structures considered --- adjoint, relaxed adjoint, hybrid, sandwich, and permutation invariant --- may be generated by appropriate compositions of a single scalar kinematic building block, the unique relaxed solution that is linear in Mandelstam invariants. The simplest of these building blocks will prove sufficient for writing higher-derivative modified color factors $\cHD$, as discussed in the following two sections.

\subsection{Color at five-points}

All five-point single-trace tree-level amplitudes have color structures expressible in terms of a $4! = 24$ element basis of traces over the generators, 
\begin{equation}
\Tr\left(T^{a_{\sigma_1}}T^{a_{\sigma_2}}T^{a_{\sigma_3}}T^{a_{\sigma_4}}T^{a_5}\right)
\end{equation}
where $\sigma$ runs over the $S_4$ permutations of external leg labels $(1,2,3,4)$. By inverting the definitions of the following color structures: 
\begin{equation}
f^{abc} = \Tr\left(\left[T^a,T^b\right]T^c\right)
\end{equation}
\begin{equation}
d^{a_1 \cdots a_n} = \frac{1}{(n-1)!} \sum_{\sigma \in S_{n-1}} \Tr(T^{a_{\sigma_1}} \cdots T^{a_{\sigma_{n-1}}} T^{a_n})
\end{equation} 
and making use of the Fierz identity, the trace basis may be converted to a basis of color structure contractions of the following forms: 
\begin{equation}
d^{abcde} \quad d^{abci}f^{ide}\quad f^{abi}d^{icj}f^{jde}\quad  f^{abi}f^{icj}f^{jde}
\end{equation}
The full basis is written out in \app{appendix:colorStructure}. Other choices of basis are possible~\cite{Bandiera:2020aqn}, including those that allow for the inclusion of $d^{abi}f^{icj}f^{jde}$ contractions (which we have exchanged here in favor of $f^{abi}d^{icj}f^{jde}$, since those relate more directly to the relaxed-adjoint structure that is the natural scalar unit-step at five-points). For notational consistency, we will introduce the following shorthand for these contractions:
\begin{equation}
\begin{split}
d^5(\langle abcde \rangle ) &\equiv d^{abcde} \\
d^4f^3(\langle abc \rangle de) &\equiv d^{abci}f^{ide} \\
f^3d^3f^3(abcde) &\equiv f^{abi}d^{icj}f^{jde} \\
f^3f^3f^3(abcde) &\equiv f^{abi}f^{icj}f^{jde}
\end{split}
\end{equation}

We may use a simple argument involving color to confirm that the algebraic structures we consider span all possible five point structures. Each algebraic structure is associated with a certain number of basis elements: there are six elements in the adjoint basis, eleven elements spanning sandwich structures (which include all relaxed structures), six elements spanning hybrid solutions, and one permutation symmetric element. This bookkeeping is in agreement with the 24 independent color traces at five points. 

\subsection{Scalar kinematics at five-points}

Scalar kinematic numerators at five points are constructed from a five-element basis of momentum invariants; one such choice, 
\begin{equation}
\label{fivePointKinBasis}
\{ (k_1\cdot k_2), (k_2\cdot k_3), (k_3\cdot k_4), (k_4\cdot k_5), (k_5\cdot k_1) \}
\end{equation}
spans all massless five-point Lorentz invariants.   In particular (lower) dimensions there may be fewer basis elements, but for generality we remain in $D$ dimensions.  This particular choice of basis maximally simplifies the propagators in the ordered amplitude $A(12345)$. 

A simple five-term ansatz then demonstrates that there is no linear solution to the adjoint algebraic constraints given in \eqns{adjointConstraints1}{adjointConstraints2}. The lowest mass-dimension non-vanishing adjoint solution purely in terms of momentum invariants arises at the cubic order in momentum invariants. Critically, however, there does exist a linear solution to relaxed adjoint constraints of \eqn{relaxedConstraints}: 
\begin{equation}
\label{linearSoln}
\begin{split}
\relaxedNum^{(1)}(12345) &= (k_1\cdot k_2) - 2(k_2\cdot k_3) - 2(k_3\cdot k_4) + (k_4\cdot k_5) + 4(k_5\cdot k_1)  \\
 &= (k_1 -  k_2)\cdot (k_5 -  k_4) \,.
 \end{split}
\end{equation}
where we label scalar numerators using a superscript denoting their order in momentum invariants, or half their mass dimension (these orders will line up exactly with order in $\ap$ corrections to super Yang-Mills or supergravity).   We note in the second line of \eqn{linearSoln} that we have made all antisymmetry properties of the terminal vertices of the half-ladder manifest.  This is  consistent with our conjectured all multiplicity expression for the unit-scalar building block given in \eqn{mPointConjecture}. With this relaxed numerator, it is trivial to generate the first independent permutation invariant in scalar weights:
\begin{equation}
\permNum^{[2]} \equiv \compPerm{\relaxedNum^{(1)} }{\relaxedNum^{(1)}}
\end{equation}
As we shall see, this linear solution to the relaxed adjoint constraints provides the unit step we need to climb to all mass-dimensions in each of our algebraic structures.

At four points, we found a ladder of scalar adjoint numerators at each order in momentum invariants through repeated composition with the linear adjoint solution $n^{ss}$. There is no linear adjoint solution at five points, and repeated composition with the simplest adjoint solution (cubic) would only generate solutions of orders 3, 6, 9, etc., missing potential adjoint solutions at order 4, 5, 7, and so forth.  In contrast, our linear \textit{relaxed} almost-adjoint building block allows for the generation of a complete ladder of scalar numerators for every building block. Given the apparent primacy of relaxed weights, we will therefore begin with a discussion of scalar kinematics and color solutions of relaxed adjoint structures.

\subsection{Relaxed building blocks}

\subsubsection{Relaxed color weights}
The relaxed almost-adjoint constraints of \eqn{relaxedConstraints} can be satisfied  by color-weights using a spanning combination of $f^3d^3f^3$-constraint satisfying color-weights which satisfy the more general sandwich constraints of \eqns{sandConstraintsSym}{sandConstraintsSix}, 
\begin{equation}
\label{relaxedColor}
c_\relaxedNum(12345) = 6 \,c_{\sandwichNum,1}  + 2\,c_{\sandwichNum,2}  -3\, c_{\sandwichNum,3}\,.
\end{equation}
Explicit forms for $c_{\sandwichNum,i}$ in terms of  $f^3d^3f^3$ color-dressings will be given subsequently in \EqnsTh{sandwich1}{sandwich2}{sandwich3}.
Additional $f^3d^3f^3$ dressings of orderings beyond $(12345)$ are necessary to ensure that $c_\relaxedNum (12345)$ satisfies the Jacobi identities that represents additional constraints on the sandwich structure automatically satisfied by $f^3d^3f^3$ alone. 

\subsubsection{Relaxed scalar kinematic weights}
Our composition rules and the linear-in-Mandelstams relaxed building block $\rOne$ give us the necessary tools with which to construct additional relaxed scalar kinematic results that satisfy the relaxed constraints given in \eqn{relaxedConstraints}. We may compose $\rOne$ with itself to generate a relaxed weight quadratic in the Mandelstam invariants:  
\begin{equation}
\relaxedNum^{(2)} = \compRel{\relaxedNum^{(1)}}{\relaxedNum^{(1)}} \, .
\end{equation}
Explicit ansatz calculation confirms that this result, $\relaxedNum^{(2)}$, is the only possible solution to the relaxed constraints at this mass dimension. 

Making use of the quadratic permutation invariant $\permNum^{[2]}$, we may write down two available structures for a third order relaxed numerator, 
\begin{equation}
\left\{\relaxedNum^{(3)}_i\right\} = \left\{  \compRel{\relaxedNum^{(1)}}{\relaxedNum^{(2)}} , ~ \rOne \permNum^{[2]}  \right\} \, .
\end{equation}
Similarly, all fourth order relaxed numerators may be generated from: 
\begin{equation}
\{\relaxedNum^{(4)}_i\} = \left\{ \compRel{\relaxedNum^{(1)}}{\relaxedNum^{(3)}_j} , ~ \compRel{\relaxedNum^{(2)}}{\relaxedNum^{(2)}} ,~  \rOne \permNum^{[3]} \right\}\,,
\end{equation}
where $\permNum^{[3]} = \compPerm{\rOne}{ \relaxedNum^{(2)}}$.  This has four total independent solutions. 

We may continue to compose to new relaxed numerators in this fashion. As the order grows, the number of possible compositions increases, but we find that it is sufficient to consider just compositions with $\rOne$ and $\relaxedNum^{(2)}$ and products of lower order results with permutation invariants (whose full definitions we will soon provide in terms of lower-order relaxed-adjoint weights),
\begin{equation}
\left\{\compRel{\relaxedNum^{(1)}}{\relaxedNum^{(m-1)}_i},~ \compRel{\relaxedNum^{(2)}}{\relaxedNum^{(m-2)}_j },~ \rOne \permNum^{[m-1]}\right\}
\end{equation}
where there may be some redundancy between the different terms written. This form allows us to build up a ladder of relaxed numerators. At ninth order, we find the compositions cease giving new structures, and the ladder closes simply to products of lower order relaxed numerators with permutation invariants, so we conjecture it is possible to express all relaxed numerator weights as follows: 
\begin{equation} 
\label{relaxedNums}
\left\{ \relaxedNum^{(m)} \right\} = \left\{
        \begin{array}{ll}
          \left\{\compRel{\relaxedNum^{(1)}}{\relaxedNum^{(m-1)}_i},~ \compRel{\relaxedNum^{(2)}}{\relaxedNum^{(m-2)}_j },~ \rOne \permNum^{[m-1]}\right\} & \quad 3 \leq m \leq 8 \,, \\
       \bigcup  \limits_{ { 1\le i \le8 ,  i+j=m}  }  \relaxedNum^{(i)}\permNum^{[j]}  & \quad m \geq 9 \,.
        \end{array}
    \right. 
\end{equation}
We have verified this constructive approach, starting with $\relaxedNum^{(1)}$ and generating all higher orders via composition, against explicit ansatz calculation through twelve powers of momentum invariants (or a mass-dimension of 24), and expect it to continue to all orders in mass-dimension.  We record the number of independent solutions at each order in Table~\ref{relaxedTable}. 

\begin{table*}
\begin{center}
\caption{Total number of linearly-independent relaxed numerator scalar solutions at each order in momentum invariants.}
\label{relaxedTable}
\begin{tabular}{ c|c|c} 
 \toprule
{\small Order} & Building Blocks & Solutions  \\
\midrule
2 & $\compRel{\relaxedNum^{(1)}}{\relaxedNum^{(1)}}$ & 1 \\
\midrule
3 & $    \left\{\compRel{\relaxedNum^{(1)}}{\relaxedNum^{(2)}_i},~ \rOne \permNum^{[2]}\right\} $ & 2 \\
\midrule
4 & $    \left\{\compRel{\relaxedNum^{(1)}}{\relaxedNum^{(3)}_i},~ \compRel{\relaxedNum^{(2)}}{\relaxedNum^{(2)}_j },~ \rOne \permNum^{[3]}\right\} $& 4 \\
\midrule
5 & $    \left\{\compRel{\relaxedNum^{(1)}}{\relaxedNum^{(4)}_i},~ \compRel{\relaxedNum^{(2)}}{\relaxedNum^{(3)}_j },~ \rOne \permNum^{[4]}\right\} $& 7 \\
\midrule
6 &$    \left\{\compRel{\relaxedNum^{(1)}}{\relaxedNum^{(5)}_i},~ \compRel{\relaxedNum^{(2)}}{\relaxedNum^{(4)}_j },~ \rOne \permNum^{[5]}\right\} $& 10 \\
\midrule
7 & $    \left\{\compRel{\relaxedNum^{(1)}}{\relaxedNum^{(6)}_i},~ \compRel{\relaxedNum^{(2)}}{\relaxedNum^{(5)}_j },~ \rOne \permNum^{[6]}\right\} $ & 17 \\
\midrule
8 &$    \left\{\compRel{\relaxedNum^{(1)}}{\relaxedNum^{(7)}_i},~ \compRel{\relaxedNum^{(2)}}{\relaxedNum^{(6)}_j },~ \rOne \permNum^{[7]}\right\} $ & 23 \\
\midrule
9 & $  \bigcup  \limits_{ { 1\le i \le8 ,  i+j=9}  }  \relaxedNum^{(i)}\permNum^{[j]}  $ & 33 \\
\midrule
10 & $  \bigcup  \limits_{ { 1\le i \le8 ,  i+j=10}  }  \relaxedNum^{(i)}\permNum^{[j]}  $ & 46 \\
\midrule
11 &  $  \bigcup  \limits_{ { 1\le i \le8 ,  i+j=11}  }  \relaxedNum^{(i)}\permNum^{[j]}  $ & 62 \\
\midrule
12 &  $  \bigcup  \limits_{ { 1\le i \le8 ,  i+j=12}  }  \relaxedNum^{(i)}\permNum^{[j]}  $ & 80 \\
\bottomrule
\end{tabular}
\end{center}
\end{table*}

\subsection{Adjoint building blocks}

\subsubsection{Adjoint color weights}
It is unsurprising that dressing each cubic vertex with $f^3f^3f^3$ satisfies adjoint constraints, but we find an additional solution given as a linear combination of hybrid $d^4f^3$ contractions: 
\begin{equation}
\label{eqn:adjointColor1}
c_{\adjNum,1}(12345) =  f^3f^3f^3(12345) 
\end{equation}
\begin{equation}
\begin{aligned}
\label{eqn:adjointColor2}
c_{\adjNum,2}(12345) =2\,d^4f^3(12345)+d^4f^3(12435)-d^4f^3(12534)\\
+2\,d^4f^3(23415)-2\,d^4f^3(23514)
\end{aligned}
\end{equation}
This second solution will prove critical for finding factorizing solutions that correspond to color-symmetric four-point contact operators (accompanied by $d^4$ color) with three-point Yang Mills insertions ($f^3$).

\subsubsection{Adjoint scalar kinematic weights}
With relaxed numerators in hand, we may proceed to construct adjoint numerators. The composition rule $\compAdj{\relaxedNum^{(a)}}{\relaxedNum^{(b)}}$  is odd under the interchange of its arguments $\relaxedNum^{(a)}$ and $\relaxedNum^{(b)}$, so 
\begin{equation}
\compAdj{\relaxedNum^{(1)}}{\relaxedNum^{(1)}}= 0
\end{equation}
This is consistent with the finding that no ansatz quadratic in momentum invariants satisfies adjoint constraints. The simplest adjoint solution, at third order in momentum invariants, is simply the composition of the linear and quadratic relaxed numerators: 
\begin{equation}
\adjNum^{(3)} = \compAdj{\relaxedNum^{(1)}}{\relaxedNum^{(2)}} = \compAdj{\relaxedNum^{(1)}}{\compRel{\relaxedNum^{(1)}}{\relaxedNum^{(1)}} 
} 
\end{equation} 
Once we have this unique mass-dimension adjoint weight, we can climb up the ladder of higher-order adjoint weights by composing with the linear relaxed scalar $\rOne$, as well as considering trivial permutation invariant products with lower-order adjoint weights,
\begin{equation}
\label{adjScalarWeights}
\adjNum^{(m)} = \left\{
        \begin{array}{ll} 
         \compAdj{\relaxedNum^{(1)}}{\adjNum^{(m-1)}}    
                                   & \quad 3 \leq m \leq 10 \text{, even} \\
        \left\{      \compAdj{\relaxedNum^{(1)}}{\adjNum^{(m-1)}}  ,   \adjNum^{(3)}\left(\permNum^{[2]}\right)^{(m-3)/2}  \right\} & \quad 3 \leq m \leq 10 \text{, odd} \\
          \bigcup  \limits_{ { 1\le i \le10 , ~ i+j=m}  } \adjNum^{(i)}\permNum^{[j]} & \quad m \geq 11
        \end{array} 
    \right.
\end{equation}
closing to simply products of lower-order adjoint results with permutation invariants at eleventh order. A summary of these results, which have been confirmed against ansatze, is given in Table~\ref{adjointTable}.

\begin{table*}
\begin{center}
\caption{Total number of linearly-independent adjoint numerator scalar solutions at each order in momentum invariants.}
\label{adjointTable}
\begin{tabular}{ c|c|c} 
 \toprule
{\small Order} & Building Blocks & Solutions  \\
\midrule
3 & $   \compAdj{\relaxedNum^{(1)}}{\adjNum^{(2)}} $ & 1 \\
\midrule
4 & $   \compAdj{\relaxedNum^{(1)}}{\adjNum^{(3)}} $  & 2 \\
\midrule
5 & $ \left\{      \compAdj{\relaxedNum^{(1)}}{\adjNum^{(4)}}  ,   \adjNum^{(3)}\left(\permNum^{[2]}\right)^{}  \right\}  $  & 5 \\
\midrule
6 &  $   \compAdj{\relaxedNum^{(1)}}{\adjNum^{(5)}} $  & 8 \\
\midrule
7 & $ \left\{      \compAdj{\relaxedNum^{(1)}}{\adjNum^{(6)}}  ,   \adjNum^{(3)}\left(\permNum^{[2]}\right)^{2}  \right\} $  & 14 \\
\midrule
8 &  $   \compAdj{\relaxedNum^{(1)}}{\adjNum^{(7)}} $ & 21 \\
\midrule
9 & $\left\{      \compAdj{\relaxedNum^{(1)}}{\adjNum^{(8)}}  ,   \adjNum^{(3)}\left(\permNum^{[2]}\right)^{3}  \right\} $  & 32 \\
\midrule
10 &  $   \compAdj{\relaxedNum^{(1)}}{\adjNum^{(9)}} $  & 45 \\
\midrule
11 & $   \bigcup  \limits_{ { 1\le i \le10 , ~ i+j=11}  } \adjNum^{(i)}\permNum^{[j]} $ & 63 \\
\midrule
12 & $   \bigcup  \limits_{ { 1\le i \le10 , ~ i+j=12}  } \adjNum^{(i)}\permNum^{[j]} $ & 84 \\
\bottomrule
\end{tabular}
\end{center}
\end{table*}

\subsection{Permutation-invariant building blocks}

\subsubsection{Permutation-invariant color weights}
Permutation-invariant color is simply given by the totally symmetric object $d^5 \equiv d^{abcde}$. We note here that theories dressed with $d^5$ color must have totally permutation-invariant ordered amplitudes $A^d(\sigma)$. We may see this by rewriting such an amplitude, comprising of symmetric $d^5$ color factorized out from permutation-invariant kinematic information $K$, in a basis of traces,
\begin{equation}
\mathcal{A}^d = d^5 \cdot K = \left(  \frac{1}{4!}\sum_{\sigma \in S_4}\Tr(\sigma 5) \right) \cdot K \, .
\end{equation}
Ordered amplitudes are simply defined as the coefficients of individual traces, 
\begin{equation}
\mathcal{A}^d =  \sum_{\sigma \in S_4}\Tr(\sigma 5) A^d (\sigma 5) \, .
\end{equation}
So we may identify: 
\begin{equation}
\label{d5ordered}
A^d (\sigma 5) = \frac{1}{4!} K \quad \forall \hspace{0.3em} \sigma \, .
\end{equation}
$K$ is permutation invariant (otherwise, the full amplitude $\mathcal{A}^d$ would not be Bose-symmetric), so the ordered amplitudes $A^d$ must also be permutation invariant. This is in contrast to amplitudes from theories dressed with standard adjoint color; for example, Yang-Mills ordered amplitudes obey the reflection identity: 
\begin{equation}
\label{reflection}
A^{\text{YM}}_m(12...m) = (-1)^m  A^{\text{YM}}_m(m...21) \, ,
\end{equation}
which, for multiplicity $m = 5$, is incompatible with the total permutation invariance of $A^d$ ordered amplitudes.

\subsubsection{Permutation-invariant scalar kinematic weights}
We may construct permutation invariants from relaxed scalar kinematic solutions using the composition rule (Eq.~\ref{eq:PIfromR}) reproduced here: 
\begin{equation}
\compPerm{\relaxedNum^j}{\relaxedNum^k}  = \sum_{g \in \Gamma_{3}}^{15} \relaxedNum^j_g \relaxedNum^k_g \,.
\end{equation} 
The primary linear relaxed solution naturally gives rise to the quadratic permutation invariant quantity $\permNum^{[2]}$: 
\begin{equation}
\permNum^{[2]} = \compPerm{\rOne}{\rOne} =  \sum_{g \in \Gamma^{(5)}_3}^{15} \rOne_g \rOne_g \, .
\end{equation}
This can be easily verified to be the unique solution to permutation-invariant constraints at second order via an explicit ansatz calculation. Similarly, the unique cubic permutation invariant is also simply a composition: 
\begin{equation}
\permNum^{[3]} = \compPerm{\rOne}{ \relaxedNum^{(2)}} \, .
\end{equation}

At fourth order, we are able to write down two structures,
\begin{equation}
\permNum^{[4]} = \left\{ \compPerm{ \rOne}{ \relaxedNum^{(3)}},\compPerm{ \relaxedNum^{(2)}} { \relaxedNum^{(2)}} \right\} \, .
\end{equation}
We pause here to note that one of the two permutation-invariant structures in $\permNum^{[4]}$ is simply the lower-order permutation invariant $\permNum^{(2)}$ squared. The other is a new solution at fourth order that is not spanned by products of lower-order permutation invariants, denoted as $\permNum^{(4)}$ and defined by: 
\begin{equation}
 \permNum^{(4)}  = \permNum^{[4]} \setminus  \left( \permNum^{(2)} \right)^2
\end{equation}
We will use this notation throughout, where $\permNum^{(m)}$ corresponds to \textit{new} permutation invariant structures found at order $m$, and $\permNum^{[m]}$ to \textit{all} permutation invariant combinations at order $m$, including products of lower order results: 
\begin{equation}
\label{eq:piNotation}
\permNum^{(m)} =  \permNum^{[m]}  \setminus  \left( \bigcup \limits_{ |\vec{v}|=m } \prod_i \permNum^{(v_i)} \right)\,
\end{equation}
where the union is over all $\vec{v}$ such that $\sum_i v_i = m$, restricting of course to non-negative integer-valued $v_i$.
For orders $m=2$ and $m=3$, there is no distinction between $\permNum^{(m)}$ and $\permNum^{[m]}$. 

We find that we can write a general form for building the permutation invariants at a given order solely from relaxed numerators of lower orders:  
\begin{equation}
\permNum^{[m]} = \bigcup  \limits_{ {  i + j = m}  } \compPerm{\relaxedNum^{(i)}}{\relaxedNum^{(j)}} = \bigcup  \limits_{ {  i + j = m}  }  \sum_{g \in \Gamma_3}^{15} \relaxedNum^{(i)}_g \relaxedNum^{(j)}_g
\end{equation} 
This construction alone turns out to be sufficient for finding all unique permutation invariant combinations of five-point momentum invariants, verified by comparison against ansatz calculations. 

We find new permutation invariant structures appearing through ninth order;  permutation invariants starting from the tenth order and higher are completely spanned by linear combinations of products of permutation invariants of lower mass dimension. The observation that permutation invariants contribute no novel information past the ninth order (or mass dimension 18) has been made before, see e.g. \cite{Boels:2013jua} and references therein. This is consistent with the notion that the ladder of relaxed numerators closes at the eighth order. Permutation invariant results are summarized in Table~\ref{PITable}. 

\begin{table*}
\begin{center}
\caption{Total number of linearly-independent permutation-invariant scalar solutions, and number of unique solutions not spanned by products of lower-order solutions, at each order in momentum invariants.}
\label{PITable}
\begin{tabular}{ c|c|c} 
 \toprule
{\small Order} & Unique Solutions $\permNum^{(m)}$ & Total Solutions $\permNum^{[m]}$  \\
\midrule
2 & 1 & 1 \\
\midrule
3 & 1 & 1 \\
\midrule
4 & 1 & 2 \\
\midrule
5 & 1 & 2 \\
\midrule
6 & 2 & 5 \\
\midrule
7 & 1 & 4 \\
\midrule
8 & 1 & 8 \\
\midrule
9 & 1 & 9 \\
\midrule
10 & 0 & 13 \\
\midrule
11 & 0 & 15 \\
\midrule
12 & 0 & 23 \\
\bottomrule
\end{tabular}
\end{center}
\end{table*}

\subsection{Hybrid building blocks}
\subsubsection{Hybrid color weights}
\label{hybridColor}
The hybrid constraints are satisfied by $d^4f^3(\langle abc \rangle de)$, but they admit an additional solution in the same surprising spirit as adjoint color, 
\begin{equation}
c_{\hybridNum,1}(12345) = d^4f^3(12345) 
\end{equation}
\begin{equation}
\begin{aligned}
c_{\hybridNum,2}(12345) = f^3f^3f^3(12345)-3\,f^3f^3f^3(12435)+f^3f^3f^3(13245)\\
-3\,f^3f^3f^3(13425)+3\,f^3f^3f^3(14235)+3\,f^3f^3f^3(14325)
\end{aligned}
\end{equation}
The two hybrid color solutions bear a striking similarity to the two adjoint color solutions, which are reproduced below: 
\begin{equation}
c_{\adjNum,1}(12345) =  f^3f^3f^3(12345) 
\end{equation}
\begin{equation}
\begin{aligned}
c_{\adjNum,2}(12345) =2\,d^4f^3(12345)+d^4f^3(12435)-d^4f^3(12534)\\
+2\,d^4f^3(23415)-2\,d^4f^3(23514)
\end{aligned}
\end{equation}
Hybrid color contains a non-trivial linear combination of $f^3f^3f^3$ structures that satisfies the hybrid constraints more naturally associated with $d^4f^3$, and adjoint color contains a non-trivial linear combination of $d^4f^3$ structures that satisfies the adjoint constraints more naturally associated with $f^3f^3f^3$. 

Such symmetry suggests the symbolic mapping between hybrid and adjoint solutions that we refer to as \textit{casting}. As previously defined in~\ref{castingDiscussion}, a weight $\hybridNum$ that satisfies hybrid constraints may be cast into an adjoint solution $\adjNum[\hybridNum]$ by writing down the following combination: 
\begin{equation}
\begin{aligned}
\adjNum[\hybridNum](12345) =  -\frac{1}{5} \Big( 2\,\hybridNum(12345)+\hybridNum(12435)-\hybridNum(12534)\\
+2\,\hybridNum(23415)-2\,\hybridNum(23514) \Big)
\end{aligned}
\end{equation}
where $c_{\adjNum,2} \propto \adjNum[d^4f^3]$ and $\adjNum[c_{\hybridNum,2}] \propto f^3f^3f^3$. Similarly, an adjoint solution $\adjNum$ may be cast into a hybrid solution $\hybridNum[\adjNum]$ as follows:
\begin{equation}
\begin{aligned}
\hybridNum[\adjNum](12345)  = \frac{1}{2} \Big( \adjNum(12345)-3\,\adjNum(12435)+\adjNum(13245)\\
-3\,\adjNum(13425)+3\,\adjNum(14235)+3\,\adjNum(14325) \Big)
\end{aligned}
\end{equation}
where $c_{\hybridNum,2} \propto \hybridNum[f^3f^3f^3]$ and $\hybridNum[c_{\adjNum,2}] \propto d^4f^3$. We also note that these casting maps are invertible, in the sense that $\adjNum[\hybridNum[a]] = a$ and $\hybridNum[\adjNum[h]] = h$. These mappings are not restricted to pure color solutions, but hold for any weights satisfying these algebraic relations (including higher-derivative color weights and even vector weights). We will find that casting gives us an easy handle on generating hybrid solutions from adjoint ones, and vice versa. 

\subsubsection{Hybrid scalar kinematic weights}
All hybrid scalar kinematic solutions may then be obtained simply as casts of adjoint scalar kinematics:
\begin{equation}
\hybridNum^{(m)} = \hybridNum\left[\adjNum^{(m)}\right]
\end{equation}
This has been verified by comparison to ansatz calculations through order $m=12$. As a consequence, the number of hybrid scalar kinematic solutions at each order is the same as the number of adjoint solutions. Note that this relation may be inverted:
\begin{equation}
\adjNum^{(m)} = \adjNum\left[\hybridNum^{(m)}\right]
\end{equation}

This one-to-one correspondence between hybrid and adjoint scalar kinematics extends to composition structure. The hybrid from relaxed composition rule $\compHybrid{\relaxedNum^{j}}{\relaxedNum^{k}} $ is odd under the interchange of its arguments, so $\compHybrid{\rOne}{\rOne} = 0$, concordant with the finding that there is no solution to imposing hybrid constraints on a quadratic ansatz \textit{and} consistent with the lack of a quadratic adjoint solution. The simplest hybrid solution, $\hybridNum^{(3)} = \hybridNum[\adjNum^{(3)}]$, is equivalently given simply by the composition of the linear and quadratic relaxed weights: 
\begin{equation}
\hybridNum^{(3)} = \compHybrid{\rOne}{\relaxedNum^{(2)}} =  \hybridNum \left[\compAdj{\rOne}{\relaxedNum^{(2)}} \right]
\end{equation} 

The same pattern that was established for generating a ladder of adjoint scalar kinematic numerators in \eqn{adjScalarWeights} thus applies to hybrid scalar kinematic numerators:
\begin{equation}
\hybridNum^{(m)} = \left\{
        \begin{array}{ll} 
         \compHybrid{\relaxedNum^{(1)}}{\hybridNum^{(m-1)}}     
                          & \quad 3 \leq m \leq 10 \text{, even} \\
        \left\{      \compHybrid{\relaxedNum^{(1)}}{\hybridNum^{(m-1)}}  ,   \hybridNum^{(3)}\left(\permNum^{[2]}\right)^{(m-3)/2}  \right\} & \quad 3 \leq m \leq 10 \text{, odd} \\
          \bigcup  \limits_{ { 1\le i \le10 , ~ i+j=m}  } \hybridNum^{(i)}\permNum^{[j]} & \quad m \geq 11
        \end{array} 
    \right.
\end{equation}
The composition structures close to products of lower order hybrid results with permutation invariants at eleventh order.

\subsection{Sandwich building blocks}
\subsubsection{Sandwich color weights}
There are three basis color solutions spanning solutions to the sandwich constraints of \eqns{sandConstraintsSym}{sandConstraintsSix}, all comprised of $f^3d^3f^3$ structures. The first is simply given by: 
\begin{equation}
\label{sandwich1}
c_{\sandwichNum,1}(12345) = f^3d^3f^3(12345)
\end{equation}
whereas the other two solutions are non-trivial linear combinations: 
\begin{equation}
\label{sandwich2}
\begin{aligned}
c_{\sandwichNum,2}(12345) = f^3d^3f^3(12534)-f^3d^3f^3(13524)+f^3d^3f^3(14523)
\end{aligned}
\end{equation}
\begin{equation}
\label{sandwich3}
\begin{aligned}
c_{\sandwichNum,3}(12345) = f^3d^3f^3(12534)-f^3d^3f^3(12435)\\
-f^3d^3f^3(13245)+f^3d^3f^3(23145)
\end{aligned}
\end{equation}
Indeed, the relaxed-adjoint color weight $c_\relaxedNum$ given in \eqn{relaxedColor} may be expressed as a linear combination of sandwich color solutions: 
\begin{equation}
\begin{split}
c_\relaxedNum(12345) &= 6 \,c_{\sandwichNum,1}  + 2\,c_{\sandwichNum,2}  -3\, c_{\sandwichNum,3}\\
&= 6\,f^3d^3f^3(12345)+3\,f^3d^3f^3(12435)-f^3d^3f^3(12534)\\
& +3\,f^3d^3f^3(13245)-2\,f^3d^3f^3(13524)\\
& +2\,f^3d^3f^3(14523)-3\,f^3d^3f^3(23145) \, ,
\end{split}
\end{equation}
as relaxed solutions also satisfy the sandwich constraint \eqn{sandConstraintsSix}. This precise combination of sandwich color weights allows $c_\relaxedNum$ to satisfy the more stringent relaxed algebraic constraints in \eqn{relaxedConstraints} of Jacobi relations on each internal edge.

\subsubsection{Sandwich scalar kinematic weights}
As $\relaxedNum_g$ weights automatically satisfy sandwich constraints, we know there is at least one solution to sandwich constraints linear in momentum invariants. Applying an ansatz to the sandwich constraints recovers uniquely $\sandwichNum^{(1)}(abcde)=\relaxedNum^{(1)}(abcde)$. For higher mass dimensions, we see additional solutions to sandwich constraints that are not compatible with Jacobi and thus do not appear as relaxed solutions. For instance, there are two unique solutions to sandwich constraints applied to a quadratic ansatz, which may be obtained via the composition of the linear relaxed solution with itself:
\begin{equation}
\sandwichNum^{(2)} = \compSandwich{\relaxedNum^{(1)}}{\relaxedNum^{(1)}}
\end{equation}
Higher orders may be obtained via a simple composition structure, which closes to products of lower order sandwich results and permutation invariants at tenth order:
\begin{equation}
\sandwichNum^{(m)} = \left\{
        \begin{array}{ll}
          \left\{  \compSandwich{\relaxedNum^{(1)}}{\sandwichNum^{(m-1)}},  \compSandwich{\sandwichNum^{(2)}}{\sandwichNum^{(m-2)}}\right\} & \quad 3 \leq m \leq 9 \\
          \bigcup  \limits_{ { 1\le i \le9 ,  i+j=m}  }  \sandwichNum^{(i)}\permNum^{[j]} & \quad m \geq 10
        \end{array}
    \right.
\end{equation}
A summary of these results is given in Table~\ref{sandwichTable}. We note here that all scalar results --- adjoint, relaxed adjoint, hybrid, sandwich, and permutation invariant --- have been generated with \textit{only one single building block}, the linear relaxed adjoint solution $\rOne$, as input. Once $\rOne$ has been found using an ansatz, all additional weights may be found constructively using composition rules. Each of these scalar kinematic and color solutions are included in an auxiliary Mathematica file~\cite{ancFiles}.

\begin{table*}
\begin{center}
\caption{Total number of linearly-independent sandwich numerator scalar solutions at each order in momentum invariants.}
\label{sandwichTable}
\begin{tabular}{ c|c|c} 
 \toprule
{\small Order} & Building Blocks & Solutions  \\
\midrule
2 &  $\compSandwich{\relaxedNum^{(1)}}{\sandwichNum^{(1)}}$  & 2 \\
\midrule
3 &  $\compSandwich{\relaxedNum^{(1)}}{\sandwichNum^{(2)}}$ & 4 \\
\midrule
4 &  $ \compSandwich{\relaxedNum^{(1)}}{\sandwichNum^{(3)}}$, $\compSandwich{\sandwichNum^{(2)}}{\sandwichNum^{(2)}}$ & 8 \\
\midrule
5 &   $\compSandwich{\relaxedNum^{(1)}}{\sandwichNum^{(4)}}$ & 14 \\
\midrule
6 &  $\compSandwich{\relaxedNum^{(1)}}{\sandwichNum^{(5)}}$ & 21 \\
\midrule
7 & $\compSandwich{\sandwichNum^{(2)}}{\sandwichNum^{(5)}}$ & 34 \\
\midrule
8 &  $\compSandwich{\relaxedNum^{(1)}}{\sandwichNum^{(7)}}$ & 49 \\
\midrule
9 &   $\compSandwich{\relaxedNum^{(1)}}{\sandwichNum^{(8)}}$ & 70 \\
\midrule
10 & $   \bigcup  \limits_{ { 1\le i \le9 ,  i+j=10}  } \sandwichNum^{(i)}\permNum^{[j]}$ & 98 \\
\midrule
11 & $  \bigcup  \limits_{ { 1\le i \le9 ,  i+j=11}  }\sandwichNum^{(i)}\permNum^{[j]}$ & 132 \\
\midrule
12 & $  \bigcup  \limits_{ { 1\le i \le9 ,  i+j=12}  }\sandwichNum^{(i)}\permNum^{[j]}$ & 173 \\
\bottomrule
\end{tabular}
\end{center}
\end{table*}

%% file: hdCorr.tex
\section{Building Higher-Derivative Color Weights}
\label{section:hdCorr}
In previous sections we identified a number of color and scalar building blocks, as well as the algebraic structures that could allow them be combined into higher-derivative adjoint-type color-weights.  We will now compose color structures with scalar kinematic numerators into \textit{higher-derivative adjoint color factors} $c^{\text{HD}}$.  These can be paired, in a double-copy sense, with adjoint vector weights $n^{\rm vec}$ to obtain higher-derivative corrections to gauge theory; we will discuss this in \sect{gaugeGravCorr}, along with how they can themselves be promoted to higher-derivative vector weights to ultimately land on higher-derivative corrections to gravity.  Here we will focus on building these higher-derivative modified color-weights.

We wish to ground this discussion in terms of full-amplitudes. Color-dual structure allows us our discussion to remain general while focusing on a simple case.  It is entirely sufficient for our purposes to  consider the pairing of $c^{\text{HD}}$ with secondary adjoint color weights $\tilde{c}_g$ to form a bi-colored theory:
\begin{equation}
\mathcal{A}_5^{\text{BC+HD}} = \sum_{g}\frac{\tilde{c}_g c^{\text{HD}}_g}{d_g}
\end{equation} 
Note that, since we in general allow $c^{\text{HD}}$ to contain color-weights that are not strictly adjoint $f^3f^3f^3$, we emphasize that we are considering a generic bi-colored theory and not necessarily a bi-adjoint theory (it is only the $\alpha^{\prime} \rightarrow 0$ limit of the theory, without higher-derivative corrections, that is strictly bi-adjoint). The interplay between scalar kinematics and more generic color structures allows us to encode non-adjoint color factors in adjoint weights $c^{\text{HD}}$ and thus in adjoint-striated double copy amplitudes. 

The higher-derivative color-weights at five-points can be clustered into two types: those associated with local operators at five points (and hence show up as contact terms), and those that factorize on physical poles and are thus associated with higher-derivative lower-multiplicity local operators.  

\subsection{Factorizing higher-derivative corrections} 
Here we consider constructing amplitudes that consistently factorize to known results from lower multiplicity.  We do so by introducing a procedure that lets us span all possible potentially factorizing solutions at a given mass-dimension, whose coefficients need simply be fixed on one particular cut.  
 
Factorizing five point amplitudes will depend on operators relevant to lower-multiplicity.  Because our mixed kinematic-color weights $c^{\text{HD}}$ are functional, we only need consider distinct topologies, not distinct labels.  As the secondary adjoint color-weight $\tilde{c}$ factorizes trivially, and in truth we are fundamentally interested in the possible contributions of $c^{\text{HD}}$ in any case, we need consider only color-ordered cuts (ordering on the secondary color factor $\tilde{c}$), which will themselves be functional. Because scalar momentum invariants vanish on for real, on-shell three-point kinematics, there are no higher-derivative scalar three-point corrections. Correspondingly, the only color-weight relevant to three points that satisfies adjoint constraints is $f^{abc}$ -- we cannot, for example, modify a $d^{abc}$ color-structure with momentum-invariants to give it the necessary antisymmetry properties (as is our method for generating adjoint $\cHD$ weights from non-adjoint color at four points and beyond).  As such, we need only consider the following ordered cut:
\begin{align}
\label{factorizingCondition}
 \lim_{s_{12}\to0} s_{12} A_5^{\text{BC+HD}}({1,2,3,4,5})  &= \sum_{s} A^{\rm bi-adj}(1,2,l^s) A^{\text{BC+HD}}(-l^{\bar s},3,4,5)\\
 &= \sum_{b} f^{a_1a_2b} A_4^{\text{BC+HD}}(-l^{b},3,4,5)\,
 \end{align}
 where, since we are considering scalar amplitudes, the generic sum over states reduces simply to a sum over the adjoint-color-index $b$ of the cut leg.

Given the local four-point higher-derivative corrections reviewed in \sect{section:fourPoints}, we must land on four-point amplitudes $A_4^{\text{BC+HD}}$ involving the color structures $f^3f^3$ and $d^4$. In order for color to factorize consistently, our factorizing five point amplitudes are thus restricted to  $c^{\text{HD}}$ involving color structures $f^3f^3f^3$ and $d^4f^3$. Correspondingly, we will consider compositions of kinematics with the general five point adjoint color numerators $c_\adjNum$ in \eqns{eqn:adjointColor1}{eqn:adjointColor2}, given in terms of $f^3f^3f^3$ and $d^4f^3$ type color weights respectively.  The necessary spanning five-point candidate $c^{\text{HD}}$ numerators at a given mass dimension are then expressed simply through repeated composition of each $c_\adjNum$ weight with an appropriate number of linear relaxed numerator $\rOne$ dressings as well as products of these two $c_\adjNum$ weights with appropriate mass-dimension scalar permutation invariants.   We have verified explicitly that this approach is sufficient to capture all factorizing solutions  through mass-dimension 18, or nine orders in momentum invariants.

Let us introduce the following notation for the nested application of composition of $\rOne$ with generic color weight $c$ to yield higher-derivative adjoint weights:
\begin{equation}
\compNest{n}{c}= \left\{
    \begin{array}{ll} 
             \compAdj{\rOne}{\compNest{n-1}{c}}  & \quad n>1 \\
          \compAdj{\rOne}{c}  & \quad n=1 
        \end{array}  \right.
\end{equation}
(When $c = c_{\adjNum}$, we may also define $\compNest{0}{c_{\adjNum}} = c_{\adjNum}$; for non-adjoint $c$, no such $n=0$ adjoint weight exists.) We find that such continued nested compositions are needed until seventh order.  Higher orders hold no new adjoint structure modulo permutation invariants: we need only consider  products of lower order nested compositions with scalar permutation invariants (themselves given ultimately solely by $\rOne$, as described earlier), 
\begin{equation}
\sum_{i+j = m}\, \compNest{i}{c_\adjNum} \, \permNum^{[j]} \quad \quad \quad \quad m > 7\, , 
\end{equation} 
in order to write down a structure that may be fixed on the cut to be consistent with lower-point results. The specific building blocks necessary to construct each order are summarized in Table~\ref{factorizingTable}. For clarity, we emphasize that these building blocks are not themselves consistently factorizing solutions; rather, they contain sufficient freedom to be fixed on \eqn{factorizingCondition} such that they factorize appropriately. The final number of fixed factorizing solutions given in Table~\ref{factorizingTable} at each order necessarily lines up with the number of Wilson coefficients accompanying solutions of each color structure $d^4$ and $f^3f^3$ in the four-point higher-derivative corrections $\mathcal{A}_4^{\text{BC+HD}}$, as generated from the four-point modified color factors $\cHD$ as given in \eqn{eqn:fourPointcHD}.  

These nested compositions and their products with permutation invariants yield all the expressions necessary for writing down factorizing amplitudes, as we have verified through mass-dimension eighteen.  We conjecture that this closed structure above seventh order will persist, requiring only a finite number of building blocks --- the two adjoint color solutions and nested compositions and permutation invariants, which all arise purely from $\rOne$. 

\begin{table*}
\begin{center}
\caption{Necessary building blocks to construct consistently factorizing adjoint modified color numerators $\cHD$, along with the final number of fixed factorizing solutions at each order.}
\label{factorizingTable}
\begin{tabular}{ c|c|c|c} 
 \toprule
{\small Order} & Building Blocks & $d^4f^3$ Solns & $f^3f^3f^3$ Solns  \\
\midrule
2 & $\compNest{2}{c_\adjNum}$, $\permNum^{(2)}\,c_\adjNum$ & 1 & 0 \\
\midrule
3 & $\compNest{3}{c_\adjNum}$, $\permNum^{(3)}\,c_\adjNum $ & 0 & 1 \\
\midrule
4 & $\compNest{4}{c_\adjNum}$ & 1 & 1 \\
\midrule
5 & $\compNest{5}{c_\adjNum}$ & 1 & 1 \\
\midrule
6 & $\compNest{6}{c_\adjNum}$ & 1 & 2 \\
\midrule
7 & $\compNest{7}{c_\adjNum}$ & 1 & 2 \\
\midrule
8 & $\sum_{i+j = 8}\compNest{i}{c_\adjNum}\, \permNum^{[j]}$ & 2 & 2 \\
\midrule
9 & $\sum_{i+j = 9}\compNest{i}{c_\adjNum}\, \permNum^{[j]}$ & 1 & 3 \\
\bottomrule
\end{tabular}
\end{center}
\end{table*}

\subsection{Local contact higher-derivative corrections}

Amplitudes arising from local five-point operators have no poles. At four points, each cubic graph is dressed with only a single propagator, meaning that only one cubic graph contributes to the residue on any given pole. Thus, for all residues to vanish (as is required for a local contact amplitude), the individual graph contributions had to be free of poles. At five points, there are two propagators per cubic graph, so multiple graph channels contribute to the same residue. As a result, we need not require that each channel be individually local, but rather that the numerators conspire so that any residues from one graph cancel with residues from other graphs across all channels. One can identify local contact amplitudes by demanding that all cuts of the functional ordered amplitude of the bicolored scalar theory vanish: 
\begin{equation} 
\label{localCondition}
\lim_{s_{i,i+1} \to 0}  s_{i,i+1}A^{\text{BC+HD}}_{5}(12345) =  0 \quad \quad \quad \quad i \in \{1,2,3,4,5\}
\end{equation}
where it is understood that $s_{5,6} \equiv s_{5,1}$. These conditions ensure there are no residues on any of the poles of $A(12345)$.  This is necessarily a subset of higher-derivative adjoint color-weights. Such local weights are needed to describe higher-derivative operators that contain no cubic or quartic interactions, but only five-point and higher vertices, as such operators cannot contribute to any amplitudes that factorize down to products of four-point and three-point corrections. 

Unlike factorizing amplitudes, local five-point contact amplitudes are not restricted to any specific color structures, so we generically expect all available color structures -- $d^5$, $d^4f^3$, $f^3f^3f^3$, and $f^3d^3f^3$ -- to appear in local higher-derivative contact amplitudes. 

The lowest order five-point contact amplitudes are found at fifth order in momentum invariants (tenth order in mass-dimension) and contain the color structures $d^4f^3$ and $f^3f^3f^3$. Only at sixth order does $f^3d^3f^3$ color becomes relevant; $d^5$ color does not emerge until ninth order. We have found that, through seventh order, nested composition structures like those used for factorizing solutions, appearing in combination with products with permutation invariants, 
\begin{equation}
\compNest{m}{c_\adjNum} + \sum_{i+j=m}\compNest{i}{c_\adjNum}p^{[j]}
\end{equation} 
may be fixed on \eqn{localCondition} to give all $d^4f^3$ and $f^3f^3f^3$ local contact amplitudes, as verified against an explicit ansatz calculation. At sixth order, the $f^3d^3f^3$ local solution may be fixed from a nested composition structure starting with sandwich color $c_{\sandwichNum} = f^3d^3f^3$ rather than adjoint color $c_\adjNum$: 
\begin{equation}
\compNest{m}{c_{\sandwichNum}} + \sum_{i+j=m}\compNest{i}{c_{\sandwichNum}}p^{[j]}
\end{equation} 
Considering the size of the common denominator for the five-point cubic graphs, imposing the condition of vanishing residues analytically quickly becomes inefficient in comparison to a new method for generating local contact amplitudes that we will describe in the next section. By uncovering novel forms of color-kinematics duality and sharpening our focus on the local contact condition, we will be able to climb several orders higher in mass-dimension to find additional higher-derivative contact amplitudes.

%% file: otherDualities.tex
\section{Local Dualities Complementing Adjoint}

\label{section:doublyDualLocal}

At five points, we have identified a number of algebraic structures which span single-trace color, and correspondingly many different color factors ($d^5$, $d^4f^3$, $f^3f^3f^3$, or $f^3d^3f^3$). For generic higher-derivative corrections, we must allow for all of these color structures to appear in our amplitudes (corresponding to the most generic traces of operators that may be written down in the action). 

We find that we can use these new algebraic structures to our advantage by considering simultaneous color-dual striations of the same full color-dressed bi-color amplitudes. Specifically, we will introduce here an efficient method for calculating five-point higher-derivative local counterterm predictions that exploit two forms of color-kinematics duality using a manifestly local construction. 

\subsection{Doubly-dual local amplitudes}

All of our higher-derivative bicolored scalar amplitudes have a manifestly adjoint double-copy structure involving two adjoint-type graph weights,
 \begin{equation}
\mathcal{A}^{\text{bi-color}} = \sum_{g \in \Gamma_3}  \frac{\cHD_g \tilde{c}_g }{d_g} \,.
\end{equation}
We encode higher-derivative corrections in the adjoint modified color factor $\cHD$ through a mixture of kinematics and color-weights. The color-weights themselves need not be adjoint; rather, the adjoint behavior of $\cHD$ is ensured by kinematics acting in the right way to complement the color-weights so that $\cHD$ obeys adjoint-type relations.  All higher-derivative corrections that can be associated with  color-weights to form adjoint-type modified color-factors are captured within $\cHD$. We will generically refer to the type of color factors appearing in $\cHD$ as $\fancyColor$, where $\fancyColor$ may be any of $d^5$, $d^4f^3$, $f^3f^3f^3$, or $f^3d^3f^3$.  

For local counterterm predictions, the scalar kinematics within $\cHD_g$ conspire between graphs to cancel out the residue on all possible poles of the amplitude, thus describing a five-point contact interaction. At four points, contact interactions were described by simply demanding that $\cHD_g$ be proportional to the unique propagator associated with graph $g$; at five points, however, this cancellation of denominators for cubic-graphs is non-trivial, as multiple graphs contribute to each pole. 

Next, we have the ubiquitous secondary adjoint color factor $\tilde{c} = \tilde{f}^3\tilde{f}^3\tilde{f}^3$ of some additional gauge group. This constitutes a simple adjoint stand-in that may be replaced in the double copy with an adjoint vector-kinematic-weight like super Yang-Mills, $\tilde{c} \rightarrow n^{\text{sYM}}$, to build higher-derivative corrections to gauge theory. These adjoint color factors $\tilde{c}$ naturally provide an adjoint-double-copy striation of the amplitude along cubic graphs, even though the generic color factors $\fancyColor$ living within $\cHD$ may not traditionally lend themselves to a cubic graph description. 

An interesting opportunity arises for $\fancyColor$ whose natural structure does not lie along cubic graphs, like $d^4f^3$ and $d^5$.  We have the freedom to choose to prioritize the $\fancyColor$ color structures, using them to provide an alternative striation of the amplitude along the graphs $\Gamma_{\fancyColor}$ relevant to their algebraic structure.  After all, it is the color structure $\fancyColor$ that survives the double copy and appears in the final gauge theory amplitude, rather than the adjoint color $\tilde{c} = \tilde{f}^3\tilde{f}^3\tilde{f}^3$. Generically, bi-color amplitudes can be written equivalently as a sum over the graphs relevant to either structure, so we can choose to rewrite $\mathcal{A}^{\text{bi-color}}$ in terms of the graphs $\Gamma_{\fancyColor}$ relevant to the color structure $\fancyColor$ inside the adjoint-type $\cHD$. Each of these graphs is dressed with an appropriate color factor $\fancyColor$ and a local function $\cFunc$ of both the higher-derivative kinematics that once resided in $\cHD$ and the remaining color $\tilde{c}$,
\begin{equation}
\mathcal{A}^{\text{bi-color}} = \sum_{g \in \Gamma_{\fancyColor}}\fancyColor_g ~ \cFunc{}_{,g}\!\left(s_{ij}, \tilde{c} \right)
\end{equation} 
The functions $\cFunc(g)$ can be fixed to two remarkable properties: they are \textit{manifestly local}, containing no propagator structure to be cancelled non-trivially between graphs, and they are \textit{color-dual} to the relevant color structures $\fancyColor$, meaning they obey the same algebraic constraints as $\fancyColor$-type color. To be clear, the functions of $\cFunc$ need not be $\cFunc$-ordered amplitudes, as we allow a summation over all graphs $\Gamma_{\fancyColor}$, not just the basis-graphs for the given algebraic structure. As such, they are rather simply functions that dress individual graphs, are manifestly local, and satisfy $\fancyColor$-type algebraic relations.

For example, for $\fancyColor = d^4f^3$, the $\cFunc(g)$ functions obey hybrid algebraic constraints (symmetry in the first three arguments, antisymmetry in the last two, and four-term Jacobi-like identities). The sum runs over the ten graphs that contain one quartic vertex and one cubic vertex. For $\fancyColor = d^5$, there is just one graph (a single quintic vertex), and its $\cFunc$ function obeys permutation invariant constraints --- a remarkably simpler description than demanding fifteen graph dressings conspiring to cancel two propagators per graph! 

Since these amplitudes enjoy two distinct color-dual striations, we refer to them as \textit{doubly-dual local amplitudes}, each admitting two equivalent descriptions: 
\begin{equation}
\mathcal{A}^{\text{bi-color}} = \sum_{g \in \Gamma_3} \frac{\tilde{c}_g ~ \cHD_g\!\left(s_{ij}, \fancyColor \right) }{d_g} = \sum_{g \in \Gamma_{\fancyColor}} \fancyColor_g ~ \cFunc{}_{,g} \!\left(s_{ij}, \tilde{c} \right)
\end{equation}
The adjoint nature of the $\tilde{c}$-striated amplitude can still be uncovered within the explicitly local $\fancyColor$-dual striation: if the $\tilde{c}$ color factors are expressed in a basis of traces, the coefficient of each trace is an ordered amplitude that obeys the $(m-3)!=2$-basis BCJ 5-point amplitude relations required of any amplitude that may be written as an adjoint double copy.

\subsection{Composing to doubly-dual solutions} 
Multiple paths exist towards finding an appropriately doubly-dual local $\cFunc(g) = \cFunc(g)\left(s_{ij}, \tilde{c} \right)$ function. While it is worth mentioning that an ansatz approach to generating $\cFunc(g)$ is more efficient than constructing an ansatz for $\cHD$ (the $\cHD$ ansatz is larger, as it must compensate for two propagators in the denominator, and, after fixing on adjoint color-kinematics, must additionally be constrained to ensure all residues vanish), both methods are surpassed by composition directly to dual $\cFunc$ functions for some of our favorite algebraic structures $\fancyColor$, as composition ensures the $\cFunc$ will obey the desired $\fancyColor$-type algebraic properties.   Numerous $\fancyColor$-dual functions can be formed at each mass-dimension, but only particular combinations of them admit the type of adjoint-double-copy structure we also require when striating the amplitude along the $\tilde{c}$ color. We arrive at these particular combinations by requiring that the adjoint $\tilde{c}$-striated ordered amplitudes satisfy the  adjoint  $2$-basis amplitude relations.  A summary of the total number of doubly-dual solutions at each order in correction for the different color-algebraic structures $\fancyColor$ may be found in Table~\ref{dualLocalTable}.

The existence of these two distinct color-kinematics dualities at play within a single amplitude thus constitutes an efficient tool for generating  local solutions. Constructing the $\fancyColor$-dual form first via composition is computationally advantageous since it is manifestly local; once a $\fancyColor$-dual expression is found, its $\tilde{c}$-ordered amplitudes may subsequently be fixed on the $2$-basis BCJ relations to impose adjoint color-kinematics duality. To be explicit, the $\fancyColor$-dual amplitude is re-striated along the adjoint color $\tilde{c}$.  Each coefficient of $\tilde{c}$, when all $\tilde{c}$ have been expressed in a in a DDM basis, must represent an adjoint-ordered amplitude if this expression is to be doubly dual,
\begin{equation}
\mathcal{A}  = \sum_{g \in \Gamma_{\fancyColor}} \fancyColor_g ~ \cFunc{}_{,g} \!\left(s_{ij}, \tilde{c} \right)
=  \sum_{\sigma \in S_3} \tilde{c}(1\sigma5) A_{\fancyColor}^{\text{HD}}(1 \sigma 5) \,.
\end{equation}
 Perhaps more familiar to many, and exactly identical, one can in turn express the $\tilde{c}$ in terms of a trace basis, and look at each trace's coefficient. The coefficient of each independent trace must represent an adjoint-ordered amplitude $A_{\fancyColor}^{\text{HD}}$ for our expression to be doubly dual:
\begin{equation}
\mathcal{A}  = \sum_{\sigma \in S_4} \textrm{Tr}(\widetilde{T}^{a_1}\widetilde{T}^{a_{\sigma_1}}\cdots \widetilde{T}^{a_{\sigma_{4}}}) A_{\fancyColor}^{\text{HD}}(1 \sigma) \,.
\end{equation} 
These ordered amplitudes contain the scalar kinematics and $\fancyColor$-type color. We then require that these ordered amplitudes $A_{\fancyColor}^{\text{HD}}$ satisfy the $2$-basis BCJ relation: i
\begin{equation}
\label{bcjRln}
s_{13} A_{\fancyColor}^{\text{HD}}(13245) - s_{35} A_{\fancyColor}^{\text{HD}}(12435) - 
(s_{34} +s_{35}) A_{\fancyColor}^{\text{HD}}(12345) = 0\,.
\end{equation}
This is sufficient to ensure that all $2$-basis BCJ relations are satisfied and that the full amplitude $\mathcal{A}$ has an adjoint color-dual representation.  Why? After establishing the $\fancyColor$-dual algebra of our $\cFunc(g)$ building blocks, our color-dressed $\mathcal{A}$ is manifestly permutation invariant.  This means that any $A_{\fancyColor}$ coefficient in a color basis either $\tilde{c}$ or it's associated trace basis will be related functionally by relabeling.  It is sufficient then to simply impose a single so-called fundamental BCJ relation. The compatibility of these ordered amplitudes with the $2$-basis field theory relations will eventually ensure that all doubly copy substitutions $\tilde{c} \rightarrow n^{\text{vec}}_{\adjNum}$ yield fully gauge-invariant vector amplitudes.

\begin{table*}
\begin{center}
\caption{Doubly dual local solutions.}
\label{dualLocalTable}
\begin{tabular}{ c|c|c|c|c|c} 
 \toprule
{\small Order} & $d^4f^3$ Solutions & $f^3f^3f^3$ Solutions & $f^3d^3f^3$ Solutions & $d^5$ Solutions & Total \\
\midrule
5 & 1 & 1 & 0 & 0 & 2 \\
\midrule
6 & 2 & 2 & 1 & 0 & 5 \\
\midrule
7 & 5 & 5 & 4 & 0 & 14 \\
\midrule
8 & 9 & 9 & 10 & 0 & 28 \\
\midrule
9 & 16 & 16 & 19 & 1 & 52 \\
\bottomrule
\end{tabular}
\end{center}
\end{table*}

For the purposes of this discussion, we will label the fully fixed functions as $\cFunc$ and the \textit{partially fixed} functions as $\hatcFunc$, where partially fixed means that they have been constrained on $\fancyColor$-type color-kinematics duality, but that they have not yet been fixed such that their $\tilde{c}$-striated ordered amplitudes $A_{\fancyColor}^{\text{HD}}$ obey the adjoint-dependent $2$-basis BCJ amplitude relations. 

\subsubsection{Doubly-dual amplitudes with permutation-invariant color}
For permutation-invariant color $\fancyColor = d^5$, finding a permutation invariant $\hatFancyFunc{\permNum}\!\left(s_{ij}, \tilde{c} \right)$ function is simple from our building blocks: 
\begin{equation} 
\label{permFromAdjoint}
\hatFancyFunc{\permNum}^{(n)} = \compPerm{\adjNum^{(n)}}{\tilde{c}^{\adjNum}}  = \sum_{g \in \Gamma_{3}}^{15} \adjNum^{(n)}_g \tilde{c}^{\adjNum}_g 
\end{equation} 
This is simply the composition of adjoint scalar kinematic numerators with adjoint color to form a permutation invariant.  We emphasize that all of thee adjoint scalar kinematic numerators are constructible via iterated composition of $\rOne$. Only this single unit-step scalar is fundamentally required.

 In general, there can be multiple such permutation invariants at any given mass dimension; these are taken to be summed over, with each given an independent coefficient, thus defining the relevant candidate solution   $\hatFancyFunc{\permNum}^{(n)}$. One then constructs the candidate bi-color amplitude as: 
\begin{equation}
\mathcal{A}^{\text{bi-color}} = d^5 \, \hatFancyFunc{\permNum}^{(n)} \!\left(s_{ij}, \tilde{c} \right) \equiv \sum_{\sigma \in S_4} \textrm{Tr}(\widetilde{T}^{a_1}\widetilde{T}^{a_{\sigma_1}}\cdots \widetilde{T}^{a_{\sigma_{4}}}) A_{\mathfrak{p}}^{\text{HD}}(1 \sigma) 
\end{equation}
These free coefficients within $\hatFancyFunc{\permNum}$ are fixed by requiring that the $\tilde{c}$-striated ordered amplitudes $A_{\mathfrak{p}}^{\text{HD}} = A_{\mathfrak{p}}^{\text{HD}}\left(s_{ij}, d^5 \right)$ satisfy the adjoint amplitude relations. Any remaining coefficients represent distinct valid local operator predictions at this mass-dimension.

We note that $\fancyFunc{\permNum}^{(n)}$ will generate the $\mathcal{O}\left( \left( \ap\right)^{n+2} \right)$ order correction to the theory.  Orders in the dimensionful parameter $\ap$ track the order in higher-derivative correction via the number of additional scalar Lorentz invariants in the amplitude relative to the uncorrected theory, as discussed in the context of \eqn{eqn:fourPointcHD}. Our manifestly local construction of $\cFunc$ containing $n$ dot-products implies that the corresponding adjoint numerators $\cHD$ must contain $(n+2)$ dot-products in order to compensate for the two propagators, versus zero such dot-products in the uncorrected bi-adjoint theory numerators. 

For $n < 7$, the $2$-basis adjoint relation of \eqn{bcjRln} admits only vanishing solutions for $\fancyFunc{\permNum}^{(n)}$; for $n = 7$, or ninth order in $\ap$, there is one solution, and at tenth order in $\ap$, there are two doubly dual local amplitudes with $d^5$ permutation invariant color. This high mass dimension makes it particularly unwieldy to give even a simple example. We will sketch an example of the procedure for identifying this lowest order non-vanishing $\fancyFunc{\permNum}^{7}$, called $\tt{fPI[9]}$ in the associated ancillary files~\cite{ancFiles}.  

As previously discussed in Table~\ref{adjointTable}, there are 14 adjoint scalar solutions $\adjNum^{7}_i$ at seven orders in Lorentz-invariant dot products.  These have been saved as a sum over distinct coefficients in ${\tt adjoint[7]}$ in the ancillary files~\cite{ancFiles}.  We may write the candidate $\hatFancyFunc{\permNum}^{(7)}$ as follows:
\begin{equation}
\hatFancyFunc{\permNum}^{(7)} = \sum_{i=1}^{14} b_i \compPerm{\tilde{c}}{ \adjNum^{7}_i} \, ,
\end{equation}
where each $b_i$ is an ansatz coefficient we give to the composition\footnote{Recall as per \eqn{permFromAdjoint} that this composition corresponds to summing over the 15 cubic graphs, dressing each with a product of both adjoint weights.} of the $i^{\text{th}}$ adjoint building block with $\tilde{c}$ to form a permutation invariant. The candidate expression for the full amplitude is then given as the product of this ansatz with $d^{abcde}$.

We may constrain this to be a doubly-dual local solution by demanding that the coefficients of $\tilde{c}$ satisfy the $2$-basis BCJ amplitude relations. Specifically, after expressing $\tilde{c}$ in a spanning DDM basis where legs one and five are fixed, $\tilde{c}(1\sigma5)$, the coefficient of each $\tilde{c}$ constitutes an ordered amplitude. We constrain this candidate expression for the ordered amplitude $A_{\permNum}^{\text{HD}}(1\sigma5)$ such that \eqn{bcjRln} holds, fixing all but one coefficient as follows:
\begin{multline}
\left\{b_1\to 0,b_2\to -\frac{25 b_{13}}{4},b_3\to \frac{15
   b_{13}}{4},b_4\to \frac{3 b_{13}}{4},b_5\to
   -\frac{b_{13}}{4},b_6\to -\frac{3 b_{13}}{4}, \right.\\
   \left. b_7\to 0,b_8\to
   0,b_9\to 0,b_{10}\to -\frac{b_{13}}{4},b_{11}\to -\frac{3
   b_{13}}{2},b_{12}\to -\frac{3 b_{13}}{4},b_{14}\to 0\right\}\,.
\end{multline}
The lone remaining parameter, $b_{13}$, can be identified as the Wilson coefficient of the corresponding local five-point operator.  In the case where there are multiple unrelated solutions, such as order ten in $\ap$, we have multiple satisfactory operators that can each rightfully have its own Wilson coefficient.

\subsubsection{Doubly-dual amplitudes with hybrid color}
For hybrid color $\fancyColor = c_{\hybridNum,1} =  d^4f^3$, we similarly construct hybrid $\fancyFunc{\hybridNum}$ functions from scalar kinematics and the secondary $\tilde{c}$ adjoint color factors. It turns out to be sufficient to consider nested compositions just like those used for factorizing amplitudes: 
\begin{equation}
\hatFancyFunc{\hybridNum}^{(n)} \equiv \compHybrid{\rOne}{\fancyFunc{\hybridNum}^{(n-1)}} 
\end{equation}
where for $n = 1$, we define $\hatFancyFunc{\hybridNum}^{(1)} \equiv \compHybrid{\rOne}{\tilde{c}_{\adjNum}}$. Again, generically we find multiple distinct solutions to the hybrid conditions at each mass dimension, which will be labeled with independent coefficients and summed over. The corresponding ordered amplitudes $A_{\hybridNum}^{\text{HD}} = A_{\hybridNum}^{\text{HD}}\left(s_{ij}, d^4f^3 \right)$ are then constrained on the 2-basis field-theory relations of \eqn{bcjRln} to find the fully fixed $\fancyFunc{\hybridNum}^{(n)}$ functions. The full amplitude is then given by: 
\begin{equation}
\mathcal{A} = \sum_{g \in \Gamma{}_{\hybridNum}} d^4f^3(g) \fancyFunc{\hybridNum}(g) = \sum_{\sigma \in S_4} \textrm{Tr}(\widetilde{T}^{a_1}\widetilde{T}^{a_{\sigma_1}}\cdots \widetilde{T}^{a_{\sigma_{4}}}) A_{\hybridNum}^{\text{HD}}(1 \sigma) 
\end{equation} 
where the first sum runs over the ten graphs $\Gamma{}_{\hybridNum}$ with one cubic vertex and one quartic vertex, and the second sum over the basis of traces at five points. This first expression makes manifest hybrid color-kinematics duality through the local hybrid-dual $\fancyFunc{\hybridNum}^{(n)}$ graph weights; the second representation exhibits adjoint color-kinematics duality through its \eqn{bcjRln} satisfying ordered amplitudes. 

For the permutation-invariant doubly dual amplitudes, we find a need to fix each individual order's $\hatcFunc$ function such that the $\tilde{c}$-ordered amplitudes satisfy the $2$-basis relations of \eqn{bcjRln}. Hybrid solutions turn out to be delightfully simpler---we can compose directly from BCJ-compatible solution to BCJ compatible solution. To see this, let us run through some examples at low mass dimension. 

Our first objects of interest are the linear nested composition, $\hatFancyFunc{\hybridNum}^{(1)} \equiv \compHybrid{\rOne}{\tilde{c}_{\adjNum}}$, and the quadratic nested composition, $\hatFancyFunc{\hybridNum}^{(2)} = \compHybrid{\rOne}{\hatFancyFunc{\hybridNum}^{(1)}}$; the corresponding ordered amplitudes in both cases admit no solutions that satisfy the 5-point fundamental BCJ relation. The cubic nested composition $\hatFancyFunc{\hybridNum}^{(3)}\equiv \compHybrid{\rOne}{\hatFancyFunc{\hybridNum}^{(2)}}$ contains eight hybrid-color-dual building blocks: both composition rules used, $\compHybrid{\rOne}{\hybridNum}$ or $\compHybrid{\rOne}{\adjNum}$, generate two independent solutions, so there is, for instance, two building blocks within $\hatFancyFunc{\hybridNum}^{(1)}$. By third order, we have $2^3=8$ building blocks, which we label as $h^{(3)}_i$.  We may write the candidate $\hatFancyFunc{\hybridNum}^{(3)}$ as follows:
\begin{equation}
\hatFancyFunc{\hybridNum}^{(3)} = \sum_{i=1}^{8} b_i h^{(3)}_i(\tilde{c},s_{jk}) \,.
\end{equation}
Here, $b_i$ are coefficients of the eight distinct hybrid building blocks $h^{(3)}_i$ that have been constructed via composition out momentum invariants and $\tilde{c}$.  These $b_i$ are to be fixed on the 2-basis BCJ relations at five-points of \eqn{bcjRln} to construct any doubly-dual solutions, if possible. We build the full amplitude by summing over all ten relevant graphs $\Gamma_{\hybridNum}$, each dressed with $d^4f^3$ and our constructed hybrid-dual $\hatFancyFunc{\hybridNum}^{(3)}$. After expressing the $\tilde{c}$ in a DDM basis $\tilde{c}(1\sigma5)$, we may interpret the coefficient of each color weight as an ordered amplitude. These ordered amplitudes must be fixed to satisfy \eqn{bcjRln} in order to represent the fully doubly-dual solution,  
\begin{equation}
\mathcal{A} = \sum_{g \in \Gamma{}_{\hybridNum}} d^4f^3(g) \fancyFunc{\hybridNum}(g) =  \sum_{\sigma \in S_3} \tilde{c}(1\sigma5) A_{\hybridNum}^{\text{HD}}(1 \sigma 5)\,.
\end{equation}
Imposing \eqn{bcjRln} on the candidate expressions for $A_{\hybridNum}^{\text{HD}}$ constrains all but one of the $b_i$ parameters, yielding a single solution at this order.  The unconstrained parameter should be understood as a stand-in for the Wilson coefficient associated with the corresponding higher-derivative local operator.  We label this solution by the number of explicit Lorentz invariants appearing alongside $\tilde{c}$ in the local hybrid weight, $\fancyFunc{\hybridNum}^{(3)}$. This gives rise to a doubly dual contact amplitude at fifth order in $\ap$, as the corresponding higher-derivative adjoint color weight $\cHD$ for this amplitude expressed over a sum of cubic graphs must contain five dot-products to compensate for the two cubic propagators. We therefore call this solution ${\tt fHybrid[5]}$ in the ancillary machine readable files~\cite{ancFiles}, where we index by the order in $\ap$.

At the next order, one could calculate $\hatFancyFunc{\hybridNum}^{(4)} \equiv \compHybrid{\rOne}{\hatFancyFunc{\hybridNum}^{(3)}}$ and then fix the ordered amplitude to generate the final solution $\fancyFunc{\hybridNum}^{(4)}$, but there is a much more direct route: composing the fully fixed cubic solution $\fancyFunc{\hybridNum}^{(3)}$ with $\rOne$: 
\begin{equation}
\fancyFunc{\hybridNum}^{(4)}  \equiv \compHybrid{\rOne}{\fancyFunc{\hybridNum}^{(3)}}
\end{equation}
This composition $\fancyFunc{\hybridNum}^{(4)}$ already yields ordered amplitudes that automatically satisfy the $(m-3)!$ relations, seemingly by virtue of using the fully fixed $\fancyFunc{\hybridNum}^{(3)}$ in the composition! 

At the next order of higher derivative correction, we may play the same sort of game, only now we have two building blocks at our disposal, $\fancyFunc{\hybridNum}^{(3)}$ and $\fancyFunc{\hybridNum}^{(4)}$: 
\begin{equation}
\fancyFunc{\hybridNum}^{(5)} \equiv  \left\{ \compHybrid{\rOne}{\fancyFunc{\hybridNum}^{(4)} },\,  \fancyFunc{\hybridNum}^{(3)} \, \mathfrak{p}^{(2)} \right\}
\end{equation}
giving rise to five independent doubly dual amplitudes at seventh order in $\ap$. We find that this pattern continues.  It is sufficient to simply consider compositions and products with permutation invariants of lower-order fully fixed $\cFunc$ functions to find higher-order fully fixed $\cFunc$ functions through at least ninth order in $\ap$,
\begin{equation}
\fancyFunc{\hybridNum}^{(n)}  \equiv  \left\{ \compHybrid{\rOne}{\fancyFunc{\hybridNum}^{(n-1)}},\, \fancyFunc{\hybridNum}^{(n-2)}\,\mathfrak{p}^{(2)} \right\} \,,
\end{equation}
without any need to fix the adjoint-striated ordered amplitudes at each order on the $(m-3)!$ constraints, as they are already satisfied! Through $\mathcal{O}\left( \left( \ap\right)^{9} \right) $, we have confirmed via explicit ansatz calculations that our composition structures and doubly dual local construction give rise to {\em all} possible contact amplitudes containing scalar higher-derivative corrections and hybrid color, with just the linear relaxed solution $\rOne$ and the relevant color structures as input building blocks.  Although we do not currently have a proof, we expect this simple pattern to continue.

\subsubsection{Doubly-dual amplitudes with adjoint color}
Once hybrid $\fancyFunc{\hybridNum}$ functions are known, obtaining adjoint $\fancyFunc{\adjNum}$ functions for $\fancyColor = c_{\adjNum,1} = f^3f^3f^3$ is rendered trivial via casting:  
\begin{equation}
\fancyFunc{\adjNum}^{(n)} = \adjNum\left[\fancyFunc{\hybridNum}^{(n)}\right]\,.
\end{equation} 
Explicitly, in terms of individual graph labels, 
\begin{equation}
\begin{aligned}
\fancyFunc{\adjNum}(12345)  =  -\frac{1}{5} \Big( 2\,\fancyFunc{\hybridNum}(12345)+\fancyFunc{\hybridNum}(12435)-\fancyFunc{\hybridNum}(12534)\\
+2\,\fancyFunc{\hybridNum}(23415)-2\,\fancyFunc{\hybridNum}(23514) \Big)
\end{aligned}
\end{equation}
The process could also be repeated in the opposite order, finding adjoint $\cFunc$ functions first via composition and then casting to hybrid solutions. This one-to-one correspondence means there are equal numbers of doubly dual contact amplitude solutions with hybrid color and with adjoint color at each order in higher-derivative correction. 

The combined methods of composition, casting, and doubly dual local construction dramatically simplify calculation -- at ninth order in $\ap$, for example, an ansatz for $\cHD$ would contain 6435 free parameters. Introducing the notion of doubly dual amplitudes would allow us to write a smaller ansatz down for $\fancyFunc{\adjNum}$, with 1980 free parameters. Using composition, we are able to avoid ansatze entirely: the hybrid solution $\fancyFunc{\hybridNum}$ is obtained directly by composing $\rOne$ with the previous order in correction, with no parameters needing to be fixed, and then the adjoint solution $\fancyFunc{\adjNum}$ is written down immediately by casting $\fancyFunc{\hybridNum}$ to adjoint. 

\subsubsection{Doubly-dual amplitudes with sandwich color}
One can proceed in much the same way for sandwich color $\fancyColor = c_{\mathfrak{s}} = f^3d^3f^3$. As  its constraints are looser (obeying a six-term identity rather than stricter three-term Jacobi identities), a more complicated composition pattern is needed to span all solutions. These compositions involve not only the $\rOne$ building block, but also the quadratic sandwich building block $\mathfrak{s}^{(2)} = \compSandwich{\rOne}{\rOne}$ built from the linear relaxed solution $\rOne$.

\subsection{Restricting to bi-adjoint and higher-derivative momentum kernels}
All of the higher-derivative correction amplitudes discussed in this and the previous section admit a double-copy description via standard KLT. In this description, the standard field-theory KLT momentum kernel (which does not itself carry any higher-derivative corrections) facilitates the marriage of ordered amplitudes involving the adjoint higher-derivative color weights $\cHD$ with ordered amplitudes involving secondary adjoint color factors $\tilde{c}$ (simply the singly-ordered bi-adjoint scalar theory ordered amplitudes). It is these secondary ordered amplitudes which can be replaced with ordered vector amplitudes to generate higher-derivative corrections to Yang-Mills, as we will discuss in the next section.  

Before we do so, let us recall to the reader that there exists a notion of local  higher-derivative momentum kernels introduced by Chi, Elvang, Herderschee, Jones, and Paranjape\footnote{See, e.g. slides and video at ref.~\cite{callumTalk}.}.  Here we identify an additional strategy one might employ towards identifying these types of local adjoint-compatible momentum kernels. We begin by using our notion of local $\fancyColor$-dual double-copy structure to write down all the ways that adjoint color weights can be combined with local higher-derivative corrections: 
\begin{equation}
\label{biadjHD}
\mathcal{A^{\text{bi-adj+HD}}} =   \sum_{\fancyColor} \sum_{g \in \Gamma_{\fancyColor}} \tcFunc(g)\!\left(s_{ij}, \tilde{c} \right)~ \cFunc(g)\!\left(s_{ij}, c \right)\,.  
\end{equation}
Note that we sum over all algebraic structures $\fancyColor$ relevant at five points. For each structure $\fancyColor$, we must dress the relevant graphs $\Gamma_{\fancyColor}$ with the product between two $\fancyColor$-dual functions,  $\tcFunc(g)$ and  $\cFunc(g)$.  Each function  is dependent upon  adjoint-type color, $c$ or $\tilde{c}$, as well as scalar kinematics that conspire with the adjoint color to allow the entire weight $\cFunc$ to be compatible with $\fancyColor$-type algebraic constraints -- in a manner compatible with adjoint-type BCJ relations,  just as these $\cFunc$ functions do in our doubly dual local amplitudes. This constitutes essentially a double copy of a doubly-dual local amplitude with itself, with the appropriate color-dual replacement $\fancyColor(g) \rightarrow \tcFunc(g)\!\left(s_{ij}, \tilde{c} \right)$, giving the most general local higher-derivative modification of the bi-adjoint scalar to admit a local double copy description. 

We then rewrite this amplitude, at each order in mass dimension, by expressing every adjoint color factor $c$ in a basis of half-ladder color weights, $c(1|\sigma|5)$, and likewise for $\tilde{c}$. This allows us to define doubly-ordered amplitudes of this theory, c.f.~related identification of doubly-ordered ref.~\cite{Carrasco:2019qwr}, by identifying them as the unique coefficients of these independent color-basis monomials: 
\begin{equation}
\mathcal{A^{\text{bi-adj+HD}}} =   \sum_{\sigma,\rho \in S_{n-2}} \tilde{c}(1|\sigma | n) A^{\text{bi-adj+HD}}(\sigma|\rho) c(1|\rho | n)  = \tilde{c}^T \cdot A^{\text{bi-adj+HD}} \cdot c\,.
\end{equation}
We now exploit the standard KLT kernel, $S$, in order to rewrite $c$ and $\tilde{c}$ in terms of their $(n-3)!$ basis of (singly-)ordered amplitudes, $c=S\cdot A$ and $\tilde{c} = S \cdot \tilde{A}$,
\begin{equation}
\mathcal{A^{\text{bi-adj+HD}}} = \tilde{ A}^T  \cdot S^T \cdot A^{\text{bi-adj+HD}} \cdot S \cdot A \, .
\end{equation}
Finally, we can now identify the most general higher-derivative kernel compatible with a dual-adjoint striation at this mass-dimension:
\begin{equation}
\label{momKernelHD}
S^{\text{HD}} =  S^T \cdot A^{\text{bi-adj+HD}} \cdot S  \, ,
\end{equation}
so that this most general local higher-derivative correction amplitude to the bi-adjoint scalar may equivalently be expressed in terms of uncorrected ordered amplitudes $A$ and this higher-derivative momentum kernel $S^{\text{HD}}$,
\begin{equation}
\mathcal{A^{\text{bi-adj+HD}}} = \tilde{ A}^T  \cdot S^{\text{HD}} \cdot A \, . 
\end{equation}
We identify some intriguing open questions regarding such an approach in our conclusion section.

%% file: gaugeGravityCorr.tex
\section{Double Copy to Gauge and Gravity Corrections}
\label{gaugeGravCorr}
\subsection{Higher-derivative gauge corrections}
We have demonstrated how composition and casting can yield an infinite tower of higher-derivative corrections in the form of adjoint-type modified color weights dressing graphs in a bi-color scalar amplitude.  The fact that Yang-Mills amplitudes are compatible with maximal supersymmetry at tree-level means that, through double-copy construction, we can trivially map bi-color higher-derivative amplitudes to an infinite tower of higher-derivative corrections to Yang-Mills -- all of them entirely compatible with maximal supersymmetry.   The bi-colored amplitudes we constructed in the previous section have the following form,
\begin{equation}
\mathcal{A}_5^{\text{BC+HD}} = \sum_{g}\frac{\tilde{c}_g c^{\text{HD}}_g}{d_g} \, .
\end{equation} 
By virtue of the fact that the higher-derivative color weights, $c^{\text{HD}}$, satisfy the adjoint algebraic relations in \eqns{adjointConstraints1}{adjointConstraints2}, we are free to replace the secondary adjoint color $\tilde{c}$ with any graph weight that depends on adjoint relations to build a gauge-invariant amplitude, like the kinematic numerators of Yang-Mills theory, or indeed any vector weight that can be expressed in terms of adjoint color-dual numerators! For example,  higher-derivative corrections compatible with maximally supersymmetric Yang-Mills can be generated by pairing $c^{\text{HD}}$ with the vector weight $n_g^{\text{sYM}}$ associated with Yang-Mills at five-points,
\begin{equation}
\label{hdVecs}
\mathcal{A}_{5}^{\text{vec}+\text{HD}} = \sum_{g}\frac{n_g^{\text{vec}} c^{\text{HD}}_g}{d_g}\,,
\end{equation}
with $n_g^{\text{vec}}=n_g^{\text{sYM}}$. 
The resulting amplitudes will be compatible with maximal supersymmetry and manifest an adjoint double copy structure, with the additional mass-dimension and generic color-weights from the higher-derivative corrections encoded in $c^{\text{HD}}$.  For an example calculation matching one of our higher-derivative vector amplitudes (generated via double copy of Yang-Mills with our second-order modified color weights $\cHD$) to a traditional Feynman rules calculation starting from higher-derivative operators in the action, see \app{appendix:operatorComparison}.

Let us point out a feature at this stage.  Because we have striated along adjoint structures, it is entirely sufficient to build the two independent ordered amplitudes, $A_5^{\text{BC+HD}}(12345)$ and $A_5^{\text{BC+HD}}(13245)$, and simply use the standard KLT relation with ordered vector amplitudes $A_5^{\text{vec}}(12354)$ and $A_5^{\text{vec}}(13254)$ to arrive at higher-derivative gauge theory amplitudes (equivalent to those obtained via graph-based adjoint double copy).  Let us emphasize application of the adjoint momentum kernel is possible even with non-adjoint color weights like $d^{abcde}$ and $f^3d^3f^3$ because we have composed those colors with scalar weights to form adjoint-type amplitudes: by virtue of the composition rules, non-adjoint color conspires with scalar kinematics to satisfy adjoint constraints.  This adjoint-structure for the $\tilde{c}$  weights guarantees that the replacement with adjoint-dual vector weights $\tilde{c} \rightarrow n^{\text{vec}}_{\adjNum}$ (such as $n^{\text{sYM}}$ results in gauge-invariant amplitudes even in the dual-striated local representation,
\begin{equation}
\label{yangMillsPromotion}
\mathcal{A}^{\text{vec+HD}} = \sum_{g \in \Gamma_{\fancyColor}} \fancyColor_g ~ \cFunc(g)\!\left(s_{ij}, n^{\text{vec}}_{\adjNum} \right)\,.
\end{equation}
It is beyond the scope of this current paper to identify all distinct adjoint vector weights, modulo scalar-weight modifications, at five-points. While there were only eight at four-points, there is more freedom at five-points.  It will be interesting and useful to identify the minimal set of vector building blocks required to build all adjoint-type higher-derivative vector corrections.  We provide some examples, including Yang-Mills, in ancillary Mathematica files~\cite{ancFiles}.

\subsection{Higher-derivative gravity corrections}
\label{section:Gravity}
The care we must take in promoting the above discussion to a double-copy construction to gravity is simply to ensure that gauge-invariance is manifest for both copies after replacing the remaining color-weights in $c^{\text{HD}}$ of \eqn{hdVecs}, or equivalent the $\fancyColor_g$ in \eqn{yangMillsPromotion}, with vector weights.  If we can maintain gauge-invariance of both copies of vector weights, our resulting expression will describe  higher-derivative corrections to the scattering of gravitons.  

For trivially modified color weights $c^{\text{HD}}(g) \propto f^3f^3f^3(g) \times \permNum$, we can simply replace $f^3f^3f^3(g)\to \tilde{n}^{\text{vec}}(g)$ in \eqn{hdVecs} as long as the $n^{\text{vec}}_g$ have been expressed in a manifestly adjoint-color-dual form.  This is because permutation invariants will not endanger the $n^{\text{vec}}$'s ability to ensure the gauge-invariance of the $\tilde{n}^\text{vec}$ copy.  As $c^{\text{HD}}$ is adjoint-type the gauge invariance of the $n^{\text{vec}}$ copy is already manifest.   

For more general modified color-weights, we must replace the color weights $\fancyColor$ within $c^{\text{HD}}(\fancyColor,s_{ij})$ with vector weights both obey the same algebra as $\fancyColor$ and require the algebra of $\fancyColor$ to build gauge-invariant quantities. In factorizing corrections, the color $\fancyColor$ within the modified color weights $\cHD$ is either of type $f^3f^3f^3$ or $d^4f^3$, so we may only replace these, respectively, with adjoint ($\adjNum_{{\text{vec}}}$) or hybrid ($\hybridNum_{{\text{vec}}}$) vector weights that respect gauge invariance of the overall amplitude. For sufficiently high mass-dimension, it is possible for the individual vector graph weights to be gauge invariant, rendering the question of gauge invariance of the full amplitude trivial.  Examples of such local dual vector weights are $\adjNum^{5}_{{\text{vec}}}$, $\hybridNum^{5}_{{\text{vec}}}$, $\sandwichNum^{6}_{{\text{vec}}}$, and $\permNum^{8}_{{\text{vec}}}$, where the superscript here refers to the number of Lorentz-invariant dot products per term in the numerator (for comparison, Yang-Mills numerators contain four dots per term, but they do not fall into this category, as the graph weights are not themselves gauge-invariant; rather, they conspire between graph channels to cancel out any gauge dependence in the overall amplitude).    Useful in both  factorizing and local gauge corrections, and the predictions associated with local gravity counterterms to be discussed, we provide these example local gauge-invariant algebraic vector weights in ancillary Mathematica files~\cite{ancFiles}.

This strategy is particularly clear in the dual-striated form of the higher-derivative vector amplitude, \eqn{yangMillsPromotion}.  If we are to promote $\fancyColor_g$ in \eqn{yangMillsPromotion} to encode spin-1 vector information via double-copy replacement, the vector weights $\tilde{n}^{\text{vec}}_{\fancyColor}$ must satisfy the same algebraic relations as $\fancyColor$ and must result in gauge-invariant amplitude expressions when dressing the graphs relevant to that algebra, such that $\mathcal{A} = \sum_{g \in \Gamma_{\fancyColor}} \fancyColor(g) \tilde{n}^{\text{vec}}_{\fancyColor}(g)$ is gauge-invariant. If such a vector weight can be identified, its corresponding diffeomorphism-invariant gravity amplitude may be written down as: 
\begin{equation}
\mathcal{A}^{\text{grav}} = \sum_{g \in \Gamma_{\fancyColor}} \tilde{n}^{\text{vec}}_{\fancyColor}(g) ~ \tilde{n}^{\text{vec}}_{\fancyColor}(g)\,,
\end{equation}
and its associated higher-derivative gravity corrections are found simply through double copy with the higher-derivative gauge theory corrections $\mathcal{A}^{\text{vec+HD}}$ given by \eqn{yangMillsPromotion}:
\begin{equation}
\mathcal{A}^{\text{grav+HD}} = \sum_{g \in \Gamma_{\fancyColor}} \tilde{n}^{\text{vec}}_{\fancyColor}(g) ~ \cFunc(g)\!\left(s_{ij}, n^{\text{vec}}_{\adjNum} \right)\,.
\end{equation}

Finally, we remark on one final form of local higher-derivative gravity counterterms that may be obtained from our doubly-dual construction: for color structure $\fancyColor$, the dual properties of $\cFunc$ ensure that we may make the replacement $\fancyColor(g) \rightarrow \tcFunc(g)\!\left(s_{ij}, \tilde{c} \right)$ to arrive at higher-derivative corrections to the bi-adjoint scalar. The most general such form of these corrections, as discussed in the context of \eqn{biadjHD}, is given by summing over all possible structures $\fancyColor$:
\begin{equation}
\mathcal{A^{\text{bi-adj+HD}}} =   \sum_{\fancyColor} \sum_{g \in \Gamma_{\fancyColor}} \tcFunc(g)\!\left(s_{ij}, \tilde{c} \right)~ \cFunc(g)\!\left(s_{ij}, c \right)\,.  
\end{equation}
For each algebraic structure $\fancyColor$, we sum over the relevant graphs g in $\Gamma_{\fancyColor}$, dressing each with two higher-derivative $\fancyColor$-dual graph weights, $\cFunc(g)$ and $\tcFunc(g)$. Since such $\cFunc$ functions are fixed so that striating the amplitude along the basis elements of the adjoint color weights $c$ yields $(m-3)!$-compatible ordered amplitudes, the replacement $c \rightarrow n^{\text{vec}}_{\adjNum}$ automatically yields gauge-invariant amplitudes (as seen previously in \eqn{yangMillsPromotion}). This compatibility with the $(m-3)!$ relations for ordered amplitudes of theories with adjoint-double-copy structure is ensured by the algebraic properties of $\fancyColor$, meaning that this compatibility is preserved when $\fancyColor$ is replaced with an appropriately color-dual $\cFunc$ function. As a result, both adjoint color weights within $\mathcal{A^{\text{bi-adj+HD}}}$, $c$ and $\tilde{c}$, may be replaced with any adjoint-color-dual vector weights to yield fully diffeomorphism-invariant higher-derivative local contact gravity amplitudes: 
\begin{equation}
\mathcal{A^{\text{GR+HD}}} =   \sum_{\fancyColor} \sum_{g \in \Gamma_{\fancyColor}} \tcFunc(g)\!\left(s_{ij}, \tilde{n}^{\text{vec}}_{\adjNum} \right)~ \cFunc(g)\!\left(s_{ij}, n^{\text{vec}}_{\adjNum} \right)\,,
\end{equation}
where these vector weights need not arise from the same theory; the only condition is that both be adjoint-color-dual. We are free to choose both vector weights to be from super-Yang-Mills, so that:  
\begin{equation}
\mathcal{A^{\text{SUGRA+HD}}} =   \sum_{\fancyColor} \sum_{g \in \Gamma_{\fancyColor}} \tcFunc(g)\!\left(s_{ij}, n^{\text{sYM}} \right)~ \cFunc(g)\!\left(s_{ij}, n^{\text{sYM}} \right)\,,
\end{equation}
generates local contact higher-derivative corrections to supergravity. We continue to label the dual graph weights $\tcFunc$ and $\cFunc$ distinctly in this case to remind the reader that, even though functions obey the same algebraic relations and are functions of the same vector weight $n^{\text{sYM}}$, they need not be the same, as they may correspond to higher-derivative corrections from different orders in mass-dimension.

%% file: fixingToStrings.tex
\section{Ladder to String Theory}
\label{section:Strings}
We expect that our new five-point color-kinematic solutions (constructively built from color-weights and ultimately the unit-step relaxed scalar solution) span the whole space of BCJ-compatible corrections to Yang-Mills and, in particular, contain the open superstring corrections. The connection can be made through the bi-colored Z-theory amplitudes, which lift, through standard field-theory double copy, field-theoretic amplitudes to string theory amplitudes at tree-level~\cite{Broedel2013tta, Huang:2016tag, Carrasco2016ldy,Mafra2016mcc, Johansson2017srf}.
\begin{equation}
\label{openZ}
A^{\textrm{open}}(\tau)=\sum_{\rho, \sigma \in S_{n-3}} Z(\tau | 1, \sigma, n, n-1) S[\sigma | \rho] A^{\mathrm{vec}}(1, \rho, n-1, n) \, , 
\end{equation}
where $S$ is the field-theory KLT momentum kernel and thus does not carry any string higher-derivative corrections. $Z$-theory encodes all the $\alpha'$ corrections to Yang-Mills that show up in the tree-level open superstring, and carries color structure from two different gauge groups: the Chan-Paton factors, which can in principle sum over all single-trace contributions, and secondary adjoint color-weights which are e.g. replaced by $n^{\text{sYM}}$ when double-copying to the open superstring amplitude. These color structures line up with $\fancyColor$ and $\tilde{c}$, respectively, in our constructed bicolored scalar amplitudes.  In terms of the modified color factors, the full $Z$-theory amplitudes can also be expressed as:
\begin{align}
\label{Zcol}
\mathcal{Z}=\sum_g \frac{\tilde{c}_g c_g^{\textrm{HD}}}{d_g} \, , 
\end{align}
so the modified color factors $\cHD$ carry the $\ap$ string corrections and the Chan-Paton color factors. Doubly-ordered amplitudes $Z(\alpha|\beta)$ can be obtained from eq. (\ref{Zcol}) by expanding both the color structures into traces:
  \begin{align}
\mathcal{Z}=\sum_{\sigma,\rho\in S_{n-1}} \textrm{Tr}(\widetilde{T}^{a_1}\widetilde{T}^{a_{\sigma_1}}\cdots \widetilde{T}^{a_{\sigma_{n-1}}})  \textrm{Tr}(T^{b_1}T^{b_{\sigma_1}}\cdots T^{b_{\sigma_{n-1}}})Z(1,\sigma |1,\rho) \, . 
  \end{align}

We have verified that up to and including order $(\alpha')^9$, our solutions  span the low-energy expansion of the open superstring at five-points. This may be checked directly by comparing our results to $Z$-theory amplitudes built using Berends-Giele recursion \cite{Mafra2016mcc}, or by fixing to the following expression for string amplitudes, which is known to high order \cite{Mafra2011nv,Schlotterer2012ny,Mafra2011nw}:
\begin{equation}
\label{openF}
A^{\text {open}}(1, \rho, n-1, n)=\sum_{\sigma \in S_{n-3}} {F_{\rho}}^{\sigma} A^{\text {vec}}(1, \sigma, n-1, n) \, .
\end{equation}
Comparing eqns. (\ref{openF}) and (\ref{openZ}) we obtain \cite{Mafra2016mcc}:
\begin{equation}
\label{feq}
{F_{\rho}}^{\sigma}=\sum_{\tau \in S_{n-3}} S[\tau | \sigma] Z(1, \rho, n-1, n |1, \tau, n, n-1) \, . 
\end{equation}
For multiplicity $n = 5$ and the choice $\rho  = \{2,3\}$, this yields equations: 
\begin{align}
{F_{23}}^{23} &= S[23 | 23] Z(12345 |12354) + S[32| 23] Z(12345 |13254) \\
{F_{23}}^{32} &= S[23 | 32] Z(12345 |12354) + S[32| 32] Z(12345 |13254)
\end{align}
These equations will allow us to constrain our results to match string theory. Our amplitudes, ordered on both group's color factors, will be used in place of the $Z$-theory doubly-ordered amplitudes in these equations, and results for $F$ will be taken directly from string theory. $F$ can be written in terms of $M$ and $P$ matrices \cite{Schlotterer2012ny,Broedel:2013tta}:
\begin{align} F = &\; 1 +\zeta_2\; P_2+\zeta_3\; M_3+\zeta_2^2\; P_4+\zeta_2\zeta_3\; P_2M_3 + \zeta_5\; M_5  \notag \\
&+ \zeta_2^3\; P_6 + \frac{1}{2}\; \zeta_3^2\; M_3 M_3 + \; \zeta_7\; M_7+\zeta_2\zeta_5\; P_2M_5  + \zeta_2^2\zeta_3\; P_4 M_3  \notag \\
& + \zeta_2^4\; P_8 + \zeta_3\zeta_5\; M_5 M_3 + \frac{1}{2}\; \zeta_2 \zeta_3^2\; P_2 M_3 M_3  + \cdots
\end{align}

We parameterize our solutions as follows: $b[n,i]$ labels the $i^{th}$ factorizing solution at $n^{th}$ order in $\ap$, and $b[n,\fancyColor,i]$ labels the $i^{th}$ local solution at $n^{th}$ order in $\ap$, with color structure $\fancyColor$ of type $\adjNum, \hybridNum, \sandwichNum,$ or $\permNum$. Here we reproduce the results through fifth order of fixing our solutions to the $\ap$ expansion of $Z$-theory (and thus equivalently fixing to open superstring theory):
\begin{align}
\left(\alpha'\right)^2:&\ b[2,1] = 24 \, \zeta_2\, ,\\
\left(\alpha'\right)^3:&\ b[3,1] = -8 \, \zeta_3\, ,\\
\left(\alpha'\right)^4:&\ b[4,1] = 48 \, \zeta_2^2, \quad b[4,2] = -16\, \zeta_2^2/5\, \\
\left(\alpha'\right)^5:&\ b[5,1] = 384\, \zeta_2 \zeta_3, \quad b[5,2] = -32\, \zeta_5, \\
 &\ b[5,\hybridNum,1] = -6650112\, \zeta_2 \zeta_3/2615,  \quad b[5,\adjNum,1] = -30960\, \zeta_5/523\, ,
\end{align}
with results up to order $(\alpha')^9$ available in the ancillary files. Note that neither $d^5$ nor $f^3d^3f^3$-type color solutions ($\permNum$ or $\sandwichNum$) contribute to the string amplitude, as their single-trace Chan-Paton coefficient amplitudes would violate the odd-point reflection identity (just as these amplitudes are incompatible with the Yang-Mills reflection identity, as discussed in the context of \eqn{reflection}).
All other building blocks are given a fixed Wilson coefficient by comparison to string theory.  Above and including sixth order in $\ap$, we frequently find that multiple linearly-independent solutions (of both the factorizing and local contact varieties) are available, as summarized in Tables \ref{factorizingTable} and \ref{dualLocalTable}.  Since string theory only picks out one particular combination of these multiple distinct solutions, there are a number of distinct (linearly-independent) operators that string theory avoids at tree level -- and whose invariant predictions are spanned by our construction at five-points. 

Some of these distinct directions in solution space may be related to higher genus.  It is possible to match to the $M'$ matrices of reference \cite{Green2013bza}, local solutions which were found to be compatible with monodromy and preserve the BCJ relations of amplitudes, but that do not show up in string amplitudes at tree level. The lowest mass dimension example occurs at $(\alpha')^7$, where we find:
\begin{eqnarray}
M_7' :\ b[7,\adjNum,1]= -\frac{1}{256} \, . 
\end{eqnarray}
This and higher mass dimensions weights, through ninth order, are collected in ancillary files~\cite{ancFiles}.

%% file: conclusion.tex
\section{Conclusion}
\label{section:Conclusion}

We have demonstrated that the wealth of higher-derivative algebraic structure at five points compatible with adjoint-type double-copy can arise from a simple linear scalar building block much like at four-points.  Beyond this we have a conjectured form for such linear blocks at all multiplicity, \eqn{mPointConjecture}.  We expect this to result in the constructive ability to write down functional expressions for a rich variety of gauge and gravity higher-derivative predictions at arbitrary multiplicity.   We demonstrate in detail how this can proceed at five-points, allowing one to methodically construct higher mass-dimension numerators from lower order solutions and find a ladder of supersymmetry-compatible higher-derivative corrections to gauge theory and gravity.  We exemplify the construction of such composition rules with four-point examples in \app{appendix:BuildingComposition}, and give explicit results for all of our five-point composition rules in \app{appendix:ExplicitComposition}.  We emphasize that the value of having such composition rules is that they allows us to build infinite towers of solutions to functional-constrains without needing to invoke an ansatz.

Adjoint-compatible higher-derivative  gauge theory amplitudes at five points arise via double-copy from bi-colored amplitudes constructed with explicit higher-derivative adjoint-compatible  color-weights:  $(\text{sYM+HD})= \text{sYM} \otimes (\text{BC+HD}) $.  This has the virtue of allowing us to impose either factorization limits or locality conditions in a simpler bicolored scalar theory. We find  factorizing solutions by composing adjoint color factors (containing both $f^3f^3f^3$ and $d^4f^3$ structures) repeatedly with $\rOne$ to give candidate expressions for $\cHD$, then fixing on factorization. We conjecture that, above seventh order, no new nested compositions are needed to generate the candidate expressions that may be fixed on factorization, as eighth and ninth order were spanned by lower order structures times permutation invariants.

Local solutions can be found by fixing the same candidate expressions instead on the condition that all possible residues of the amplitude vanish. We find that the local bicolored scalar amplitudes are more efficiently constructed by making manifest a novel form of color-kinematics duality, in which each amplitude is manifestly local and displays two different algebraic striations between color and kinematics. These amplitudes like the factorizing solutions, contain two color structures: color factors $\fancyColor$ that will survive in the gauge theory amplitude, of the form $f^3f^3f^3$, $d^4f^3$, $f^3d^3f^3$, or $d^5$, as well as a secondary standard adjoint color factor $\tilde{c}_{\adjNum}$, which is replaced with $n^{\text{vec}}$ in the double copy to gauge theory corrections. Coefficients of a basis element of $\tilde{c}_{\adjNum}$ are ordered amplitudes that consist of scalar kinematics and $\fancyColor$ color and obey the $(m-3)!$-basis BCJ relations, ensuring compatibility with adjoint color-kinematics duality. Coefficients of the color factors $\fancyColor$, $\cFunc(\tilde{c}_{\adjNum}, s_{ij})$,  contain the secondary gauge group, and are manifestly local objects that satisfy the same algebraic constraints as the particular color structure of $\fancyColor$. For instance, if $\fancyColor_g = d^4f^3(g)$, its coefficient functions $\cFunc{}_{,g}$ obeys the same mixed vertex (anti)symmetry and four point identities as $d^4f^3$. The amplitude is thus \textit{doubly dual}: it exhibits both adjoint and $d^4f^3$ color-kinematics duality, depending upon striation.

Our  bicolored scalar solutions can be trivially recycled to maximally supersymmetric gauge theory corrections through double copy with super Yang-Mills compatible graph-weights.  Double copy of gauge theory corrections with another gauge theory to generate higher derivative supersymmetry-compatible gravity amplitudes is straightforward in some cases and more complicated in others. In all amplitudes, the replacement of the second gauge group's adjoint color factor $\tilde{c}_{\adjNum} \rightarrow n^{\text{vec}}_{\adjNum}$ is always allowed; but care must be taken when replacing $\fancyColor$ with dual vector kinematics that gauge invariance of the overall amplitude is preserved. Color structures like $d^4f^3$ demand dual gauge theory solutions; we have not yet identified consistently factorizing $d^4f^3$-dual vector numerators. Further exploration of gauge theory solutions could yield additional gravity solutions of interest. We additionally find that there is a second avenue for generating local gravity corrections from our doubly dual amplitudes by dressing each graph with two copies of $\cFunc(n^{\text{vec}}_{\adjNum}, s_{ij})$, which is guaranteed to be gauge invariant. 

We verify our proposition that  these results --- both factorizing and local --- span the low-energy expansion of the five point open superstring amplitude through $\mathcal{O}\left( \ap^9 \right)$. We note that we find additional local solutions to those appearing in the open superstring at tree-level, both for color structures compatible with string theory ($f^3f^3f^3$ and $d^4f^3$) and those forbidden by monodromy ($f^3d^3f^3$ and $d^5$).

When we restrict ourselves to manifestly bi-adjoint higher-derivative color-weights, we suggest a procedure for constructing a modified KLT kernel, \eqn{momKernelHD}, which encodes all higher-derivative corrections compatible with two adjoint-type algebraic structures.  Our approach of summing over all double-copies of algebraic structures given in terms of adjoint-color-weights and momentum invariants in \eqn{biadjHD} guarantees that there can be no missing form of higher-derivative correction, but it does not guarantee a lack of redundancy.  Indeed, the existence of casting between e.g. $f^3f^3f^3$ and $d^4f^3$ algebraic structures suggests a potentially large amount of redundancy.  This leaves us with a number of interesting open questions that would be exciting to pursue: Is there an efficient way of isolating the unique contributions?  Is there an analogy to the unit-step linear relaxed numerator that can be composed with KLT to generate all KLT-type higher-derivative corrections? 

We have not provided explicit operators for any of the higher-derivative corrections we present here, save  a pedagogical example in \app{appendix:operatorComparison}. We note that identifying operators associated with particular predictions can be accomplished by giving an appropriate mass-dimension ansatz to an action and constraining it on Feynman rules.  Yet this is both laborious and tedious, and only grows more so as multiplicity and higher-mass-dimension increases.  While arguably the gauge-invariant prediction is ultimately what matters, we find this situation limiting.  It would be interesting to see if Hilbert-series methods, see e.g.~\cite{Henning:2015alf, Lehman:2015via, Lehman:2015coa, Henning:2015daa, Henning:2017fpj}, can be applied to make constructive double-copy structure manifest. We expect that any approach towards identifying efficient means of generating higher-derivative operators relevant to these higher-derivative predictions will also aid in clarifying open questions regarding double-copy construction and the web-of-theories. 

The fact that the encoding of higher derivative supersymmetry-compatible corrections to gauge theory and gravity within scalar theories, constructed from simple color and scalar kinematic building blocks and composition rules, continues at five points indicates the potential for using these techniques at both higher multiplicity and at loop level. In particular, the five-point results of this paper could be used through unitarity  to construct an infinite tower of higher derivative multiloop corrections, as unitarity cuts of two loop integrands may be fixed on the five point tree level results presented here and in the auxiliary files~\cite{ancFiles}. This method shows promise at loop level, since capturing all higher derivative corrections in modified color factors would allow us to trivially recycle hard-won multiloop gauge theory integrand numerators into their higher derivative corrections:  
\begin{equation}
\mathcal{A}_{L-\text{loops}}^{\text{sYM+HD}} = \sum_g \frac{1}{S_g} \int \prod_i^L \frac{d^D l_i}{(2\pi)^D}  \,\,\frac{n^{\text{sYM}}_L(g) \cHD_L(g)}{d(g)}
\end{equation}  
Adjoint-dual loop level integrand numerators $n^{\text{sYM}}_L$ have been found through four loops \cite{SimplifyingBCJ}. We conjecture that, for suitable integrand numerators $\cHD_L$, colors factors could be appropriately replaced to generate higher derivative gravity integrands via double copy. 

There are also many opportunities for further work at tree level. At four points, we found that only four tensor structures in addition to Yang-Mills were necessary to generate all corrections within the bosonic open string; non-supersymmetric vector numerators at five points that could yield more general corrections remain to be explored. Of further interest are other color algebraic structures: in particular, multitrace amplitudes and predictions concerning states charged in the fundamental were recently  shown  to be expressible in terms of modified color factors for the NLSM and higher derivatives \cite{Low:2019wuv,Low:2020ubn}. In addition to expanding this framework to be compatible with fundamental color-kinematics duality, extending it to include massive particles could allow for the description of corrections to QCD containing standard model fermions, which should be of interest in the context of standard model effective field theory. 

%% file: buildingCompAppendix.tex
\section{Finding Composition Rules}
\label{appendix:BuildingComposition}

\allowdisplaybreaks[1]

At four points, we may generate a new adjoint numerator $n$ from the composition of two known adjoint numerators $j$ and $k$: 
\begin{equation}
n_{\adjNum}(1234) =\compAdj{j_{\adjNum}}{k_{\adjNum}} = j_{\adjNum}(4123)k_{\adjNum}(4123) - j_{\adjNum}(4231)k_{\adjNum}(4231)
\end{equation}
Although it is not difficult to verify that this form ensures that $n_{\adjNum}$ satisfies antisymmetry and Jacobi, we will construct it explicitly here, so it is clear how more complicated compositions may be generated at five points. 

To begin, let us make a judicious choice of basis for four point graphs, $g(4123)$ and $g(4231)$, corresponding to the $t$ and $u$ channel graphs respectively. We start by writing down an ansatz for the composed numerator $n$ as the outer product of these basis sets for numerators $j_{\adjNum}$ and $k_{\adjNum}$: 
\begin{equation}
\begin{split}
n_{\adjNum}(1234) = \alpha j_{\adjNum}(4123)k_{\adjNum}(4123) + \beta j_{\adjNum}(4231)k_{\adjNum}(4123) \\
+ \gamma j_{\adjNum}(4123)k_{\adjNum}(4231) + \delta j_{\adjNum}(4231)k_{\adjNum}(4231)
\end{split}
\end{equation}
First, impose antisymmetry on $n_{\adjNum}$, using the antisymmetry of $j_{\adjNum}$ and $k_{\adjNum}$ to return to the basis and constrain the ansatz: 
\begin{equation}
\begin{split}
n_{\adjNum}(2134) = \alpha j_{\adjNum}(4213)k_{\adjNum}(4213) + \beta j_{\adjNum}(4132)k_{\adjNum}(4213) \\
+ \gamma j_{\adjNum}(4213)k_{\adjNum}(4132) + \delta j_{\adjNum}(4132)k_{\adjNum}(4132)
\end{split}
\end{equation}
\begin{equation}
\begin{split}
n_{\adjNum}(2134) = \alpha j_{\adjNum}(4231)k_{\adjNum}(4231) + \beta j_{\adjNum}(4123)k_{\adjNum}(4231) \\
+ \gamma j_{\adjNum}(4231)k_{\adjNum}(4123) + \delta j_{\adjNum}(4123)k_{\adjNum}(4123)
\end{split}
\end{equation}
\begin{equation}
\begin{split}
n_{\adjNum}(2134) + n_{\adjNum}(1234) = (\alpha + \delta) j_{\adjNum}(4231)k_{\adjNum}(4231) + (\beta + \gamma) j_{\adjNum}(4123)k_{\adjNum}(4231) \\
+ (\beta + \gamma)  j_{\adjNum}(4231)k_{\adjNum}(4123) + (\beta + \gamma)  j_{\adjNum}(4123)k_{\adjNum}(4123)
\end{split}
\end{equation}
\begin{equation}
n_{\adjNum}(2134)+n_{\adjNum}(1234) = 0 \implies \delta = -\alpha, \quad \gamma = -\beta
\end{equation}
Thus, antisymmetry reduces our ansatz down to two parameters: 
\begin{equation}
\begin{split}
n_{\adjNum}(1234) = \alpha[j_{\adjNum}(4123)k_{\adjNum}(4123) - j_{\adjNum}(4231)k_{\adjNum}(4231)] \\
+ \beta [j_{\adjNum}(4231)k_{\adjNum}(4123) - j_{\adjNum}(4123)k_{\adjNum}(4231)]
\end{split}
\end{equation}
Now we must impose Jacobi: 
\begin{equation}
n_{\adjNum}(1234) = n_{\adjNum}(4123) + n_{\adjNum}(4231)
\end{equation}
\begin{equation}
\begin{split}
\alpha[j_{\adjNum}(4123)k_{\adjNum}(4123) - j_{\adjNum}(4231)k_{\adjNum}(4231)] + \beta [j_{\adjNum}(4231)k_{\adjNum}(4123) - j_{\adjNum}(4123)k_{\adjNum}(4231)] \\ 
= \alpha[j_{\adjNum}(1234)k_{\adjNum}(1234) - j_{\adjNum}(4231)k_{\adjNum}(4231)] + \beta [-j_{\adjNum}(4231)k_{\adjNum}(1234) + j_{\adjNum}(1234)k_{\adjNum}(4231)]\\
+ \alpha[j_{\adjNum}(4123)k_{\adjNum}(4123) - j_{\adjNum}(1234)k_{\adjNum}(1234)] + \beta [-j_{\adjNum}(1234)k_{\adjNum}(4123) + j_{\adjNum}(4123)k_{\adjNum}(1234)]
\end{split}
\end{equation}
\begin{equation}
\begin{split}
0 = \beta [j_{\adjNum}(4231)k_{\adjNum}(4123) - j_{\adjNum}(4123)k_{\adjNum}(4231)+ j_{\adjNum}(4231)k_{\adjNum}(1234) \\
-  j_{\adjNum}(1234)k_{\adjNum}(4231) +j_{\adjNum}(1234)k_{\adjNum}(4123) - j_{\adjNum}(4123)k_{\adjNum}(1234)]
\end{split}
\end{equation}
Using the fact that both $j_{\adjNum}$ and $k_{\adjNum}$ themselves obey Jacobi allows us to express this relation in terms of basis elements, and we find $\beta = 0$, leaving us with the simple composition rule: 
\begin{equation}
n_{\adjNum}(1234) = j_{\adjNum}(4123)k_{\adjNum}(4123) - j_{\adjNum}(4231)k_{\adjNum}(4231)
\end{equation}

We proceed to find compositions at multiplicity five in a similar manner. The general form of a composition rule that takes two numerators of given algebraic structures $\mathfrak{e}$ and $\mathfrak{f}$ and composes them into a new numerator desired algebraic structure $\mathfrak{g}$ is: 
\begin{equation}
\mathfrak{g}(12345) = \compArb{\mathfrak{e}}{\mathfrak{f}}{\mathfrak{g}}  = \sum_{g \in \rho_\mathfrak{e}}\sum_{g^{\prime} \in \rho_\mathfrak{f}} \beta_{g,g^{\prime}}^{\mathfrak{efg}} \mathfrak{e}(g)\mathfrak{f}(g^{\prime})
\end{equation}
where the sums run over the basis graphs $\rho$ for the given type of numerator. The coefficients $\beta_{g,g^{\prime}}^{\mathfrak{f},\mathfrak{g},\mathfrak{e}}$ are chosen such that $n_{\mathfrak{e}}$ obeys the $\mathfrak{e}$-type algebraic constraints, simply by virtue of the algebraic properties of $j_{\mathfrak{f}}$ and $k_{\mathfrak{g}}$.

%% file: colorStructureAppendix.tex
\section{Color Structure Basis at Five-Points}
\label{appendix:colorStructure}
We choose our 24 element basis of color structures at five points (in agreement with the $4!$ trace basis elements) to be given as follows: first, there is the single permutation invariant structure $d^5$. Then, antisymmetry and Jacobi constrain us to a basis of six adjoint structures $f^3f^3f^3$; we choose the $(m-2)!$ DDM basis $f^3f^3f^3(1\sigma5)$: 
\begin{equation}
\label{ddmBasis}
\begin{split}
f^3f^3f^3(12345) \quad f^3f^3f^3(12435) \quad f^3f^3f^3(13245) \\
f^3f^3f^3(13425) \quad f^3f^3f^3(14235) \quad f^3f^3f^3(14325)
\end{split}
\end{equation}
Based on the hybrid algebraic constraints of \eqns{eqn:hybridSymmetries}{eqn:hybridJacobi}, we land on a basis of six $d^4f^3$ color structures: 
\begin{equation}
\label{hybridBasis}
\begin{split}
& d^4f^3(12345) \quad d^4f^3(12435) \quad d^4f^3(12534) \\
& d^4f^3(14523) \quad d^4f^3(23415) \quad d^4f^3(23514) \\
\end{split}
\end{equation}
Finally, the sandwich constraints given in \eqns{sandConstraintsSym}{sandConstraintsSix} give an eleven-element basis of $f^3d^3f^3$ color structures: 
\begin{equation}
\label{sandwichBasis}
\begin{split}
& f^3d^3f^3(12345) \quad f^3d^3f^3(12435) \quad f^3d^3f^3(12534) \\
& f^3d^3f^3(13245) \quad f^3d^3f^3(13425) \quad f^3d^3f^3(13524) \\
& f^3d^3f^3(14235) \quad f^3d^3f^3(14325) \quad f^3d^3f^3(14523) \\
& f^3d^3f^3(23145) \quad f^3d^3f^3(24135) \quad 
\end{split}
\end{equation}
We have made the choice to prioritize $f^3d^3f^3$ structures in our basis over $d^3f^3f^3$; these structures may be expressed completely in terms of the $f^3d^3f^3$ structures as follows: 
\begin{equation}
d^3f^3f^3(abcde) = f^3d^3f^3(acbde)+ f^3d^3f^3(bcade) \, . 
\end{equation}

%% file: operatorAppendix.tex
\section{Operator Comparison at Five-Points}
\label{appendix:operatorComparison}

In this appendix, we briefly discuss an example of lining up our amplitudes with results from a traditional operator approach using Feynman rules. We will use as our example the lowest order higher-derivative correction to super-Yang-Mills theory at five points, which is a factorizing correction carrying $d^4f^3$ color at second order in $\ap$. The relevant action for this correction is given by: 
\begin{equation}
\mathcal{S} = \int d^d x \left(\mathcal{L}_{\text{YM}} + \ap^2 F^4_{\text{SUSY}} \right) \, 
\end{equation}
where the operators are defined as follows: 
\begin{equation}
\mathcal{L}_{\text{YM}} \equiv -\frac{1}{4} \Tr \left[ F_{\mu\nu}F^{\mu\nu}  \right] \, 
\end{equation} 
\begin{equation}
F^4_{\text{SUSY}} \equiv \Tr \left[ F_{\mu}^{\nu} F_{\nu}^{\rho} F_{\rho}^{\sigma} F_{\sigma}^{\mu}  + 2 F_{\mu}^{\nu} F_{\rho}^{\sigma} F_{\nu}^{\rho} F_{\sigma}^{\mu} - \tfrac{1}{4}F_{\mu\nu}F_{\rho\sigma}F^{\mu\nu}F^{\rho\sigma} - \tfrac{1}{2}F_{\mu\nu}F^{\mu\nu}F_{\rho\sigma}F^{\rho\sigma}    \right] \, . 
\end{equation} 
\begin{equation}
F_{\mu\nu} \equiv \partial_{\mu} A_{\nu} - \partial_{\nu}A_{\mu} + \frac{i}{\sqrt{2}} g \left(A_{\mu}A_{\nu} - A_{\nu}A_{\mu} \right)  \, .
\end{equation}

The Feynman rules calculation for this five-point correction contains two parts: one five-point contact contribution arising from the five-field terms within $F^4_{\text{SUSY}}$, and one factorizing contribution arising from the four-field terms of $F^4_{\text{SUSY}}$ sewn together with a three-point Yang-Mills insertion. Schematically, with $F^4$-vertices drawn with filled (teal) circles and standard Yang-Mills vertices as open (pink) circles,
\begin{equation}
\mathcal{A}_{\text{action}} = \contactFiveNew{}{}{}{}{} + \sum_{g \in \Gamma_{\hybridNum}}^{10}  \quadCubeFiveNew{}{}{}{}{} 
\end{equation} 
The factorizing contribution must be found by dressing the ten hybrid graphs $\Gamma_{\hybridNum}$ at five points: the quartic vertex is dressed with the off-shell four-point vertex from the $F^4_{\text{SUSY}}$ operator, the cubic vertex is dressed with the off-shell three-point Yang-Mills vertex, and each graph is dressed with its associated propagator. The contact contribution dresses the single quintic graph at five points. Both of these contributions are necessary for gauge invariance of the overall amplitude. 

Our method for constructing this correction involves an adjoint double copy between Yang Mills and our higher-derivative bicolor theory: each cubic graph is dressed with an adjoint Yang-Mills numerator and the second order adjoint modified color weight:
\begin{equation}
\mathcal{A}_{\text{double-copy}} = \sum_{g \in \Gamma_3}^{15} \frac{n^{\text{YM}}_g c^{\text{HD(2)}}_g}{d_g} \, .
\end{equation}
The explicit form of this higher-derivative color weight, generated via compositions fixed to factorize appropriately down to four-point higher-derivative corrections times the three-point uncorrected bi-adjoint scalar, is given in the ancillary files and reproduced here: 
\begin{equation}
\begin{split}
c^{\text{HD(2)}}(12345) &= \frac{2}{5} \, b[2,1] \, \Big( d^4f^3(14523) s_{45} \left(-s_{12} + 2 s_{34} + s_{45}\right) \\
&+  d^4f^3(12345) \left(s_{12}^2 + s_{12} (9 s_{23} - 6 s_{45}) - 
    s_{45} (5 s_{23} + s_{45})\right) \\
    &+ d^4f^3(12435) \left((5 s_{15} - 5 s_{23} - 8 s_{34} - 
       2 s_{45}) s_{45} + 
    s_{12} (5 s_{15} + s_{23} + 4 s_{45})\right) \\
    &+ d^4f^3(23514) \left(s_{12}^2 - s_{45} (9 s_{34} + s_{45}) + 
    s_{12} (5 s_{34} + 6 s_{45})\right) \\
    &+ d^4f^3(23415) \left(-6 s_{12}^2 + s_{12} (5 s_{34} + 8 s_{45}) - 
    s_{45} (9 s_{34} + 8 s_{45})\right) \\
    &+ d^4f^3(12534) \big( (5 s_{15} - 8 s_{34} - s_{45}) s_{45} + 
    s_{12} (5 s_{15} - 6 s_{23} + 9 s_{45}) \big) \Big) \, . 
\end{split}
\end{equation}
We find that fixing $b[2,1] = 6\sqrt{2} \, i \, g$ allows for agreement between our amplitude calculation and the Feynman rules calculation: 
\begin{equation}
\mathcal{A}_{\text{double-copy}} = \mathcal{A}_{\text{action}} \, ,
\end{equation}
verifying the compatibility of our double copy construction method with higher-derivative color weights with traditional operator calculations.

%% file: explicitCompAppendix.tex
\section{Relevant Composition at Five-Points}
\label{appendix:ExplicitComposition}

\begin{align*}
\compRel{j_{\relaxedNum}}{k_{\relaxedNum}} =   -2j_{\relaxedNum}(12435)k_{\relaxedNum}(13245)  \
-2j_{\relaxedNum}(13245)k_{\relaxedNum}(12435)  \
-2j_{\relaxedNum}(13425)k_{\relaxedNum}(13425)  \\  \
-2j_{\relaxedNum}(13425)k_{\relaxedNum}(14235)  \
-2j_{\relaxedNum}(14235)k_{\relaxedNum}(13425)  \
-2j_{\relaxedNum}(14235)k_{\relaxedNum}(14235)  \\  \
+3j_{\relaxedNum}(12435)k_{\relaxedNum}(13425)  \
+3j_{\relaxedNum}(12435)k_{\relaxedNum}(14235)  \
+3j_{\relaxedNum}(13245)k_{\relaxedNum}(13425)  \\  \
+3j_{\relaxedNum}(13245)k_{\relaxedNum}(14235)  \
+3j_{\relaxedNum}(13425)k_{\relaxedNum}(12435)  \
+3j_{\relaxedNum}(13425)k_{\relaxedNum}(13245)  \\  \
+3j_{\relaxedNum}(14235)k_{\relaxedNum}(12435)  \
+3j_{\relaxedNum}(14235)k_{\relaxedNum}(13245)  \
-4j_{\relaxedNum}(12435)k_{\relaxedNum}(12435)  \\  \
-4j_{\relaxedNum}(13245)k_{\relaxedNum}(13245)  \
+4j_{\relaxedNum}(14325)k_{\relaxedNum}(14325)  \
-j_{\relaxedNum}(13425)k_{\relaxedNum}(14325)  \\  \
-j_{\relaxedNum}(14235)k_{\relaxedNum}(14325)  \
-j_{\relaxedNum}(14325)k_{\relaxedNum}(13425)  \
-j_{\relaxedNum}(14325)k_{\relaxedNum}(14235) 
\end{align*}

\begin{align*}
\compAdj{j_{\relaxedNum}}{k_{\relaxedNum}} =   -2j_{\relaxedNum}(12435)k_{\relaxedNum}(14325)  \
+2j_{\relaxedNum}(13245)k_{\relaxedNum}(14325)  \
+2j_{\relaxedNum}(14325)k_{\relaxedNum}(12435)  \\  \
-2j_{\relaxedNum}(14325)k_{\relaxedNum}(13245)  \
+3j_{\relaxedNum}(13425)k_{\relaxedNum}(14325)  \
-3j_{\relaxedNum}(14235)k_{\relaxedNum}(14325)  \\  \
-3j_{\relaxedNum}(14325)k_{\relaxedNum}(13425)  \
+3j_{\relaxedNum}(14325)k_{\relaxedNum}(14235)  \
+5j_{\relaxedNum}(12435)k_{\relaxedNum}(14235)  \\  \
-5j_{\relaxedNum}(13245)k_{\relaxedNum}(13425)  \
+5j_{\relaxedNum}(13425)k_{\relaxedNum}(13245)  \
-5j_{\relaxedNum}(14235)k_{\relaxedNum}(12435)  \\  \
-6j_{\relaxedNum}(13425)k_{\relaxedNum}(14235)  \
+6j_{\relaxedNum}(14235)k_{\relaxedNum}(13425)  \
-j_{\relaxedNum}(12435)k_{\relaxedNum}(13425)  \\  \
+j_{\relaxedNum}(13245)k_{\relaxedNum}(14235)  \
+j_{\relaxedNum}(13425)k_{\relaxedNum}(12435)  \
-j_{\relaxedNum}(14235)k_{\relaxedNum}(13245) 
\end{align*}

\begin{align*}
\compAdj{j_{\relaxedNum}}{k_{\adjNum}} = a_1\Big( 2j_{\relaxedNum}(12435)k_{\adjNum}(13425)  \
+2j_{\relaxedNum}(13245)k_{\adjNum}(14235)  \\  \
+2j_{\relaxedNum}(13425)k_{\adjNum}(12435)  \
+2j_{\relaxedNum}(13425)k_{\adjNum}(13245)  \\  \
+2j_{\relaxedNum}(14235)k_{\adjNum}(12435)  \
+2j_{\relaxedNum}(14235)k_{\adjNum}(13245)  \\  \
-2j_{\relaxedNum}(14325)k_{\adjNum}(12435)  \
-2j_{\relaxedNum}(14325)k_{\adjNum}(13245)  \\  \
-3j_{\relaxedNum}(12435)k_{\adjNum}(12345)  \
-3j_{\relaxedNum}(13245)k_{\adjNum}(12345)  \\  \
+4j_{\relaxedNum}(14325)k_{\adjNum}(12345)  \
-j_{\relaxedNum}(12435)k_{\adjNum}(12435)  \\  \
+j_{\relaxedNum}(12435)k_{\adjNum}(13245)  \
-j_{\relaxedNum}(12435)k_{\adjNum}(14325)  \\  \
+j_{\relaxedNum}(13245)k_{\adjNum}(12435)  \
-j_{\relaxedNum}(13245)k_{\adjNum}(13245)  \\  \
-j_{\relaxedNum}(13245)k_{\adjNum}(14325)  \
+j_{\relaxedNum}(13425)k_{\adjNum}(12345)  \\  \
-j_{\relaxedNum}(13425)k_{\adjNum}(13425)  \
-j_{\relaxedNum}(13425)k_{\adjNum}(14235)  \\  \
+j_{\relaxedNum}(14235)k_{\adjNum}(12345)  \
-j_{\relaxedNum}(14235)k_{\adjNum}(13425)  \\  \
-j_{\relaxedNum}(14235)k_{\adjNum}(14235)  \
+j_{\relaxedNum}(14325)k_{\adjNum}(13425)  \\  \
+j_{\relaxedNum}(14325)k_{\adjNum}(14235) \Big) \\ + a_2\Big( \
2j_{\relaxedNum}(14325)k_{\adjNum}(12345)  \
-j_{\relaxedNum}(12435)k_{\adjNum}(12345)  \\  \
+j_{\relaxedNum}(12435)k_{\adjNum}(13425)  \
-j_{\relaxedNum}(12435)k_{\adjNum}(14235)  \\  \
-j_{\relaxedNum}(13245)k_{\adjNum}(12345)  \
-j_{\relaxedNum}(13245)k_{\adjNum}(13425)  \\  \
+j_{\relaxedNum}(13245)k_{\adjNum}(14235)  \
+j_{\relaxedNum}(13425)k_{\adjNum}(12435)  \\  \
+j_{\relaxedNum}(13425)k_{\adjNum}(13245)  \
-j_{\relaxedNum}(13425)k_{\adjNum}(14325)  \\  \
+j_{\relaxedNum}(14235)k_{\adjNum}(12435)  \
+j_{\relaxedNum}(14235)k_{\adjNum}(13245)  \\  \
-j_{\relaxedNum}(14235)k_{\adjNum}(14325)  \
-j_{\relaxedNum}(14325)k_{\adjNum}(12435)  \\  \
-j_{\relaxedNum}(14325)k_{\adjNum}(13245)  \
+j_{\relaxedNum}(14325)k_{\adjNum}(14325) \Big)
\end{align*}

\begin{align*}
\compHybrid{j_{\relaxedNum}}{k_{\relaxedNum}} =  -2j_{\relaxedNum}(12435)k_{\relaxedNum}(14235)  \
+2j_{\relaxedNum}(12435)k_{\relaxedNum}(14325)  \
+2j_{\relaxedNum}(14235)k_{\relaxedNum}(12435)  \\  \
-2j_{\relaxedNum}(14325)k_{\relaxedNum}(12435)  \
+j_{\relaxedNum}(12435)k_{\relaxedNum}(13245)  \
-j_{\relaxedNum}(12435)k_{\relaxedNum}(13425)  \\  \
-j_{\relaxedNum}(13245)k_{\relaxedNum}(12435)  \
-j_{\relaxedNum}(13245)k_{\relaxedNum}(13425)  \
+j_{\relaxedNum}(13425)k_{\relaxedNum}(12435)  \\  \
+j_{\relaxedNum}(13425)k_{\relaxedNum}(13245)  \
+j_{\relaxedNum}(13425)k_{\relaxedNum}(14235)  \
-j_{\relaxedNum}(13425)k_{\relaxedNum}(14325)  \\  \
-j_{\relaxedNum}(14235)k_{\relaxedNum}(13425)  \
+j_{\relaxedNum}(14325)k_{\relaxedNum}(13425) 
\end{align*}

\begin{align*}
\compHybrid{j_{\relaxedNum}}{k_{\hybridNum}} = a_1\Big( 2j_{\relaxedNum}(14235)k_{\hybridNum}(13524)  \
-j_{\relaxedNum}(12435)k_{\hybridNum}(12534)  \\  \
+j_{\relaxedNum}(12435)k_{\hybridNum}(14523)  \
-j_{\relaxedNum}(12435)k_{\hybridNum}(23514)  \\  \
-j_{\relaxedNum}(12435)k_{\hybridNum}(34512)  \
-j_{\relaxedNum}(13245)k_{\hybridNum}(12534)  \\  \
-j_{\relaxedNum}(13245)k_{\hybridNum}(13524)  \
+j_{\relaxedNum}(13425)k_{\hybridNum}(12534)  \\  \
-j_{\relaxedNum}(13425)k_{\hybridNum}(14523)  \
-j_{\relaxedNum}(13425)k_{\hybridNum}(23514)  \\  \
-j_{\relaxedNum}(13425)k_{\hybridNum}(24513)  \
+j_{\relaxedNum}(14235)k_{\hybridNum}(12534)  \\  \
+j_{\relaxedNum}(14235)k_{\hybridNum}(23514)  \
+j_{\relaxedNum}(14325)k_{\hybridNum}(12534)  \\  \
+j_{\relaxedNum}(14325)k_{\hybridNum}(23514) \Big) \\ + a_2\Big( \
2j_{\relaxedNum}(13245)k_{\hybridNum}(24513)  \
+2j_{\relaxedNum}(13245)k_{\hybridNum}(34512)  \\  \
-2j_{\relaxedNum}(13425)k_{\hybridNum}(34512)  \
+2j_{\relaxedNum}(14235)k_{\hybridNum}(14523)  \\  \
-2j_{\relaxedNum}(14235)k_{\hybridNum}(34512)  \
-2j_{\relaxedNum}(14325)k_{\hybridNum}(14523)  \\  \
+2j_{\relaxedNum}(14325)k_{\hybridNum}(34512)  \
+j_{\relaxedNum}(12435)k_{\hybridNum}(13524)  \\  \
-j_{\relaxedNum}(12435)k_{\hybridNum}(14523)  \
+j_{\relaxedNum}(12435)k_{\hybridNum}(34512)  \\  \
+j_{\relaxedNum}(13245)k_{\hybridNum}(23514)  \
+j_{\relaxedNum}(13425)k_{\hybridNum}(12534)  \\  \
+j_{\relaxedNum}(13425)k_{\hybridNum}(14523)  \
-j_{\relaxedNum}(13425)k_{\hybridNum}(23514)  \\  \
-j_{\relaxedNum}(13425)k_{\hybridNum}(24513)  \
-j_{\relaxedNum}(14235)k_{\hybridNum}(12534)  \\  \
-j_{\relaxedNum}(14235)k_{\hybridNum}(23514)  \
-j_{\relaxedNum}(14325)k_{\hybridNum}(13524) \Big)
\end{align*}

\begin{align*}
\compSandwich{j_{\relaxedNum}}{k_{\relaxedNum}} = a_1\Big( -2j_{\relaxedNum}(12435)k_{\relaxedNum}(13245)  \
-2j_{\relaxedNum}(13245)k_{\relaxedNum}(12435)  \\  \
-2j_{\relaxedNum}(13425)k_{\relaxedNum}(13425)  \
-2j_{\relaxedNum}(13425)k_{\relaxedNum}(14235)  \\  \
-2j_{\relaxedNum}(14235)k_{\relaxedNum}(13425)  \
-2j_{\relaxedNum}(14235)k_{\relaxedNum}(14235)  \\  \
+j_{\relaxedNum}(12435)k_{\relaxedNum}(13425)  \
+j_{\relaxedNum}(12435)k_{\relaxedNum}(14235)  \\  \
+j_{\relaxedNum}(13245)k_{\relaxedNum}(13425)  \
+j_{\relaxedNum}(13245)k_{\relaxedNum}(14235)  \\  \
+j_{\relaxedNum}(13425)k_{\relaxedNum}(12435)  \
+j_{\relaxedNum}(13425)k_{\relaxedNum}(13245)  \\  \
+j_{\relaxedNum}(13425)k_{\relaxedNum}(14325)  \
+j_{\relaxedNum}(14235)k_{\relaxedNum}(12435)  \\  \
+j_{\relaxedNum}(14235)k_{\relaxedNum}(13245)  \
+j_{\relaxedNum}(14235)k_{\relaxedNum}(14325)  \\  \
+j_{\relaxedNum}(14325)k_{\relaxedNum}(13425)  \
+j_{\relaxedNum}(14325)k_{\relaxedNum}(14235) \Big) \\ + a_2\Big( \
-j_{\relaxedNum}(12435)k_{\relaxedNum}(12435)  \
-j_{\relaxedNum}(12435)k_{\relaxedNum}(13245)  \\  \
+j_{\relaxedNum}(12435)k_{\relaxedNum}(13425)  \
+j_{\relaxedNum}(12435)k_{\relaxedNum}(14235)  \\  \
-j_{\relaxedNum}(13245)k_{\relaxedNum}(12435)  \
-j_{\relaxedNum}(13245)k_{\relaxedNum}(13245)  \\  \
+j_{\relaxedNum}(13245)k_{\relaxedNum}(13425)  \
+j_{\relaxedNum}(13245)k_{\relaxedNum}(14235)  \\  \
+j_{\relaxedNum}(13425)k_{\relaxedNum}(12435)  \
+j_{\relaxedNum}(13425)k_{\relaxedNum}(13245)  \\  \
-j_{\relaxedNum}(13425)k_{\relaxedNum}(13425)  \
-j_{\relaxedNum}(13425)k_{\relaxedNum}(14235)  \\  \
+j_{\relaxedNum}(14235)k_{\relaxedNum}(12435)  \
+j_{\relaxedNum}(14235)k_{\relaxedNum}(13245)  \\  \
-j_{\relaxedNum}(14235)k_{\relaxedNum}(13425)  \
-j_{\relaxedNum}(14235)k_{\relaxedNum}(14235)  \\  \
+j_{\relaxedNum}(14325)k_{\relaxedNum}(14325) \Big)
\end{align*}

\begin{align*}
\compHybrid{j_{\relaxedNum}}{k_{\adjNum}} = a_1\Big( -2j_{\relaxedNum}(12435)k_{\adjNum}(12345)  \
-2j_{\relaxedNum}(12435)k_{\adjNum}(14325)  \\  \
-2j_{\relaxedNum}(14235)k_{\adjNum}(13425)  \
+2j_{\relaxedNum}(14325)k_{\adjNum}(12345)  \\  \
-2j_{\relaxedNum}(14325)k_{\adjNum}(12435)  \
+2j_{\relaxedNum}(14325)k_{\adjNum}(14235)  \\  \
-3j_{\relaxedNum}(13425)k_{\adjNum}(13245)  \
+3j_{\relaxedNum}(14235)k_{\adjNum}(14325)  \\  \
+j_{\relaxedNum}(12435)k_{\adjNum}(13245)  \
+j_{\relaxedNum}(12435)k_{\adjNum}(13425)  \\  \
+j_{\relaxedNum}(13245)k_{\adjNum}(12345)  \
+j_{\relaxedNum}(13245)k_{\adjNum}(13245)  \\  \
-j_{\relaxedNum}(13245)k_{\adjNum}(14235)  \
-j_{\relaxedNum}(13245)k_{\adjNum}(14325)  \\  \
+j_{\relaxedNum}(13425)k_{\adjNum}(12435)  \
-j_{\relaxedNum}(13425)k_{\adjNum}(14235)  \\  \
+j_{\relaxedNum}(13425)k_{\adjNum}(14325)  \
-j_{\relaxedNum}(14235)k_{\adjNum}(12345)  \\  \
-j_{\relaxedNum}(14235)k_{\adjNum}(12435)  \
+j_{\relaxedNum}(14235)k_{\adjNum}(13245)  \\  \
+j_{\relaxedNum}(14235)k_{\adjNum}(14235)  \
-j_{\relaxedNum}(14325)k_{\adjNum}(13425) \Big) \\ + a_2\Big( \
2j_{\relaxedNum}(12435)k_{\adjNum}(12435)  \
-2j_{\relaxedNum}(12435)k_{\adjNum}(14235)  \\  \
+2j_{\relaxedNum}(13245)k_{\adjNum}(12435)  \
+2j_{\relaxedNum}(13245)k_{\adjNum}(13425)  \\  \
-2j_{\relaxedNum}(14235)k_{\adjNum}(13425)  \
-2j_{\relaxedNum}(14325)k_{\adjNum}(13245)  \\  \
+2j_{\relaxedNum}(14325)k_{\adjNum}(14325)  \
-3j_{\relaxedNum}(14235)k_{\adjNum}(12345)  \\  \
-3j_{\relaxedNum}(14235)k_{\adjNum}(12435)  \
+3j_{\relaxedNum}(14235)k_{\adjNum}(13245)  \\  \
+3j_{\relaxedNum}(14235)k_{\adjNum}(14235)  \
-4j_{\relaxedNum}(12435)k_{\adjNum}(12345)  \\  \
+4j_{\relaxedNum}(14325)k_{\adjNum}(12345)  \
-5j_{\relaxedNum}(13425)k_{\adjNum}(13245)  \\  \
+j_{\relaxedNum}(12435)k_{\adjNum}(13245)  \
+j_{\relaxedNum}(12435)k_{\adjNum}(13425)  \\  \
+j_{\relaxedNum}(13245)k_{\adjNum}(12345)  \
+j_{\relaxedNum}(13245)k_{\adjNum}(13245)  \\  \
-j_{\relaxedNum}(13245)k_{\adjNum}(14235)  \
-j_{\relaxedNum}(13245)k_{\adjNum}(14325)  \\  \
-j_{\relaxedNum}(13425)k_{\adjNum}(12435)  \
+j_{\relaxedNum}(13425)k_{\adjNum}(14235)  \\  \
-j_{\relaxedNum}(13425)k_{\adjNum}(14325)  \
+j_{\relaxedNum}(14235)k_{\adjNum}(14325)  \\  \
-j_{\relaxedNum}(14325)k_{\adjNum}(13425) \Big)
\end{align*}

\begin{align*}
\compPerm{j_{\relaxedNum}}{k_{\relaxedNum}} =  -2j_{\relaxedNum}(12435)k_{\relaxedNum}(13425)  \
-2j_{\relaxedNum}(12435)k_{\relaxedNum}(14235)  \
-2j_{\relaxedNum}(13245)k_{\relaxedNum}(13425)  \\  \
-2j_{\relaxedNum}(13245)k_{\relaxedNum}(14235)  \
-2j_{\relaxedNum}(13425)k_{\relaxedNum}(12435)  \
-2j_{\relaxedNum}(13425)k_{\relaxedNum}(13245)  \\  \
+2j_{\relaxedNum}(13425)k_{\relaxedNum}(14235)  \
-2j_{\relaxedNum}(13425)k_{\relaxedNum}(14325)  \
-2j_{\relaxedNum}(14235)k_{\relaxedNum}(12435)  \\  \
-2j_{\relaxedNum}(14235)k_{\relaxedNum}(13245)  \
+2j_{\relaxedNum}(14235)k_{\relaxedNum}(13425)  \
-2j_{\relaxedNum}(14235)k_{\relaxedNum}(14325)  \\  \
-2j_{\relaxedNum}(14325)k_{\relaxedNum}(13425)  \
-2j_{\relaxedNum}(14325)k_{\relaxedNum}(14235)  \
+3j_{\relaxedNum}(12435)k_{\relaxedNum}(12435)  \\  \
+3j_{\relaxedNum}(13245)k_{\relaxedNum}(13245)  \
+3j_{\relaxedNum}(14325)k_{\relaxedNum}(14325)  \
+4j_{\relaxedNum}(13425)k_{\relaxedNum}(13425)  \\  \
+4j_{\relaxedNum}(14235)k_{\relaxedNum}(14235)  \
+j_{\relaxedNum}(12435)k_{\relaxedNum}(13245)  \
+j_{\relaxedNum}(12435)k_{\relaxedNum}(14325)  \\  \
+j_{\relaxedNum}(13245)k_{\relaxedNum}(12435)  \
+j_{\relaxedNum}(13245)k_{\relaxedNum}(14325)  \
+j_{\relaxedNum}(14325)k_{\relaxedNum}(12435)  \\  \
+j_{\relaxedNum}(14325)k_{\relaxedNum}(13245) 
\end{align*}

\begin{align*}
\compPerm{j_{\adjNum}}{k_{\adjNum}} = -2j_{\adjNum}(12345)k_{\adjNum}(12435)  \
-2j_{\adjNum}(12345)k_{\adjNum}(13245)  \
+2j_{\adjNum}(12345)k_{\adjNum}(14325)  \\  \
-2j_{\adjNum}(12435)k_{\adjNum}(12345)  \
+2j_{\adjNum}(12435)k_{\adjNum}(13425)  \
-2j_{\adjNum}(12435)k_{\adjNum}(14235)  \\  \
-2j_{\adjNum}(13245)k_{\adjNum}(12345)  \
-2j_{\adjNum}(13245)k_{\adjNum}(13425)  \
+2j_{\adjNum}(13245)k_{\adjNum}(14235)  \\  \
+2j_{\adjNum}(13425)k_{\adjNum}(12435)  \
-2j_{\adjNum}(13425)k_{\adjNum}(13245)  \
-2j_{\adjNum}(13425)k_{\adjNum}(14325)  \\  \
-2j_{\adjNum}(14235)k_{\adjNum}(12435)  \
+2j_{\adjNum}(14235)k_{\adjNum}(13245)  \
-2j_{\adjNum}(14235)k_{\adjNum}(14325)  \\  \
+2j_{\adjNum}(14325)k_{\adjNum}(12345)  \
-2j_{\adjNum}(14325)k_{\adjNum}(13425)  \
-2j_{\adjNum}(14325)k_{\adjNum}(14235)  \\  \
+5j_{\adjNum}(12345)k_{\adjNum}(12345)  \
+5j_{\adjNum}(12435)k_{\adjNum}(12435)  \
+5j_{\adjNum}(13245)k_{\adjNum}(13245)  \\  \
+5j_{\adjNum}(13425)k_{\adjNum}(13425)  \
+5j_{\adjNum}(14235)k_{\adjNum}(14235)  \
+5j_{\adjNum}(14325)k_{\adjNum}(14325)  \\  \
-j_{\adjNum}(12345)k_{\adjNum}(13425)  \
-j_{\adjNum}(12345)k_{\adjNum}(14235)  \
-j_{\adjNum}(12435)k_{\adjNum}(13245)  \\  \
-j_{\adjNum}(12435)k_{\adjNum}(14325)  \
-j_{\adjNum}(13245)k_{\adjNum}(12435)  \
-j_{\adjNum}(13245)k_{\adjNum}(14325)  \\  \
-j_{\adjNum}(13425)k_{\adjNum}(12345)  \
-j_{\adjNum}(13425)k_{\adjNum}(14235)  \
-j_{\adjNum}(14235)k_{\adjNum}(12345)  \\  \
-j_{\adjNum}(14235)k_{\adjNum}(13425)  \
-j_{\adjNum}(14325)k_{\adjNum}(12435)  \
-j_{\adjNum}(14325)k_{\adjNum}(13245)
\end{align*}

\begin{align*}
\compSandwich{j_{\relaxedNum}}{k_{\adjNum}} = a_1\Big( j_{\relaxedNum}(12435)k_{\adjNum}(12345)  \
-j_{\relaxedNum}(12435)k_{\adjNum}(12435)  \\  \
-j_{\relaxedNum}(12435)k_{\adjNum}(13245)  \
-j_{\relaxedNum}(13245)k_{\adjNum}(12345)  \\  \
+j_{\relaxedNum}(13245)k_{\adjNum}(12435)  \
+j_{\relaxedNum}(13245)k_{\adjNum}(13245)  \\  \
+j_{\relaxedNum}(13425)k_{\adjNum}(12345)  \
+j_{\relaxedNum}(13425)k_{\adjNum}(12435)  \\  \
-j_{\relaxedNum}(13425)k_{\adjNum}(13245)  \
-j_{\relaxedNum}(14235)k_{\adjNum}(12345)  \\  \
+j_{\relaxedNum}(14235)k_{\adjNum}(12435)  \
-j_{\relaxedNum}(14235)k_{\adjNum}(13245)  \\  \
-j_{\relaxedNum}(14325)k_{\adjNum}(12435)  \
+j_{\relaxedNum}(14325)k_{\adjNum}(13245) \Big) \\ + a_2\Big( \
-j_{\relaxedNum}(12435)k_{\adjNum}(12345)  \
+j_{\relaxedNum}(12435)k_{\adjNum}(12435)  \\  \
+j_{\relaxedNum}(12435)k_{\adjNum}(13245)  \
-j_{\relaxedNum}(12435)k_{\adjNum}(14325)  \\  \
+j_{\relaxedNum}(13245)k_{\adjNum}(12345)  \
-j_{\relaxedNum}(13245)k_{\adjNum}(12435)  \\  \
-j_{\relaxedNum}(13245)k_{\adjNum}(13245)  \
+j_{\relaxedNum}(13245)k_{\adjNum}(14325)  \\  \
+j_{\relaxedNum}(13425)k_{\adjNum}(12345)  \
+j_{\relaxedNum}(13425)k_{\adjNum}(13425)  \\  \
-j_{\relaxedNum}(13425)k_{\adjNum}(14235)  \
-j_{\relaxedNum}(14235)k_{\adjNum}(12345)  \\  \
+j_{\relaxedNum}(14235)k_{\adjNum}(13425)  \
-j_{\relaxedNum}(14235)k_{\adjNum}(14235)  \\  \
-j_{\relaxedNum}(14325)k_{\adjNum}(13425)  \
+j_{\relaxedNum}(14325)k_{\adjNum}(14235) \Big)
\end{align*}

\begin{align*}
\compSandwich{j_{\sandwichNum}}{k_{\adjNum}} = a_1\Big(3j_{\sandwichNum}(12435)k_{\adjNum}(13245)  \
-3j_{\sandwichNum}(12435)k_{\adjNum}(14235)  \\  \
-3j_{\sandwichNum}(12534)k_{\adjNum}(12345)  \
+3j_{\sandwichNum}(12534)k_{\adjNum}(13245)  \\  \
+3j_{\sandwichNum}(12534)k_{\adjNum}(13425)  \
-3j_{\sandwichNum}(12534)k_{\adjNum}(14325)  \\  \
-3j_{\sandwichNum}(13245)k_{\adjNum}(12435)  \
+3j_{\sandwichNum}(13245)k_{\adjNum}(13425)  \\  \
+3j_{\sandwichNum}(23145)k_{\adjNum}(12345)  \
-3j_{\sandwichNum}(23145)k_{\adjNum}(12435)  \\  \
-3j_{\sandwichNum}(23145)k_{\adjNum}(14235)  \
+3j_{\sandwichNum}(23145)k_{\adjNum}(14325)  \\  \
-j_{\sandwichNum}(12435)k_{\adjNum}(12345)  \
+j_{\sandwichNum}(12435)k_{\adjNum}(12435)  \\  \
-j_{\sandwichNum}(12435)k_{\adjNum}(13425)  \
+j_{\sandwichNum}(12435)k_{\adjNum}(14325)  \\  \
+j_{\sandwichNum}(12534)k_{\adjNum}(12435)  \
+j_{\sandwichNum}(12534)k_{\adjNum}(14235)  \\  \
+j_{\sandwichNum}(13245)k_{\adjNum}(12345)  \
-j_{\sandwichNum}(13245)k_{\adjNum}(13245)  \\  \
+j_{\sandwichNum}(13245)k_{\adjNum}(14235)  \
-j_{\sandwichNum}(13245)k_{\adjNum}(14325)  \\  \
-j_{\sandwichNum}(23145)k_{\adjNum}(13245)  \
-j_{\sandwichNum}(23145)k_{\adjNum}(13425) \Big) \\ + a_2\Big( \
-2j_{\sandwichNum}(12435)k_{\adjNum}(13425)  \
+2j_{\sandwichNum}(12435)k_{\adjNum}(14325)  \\  \
+2j_{\sandwichNum}(12534)k_{\adjNum}(14235)  \
+2j_{\sandwichNum}(13245)k_{\adjNum}(14235)  \\  \
-2j_{\sandwichNum}(13245)k_{\adjNum}(14325)  \
-2j_{\sandwichNum}(23145)k_{\adjNum}(13425)  \\  \
+3j_{\sandwichNum}(12435)k_{\adjNum}(13245)  \
+3j_{\sandwichNum}(12534)k_{\adjNum}(13245)  \\  \
-3j_{\sandwichNum}(13245)k_{\adjNum}(12435)  \
-3j_{\sandwichNum}(13425)k_{\adjNum}(12345)  \\  \
-3j_{\sandwichNum}(13524)k_{\adjNum}(12345)  \
+3j_{\sandwichNum}(14235)k_{\adjNum}(12345)  \\  \
-3j_{\sandwichNum}(23145)k_{\adjNum}(12435)  \
+3j_{\sandwichNum}(24135)k_{\adjNum}(12345)  \\  \
+j_{\sandwichNum}(12435)k_{\adjNum}(12345)  \
-j_{\sandwichNum}(12435)k_{\adjNum}(12435)  \\  \
-j_{\sandwichNum}(12534)k_{\adjNum}(12435)  \
-j_{\sandwichNum}(13245)k_{\adjNum}(12345)  \\  \
+j_{\sandwichNum}(13245)k_{\adjNum}(13245)  \
+j_{\sandwichNum}(23145)k_{\adjNum}(13245) \Big) \\ + a_3\Big( \
j_{\sandwichNum}(12435)k_{\adjNum}(12345)  \
-j_{\sandwichNum}(12435)k_{\adjNum}(12435)  \\  \
-j_{\sandwichNum}(12435)k_{\adjNum}(13425)  \
+j_{\sandwichNum}(12435)k_{\adjNum}(14325)  \\  \
-j_{\sandwichNum}(12534)k_{\adjNum}(12435)  \
+j_{\sandwichNum}(12534)k_{\adjNum}(13245)  \\  \
+j_{\sandwichNum}(12534)k_{\adjNum}(14235)  \
-j_{\sandwichNum}(13245)k_{\adjNum}(12345)  \\  \
+j_{\sandwichNum}(13245)k_{\adjNum}(13245)  \
+j_{\sandwichNum}(13245)k_{\adjNum}(14235)  \\  \
-j_{\sandwichNum}(13245)k_{\adjNum}(14325)  \
-j_{\sandwichNum}(13425)k_{\adjNum}(12345)  \\  \
+j_{\sandwichNum}(13425)k_{\adjNum}(12435)  \
-j_{\sandwichNum}(13524)k_{\adjNum}(13245)  \\  \
+j_{\sandwichNum}(14235)k_{\adjNum}(12345)  \
-j_{\sandwichNum}(14235)k_{\adjNum}(13245)  \\  \
-j_{\sandwichNum}(14523)k_{\adjNum}(12435)  \
+j_{\sandwichNum}(14523)k_{\adjNum}(13245)  \\  \
-j_{\sandwichNum}(23145)k_{\adjNum}(13425)  \
+j_{\sandwichNum}(24135)k_{\adjNum}(13245) \Big) \\ + a_4\Big( \
-2j_{\sandwichNum}(12435)k_{\adjNum}(14325)  \
-2j_{\sandwichNum}(12534)k_{\adjNum}(12435)  \\  \
+2j_{\sandwichNum}(13245)k_{\adjNum}(14325)  \
+2j_{\sandwichNum}(23145)k_{\adjNum}(13425)  \\  \
-3j_{\sandwichNum}(12534)k_{\adjNum}(12345)  \
-3j_{\sandwichNum}(12534)k_{\adjNum}(13425)  \\  \
+3j_{\sandwichNum}(13425)k_{\adjNum}(12345)  \
-3j_{\sandwichNum}(13425)k_{\adjNum}(13245)  \\  \
+3j_{\sandwichNum}(13425)k_{\adjNum}(13425)  \
-3j_{\sandwichNum}(13425)k_{\adjNum}(14235)  \\  \
+3j_{\sandwichNum}(13425)k_{\adjNum}(14325)  \
+3j_{\sandwichNum}(13524)k_{\adjNum}(12435)  \\  \
+3j_{\sandwichNum}(13524)k_{\adjNum}(13425)  \
-3j_{\sandwichNum}(13524)k_{\adjNum}(14235)  \\  \
-3j_{\sandwichNum}(13524)k_{\adjNum}(14325)  \
-3j_{\sandwichNum}(14235)k_{\adjNum}(12345)  \\  \
+3j_{\sandwichNum}(14235)k_{\adjNum}(12435)  \
+3j_{\sandwichNum}(14235)k_{\adjNum}(13425)  \\  \
-3j_{\sandwichNum}(14235)k_{\adjNum}(14235)  \
-3j_{\sandwichNum}(14235)k_{\adjNum}(14325)  \\  \
-3j_{\sandwichNum}(14523)k_{\adjNum}(12435)  \
+3j_{\sandwichNum}(14523)k_{\adjNum}(13245)  \\  \
+3j_{\sandwichNum}(23145)k_{\adjNum}(12345)  \
-3j_{\sandwichNum}(23145)k_{\adjNum}(12435)  \\  \
-3j_{\sandwichNum}(23145)k_{\adjNum}(14235)  \
-3j_{\sandwichNum}(24135)k_{\adjNum}(12435)  \\  \
-3j_{\sandwichNum}(24135)k_{\adjNum}(13425)  \
+3j_{\sandwichNum}(24135)k_{\adjNum}(14235)  \\  \
+3j_{\sandwichNum}(24135)k_{\adjNum}(14325)  \
-4j_{\sandwichNum}(12435)k_{\adjNum}(12345)  \\  \
+4j_{\sandwichNum}(12534)k_{\adjNum}(14235)  \
+4j_{\sandwichNum}(13245)k_{\adjNum}(12345)  \\  \
+6j_{\sandwichNum}(12435)k_{\adjNum}(13245)  \
+6j_{\sandwichNum}(12534)k_{\adjNum}(13245)  \\  \
-6j_{\sandwichNum}(13245)k_{\adjNum}(12435)  \
-6j_{\sandwichNum}(14523)k_{\adjNum}(13425)  \\  \
+6j_{\sandwichNum}(14523)k_{\adjNum}(14235)  \
+j_{\sandwichNum}(12435)k_{\adjNum}(12435)  \\  \
-j_{\sandwichNum}(12435)k_{\adjNum}(13425)  \
-j_{\sandwichNum}(13245)k_{\adjNum}(13245)  \\  \
+j_{\sandwichNum}(13245)k_{\adjNum}(14235)  \
-j_{\sandwichNum}(23145)k_{\adjNum}(13245) \Big)
\end{align*}

\begin{equation}
\begin{split}
g_{1}=(12345) \quad g_{2}=(12435) \quad g_{3}=(12534) \quad \\
g_{4}=(13245) \quad g_{5}=(13425) \quad g_{6}=(13524) \quad \\
g_{7}=(14235) \quad g_{8}=(14325) \quad g_{9}=(14523) \quad \\
g_{10}=(23145) \quad g_{11}=(24135) \quad 
\end{split}
\end{equation}

\begin{align*}
\compSandwich{j_{\relaxedNum}}{k_{\sandwichNum}} =a_1\Big( 2j_{\relaxedNum}(g_{5})k_{\sandwichNum}(g_{2})  \
-2j_{\relaxedNum}(g_{5})k_{\sandwichNum}(g_{4})  \
-2j_{\relaxedNum}(g_{7})k_{\sandwichNum}(g_{2})  \
+2j_{\relaxedNum}(g_{7})k_{\sandwichNum}(g_{4})  \
-3j_{\relaxedNum}(g_{5})k_{\sandwichNum}(g_{3})  \\  \
-3j_{\relaxedNum}(g_{7})k_{\sandwichNum}(g_{10})  \
+j_{\relaxedNum}(g_{2})k_{\sandwichNum}(g_{1})  \
+j_{\relaxedNum}(g_{2})k_{\sandwichNum}(g_{3})  \
-j_{\relaxedNum}(g_{2})k_{\sandwichNum}(g_{4})  \
+j_{\relaxedNum}(g_{4})k_{\sandwichNum}(g_{1})  \\  \
+j_{\relaxedNum}(g_{4})k_{\sandwichNum}(g_{10})  \
-j_{\relaxedNum}(g_{4})k_{\sandwichNum}(g_{2})  \
-j_{\relaxedNum}(g_{5})k_{\sandwichNum}(g_{1})  \
+j_{\relaxedNum}(g_{5})k_{\sandwichNum}(g_{10})  \
-j_{\relaxedNum}(g_{7})k_{\sandwichNum}(g_{1})  \\  \
+j_{\relaxedNum}(g_{7})k_{\sandwichNum}(g_{3})  \
-j_{\relaxedNum}(g_{8})k_{\sandwichNum}(g_{1})  \
+j_{\relaxedNum}(g_{8})k_{\sandwichNum}(g_{2})  \
+j_{\relaxedNum}(g_{8})k_{\sandwichNum}(g_{4}) \Big) \\ + a_2\Big( \
2j_{\relaxedNum}(g_{2})k_{\sandwichNum}(g_{6})  \
+2j_{\relaxedNum}(g_{5})k_{\sandwichNum}(g_{3})  \
-2j_{\relaxedNum}(g_{5})k_{\sandwichNum}(g_{6})  \
+2j_{\relaxedNum}(g_{5})k_{\sandwichNum}(g_{9})  \
-2j_{\relaxedNum}(g_{7})k_{\sandwichNum}(g_{6})  \\  \
+2j_{\relaxedNum}(g_{7})k_{\sandwichNum}(g_{9})  \
-j_{\relaxedNum}(g_{2})k_{\sandwichNum}(g_{3})  \
-j_{\relaxedNum}(g_{2})k_{\sandwichNum}(g_{4})  \
+j_{\relaxedNum}(g_{2})k_{\sandwichNum}(g_{7})  \
-j_{\relaxedNum}(g_{2})k_{\sandwichNum}(g_{9})  \\  \
+j_{\relaxedNum}(g_{4})k_{\sandwichNum}(g_{11})  \
-j_{\relaxedNum}(g_{4})k_{\sandwichNum}(g_{2})  \
-j_{\relaxedNum}(g_{4})k_{\sandwichNum}(g_{3})  \
+j_{\relaxedNum}(g_{4})k_{\sandwichNum}(g_{5})  \
+j_{\relaxedNum}(g_{4})k_{\sandwichNum}(g_{6})  \\  \
-j_{\relaxedNum}(g_{4})k_{\sandwichNum}(g_{9})  \
-j_{\relaxedNum}(g_{5})k_{\sandwichNum}(g_{10})  \
+j_{\relaxedNum}(g_{5})k_{\sandwichNum}(g_{4})  \
-j_{\relaxedNum}(g_{5})k_{\sandwichNum}(g_{5})  \
-j_{\relaxedNum}(g_{5})k_{\sandwichNum}(g_{7})  \\  \
+j_{\relaxedNum}(g_{7})k_{\sandwichNum}(g_{2})  \
+j_{\relaxedNum}(g_{7})k_{\sandwichNum}(g_{3})  \
-j_{\relaxedNum}(g_{7})k_{\sandwichNum}(g_{5})  \
-j_{\relaxedNum}(g_{7})k_{\sandwichNum}(g_{7})  \
+j_{\relaxedNum}(g_{8})k_{\sandwichNum}(g_{5})  \\  \
+j_{\relaxedNum}(g_{8})k_{\sandwichNum}(g_{7}) \Big) \\ + a_3\Big( \
-j_{\relaxedNum}(g_{2})k_{\sandwichNum}(g_{1})  \
+j_{\relaxedNum}(g_{2})k_{\sandwichNum}(g_{3})  \
-j_{\relaxedNum}(g_{2})k_{\sandwichNum}(g_{6})  \
+j_{\relaxedNum}(g_{2})k_{\sandwichNum}(g_{8})  \
+j_{\relaxedNum}(g_{2})k_{\sandwichNum}(g_{9})  \\  \
-j_{\relaxedNum}(g_{4})k_{\sandwichNum}(g_{1})  \
+j_{\relaxedNum}(g_{4})k_{\sandwichNum}(g_{3})  \
-j_{\relaxedNum}(g_{4})k_{\sandwichNum}(g_{6})  \
+j_{\relaxedNum}(g_{4})k_{\sandwichNum}(g_{8})  \
+j_{\relaxedNum}(g_{4})k_{\sandwichNum}(g_{9})  \\  \
+j_{\relaxedNum}(g_{5})k_{\sandwichNum}(g_{1})  \
-j_{\relaxedNum}(g_{5})k_{\sandwichNum}(g_{3})  \
+j_{\relaxedNum}(g_{5})k_{\sandwichNum}(g_{6})  \
-j_{\relaxedNum}(g_{5})k_{\sandwichNum}(g_{8})  \
-j_{\relaxedNum}(g_{5})k_{\sandwichNum}(g_{9})  \\  \
+j_{\relaxedNum}(g_{7})k_{\sandwichNum}(g_{1})  \
-j_{\relaxedNum}(g_{7})k_{\sandwichNum}(g_{3})  \
+j_{\relaxedNum}(g_{7})k_{\sandwichNum}(g_{6})  \
-j_{\relaxedNum}(g_{7})k_{\sandwichNum}(g_{8})  \
-j_{\relaxedNum}(g_{7})k_{\sandwichNum}(g_{9})  \\  \
+j_{\relaxedNum}(g_{8})k_{\sandwichNum}(g_{8}) \Big) \\ + a_4\Big( \
-j_{\relaxedNum}(g_{2})k_{\sandwichNum}(g_{3})  \
+j_{\relaxedNum}(g_{2})k_{\sandwichNum}(g_{6})  \
-j_{\relaxedNum}(g_{2})k_{\sandwichNum}(g_{9})  \
-j_{\relaxedNum}(g_{4})k_{\sandwichNum}(g_{3})  \
+j_{\relaxedNum}(g_{4})k_{\sandwichNum}(g_{6})  \\  \
-j_{\relaxedNum}(g_{4})k_{\sandwichNum}(g_{9})  \
+j_{\relaxedNum}(g_{5})k_{\sandwichNum}(g_{3})  \
-j_{\relaxedNum}(g_{5})k_{\sandwichNum}(g_{6})  \
+j_{\relaxedNum}(g_{5})k_{\sandwichNum}(g_{9})  \
+j_{\relaxedNum}(g_{7})k_{\sandwichNum}(g_{3})  \\  \
-j_{\relaxedNum}(g_{7})k_{\sandwichNum}(g_{6})  \
+j_{\relaxedNum}(g_{7})k_{\sandwichNum}(g_{9})  \
+j_{\relaxedNum}(g_{8})k_{\sandwichNum}(g_{3})  \
-j_{\relaxedNum}(g_{8})k_{\sandwichNum}(g_{6})  \
+j_{\relaxedNum}(g_{8})k_{\sandwichNum}(g_{9}) \Big) \\ + a_5\Big( \
-2j_{\relaxedNum}(g_{5})k_{\sandwichNum}(g_{2})  \
-2j_{\relaxedNum}(g_{7})k_{\sandwichNum}(g_{4})  \
-j_{\relaxedNum}(g_{2})k_{\sandwichNum}(g_{1})  \
-j_{\relaxedNum}(g_{2})k_{\sandwichNum}(g_{10})  \
+j_{\relaxedNum}(g_{2})k_{\sandwichNum}(g_{2})  \\  \
-j_{\relaxedNum}(g_{4})k_{\sandwichNum}(g_{1})  \
-j_{\relaxedNum}(g_{4})k_{\sandwichNum}(g_{3})  \
+j_{\relaxedNum}(g_{4})k_{\sandwichNum}(g_{4})  \
+j_{\relaxedNum}(g_{5})k_{\sandwichNum}(g_{1})  \
-j_{\relaxedNum}(g_{5})k_{\sandwichNum}(g_{10})  \\  \
+j_{\relaxedNum}(g_{5})k_{\sandwichNum}(g_{3})  \
+j_{\relaxedNum}(g_{7})k_{\sandwichNum}(g_{1})  \
+j_{\relaxedNum}(g_{7})k_{\sandwichNum}(g_{10})  \
-j_{\relaxedNum}(g_{7})k_{\sandwichNum}(g_{3})  \
+j_{\relaxedNum}(g_{8})k_{\sandwichNum}(g_{1})  \\  \
+j_{\relaxedNum}(g_{8})k_{\sandwichNum}(g_{10})  \
+j_{\relaxedNum}(g_{8})k_{\sandwichNum}(g_{3}) \Big) \\ + a_6\Big( \
-2j_{\relaxedNum}(g_{5})k_{\sandwichNum}(g_{9})  \
-2j_{\relaxedNum}(g_{7})k_{\sandwichNum}(g_{3})  \
-2j_{\relaxedNum}(g_{7})k_{\sandwichNum}(g_{9})  \
-j_{\relaxedNum}(g_{2})k_{\sandwichNum}(g_{1})  \
-j_{\relaxedNum}(g_{2})k_{\sandwichNum}(g_{10})  \\  \
+j_{\relaxedNum}(g_{2})k_{\sandwichNum}(g_{11})  \
+j_{\relaxedNum}(g_{2})k_{\sandwichNum}(g_{3})  \
+j_{\relaxedNum}(g_{2})k_{\sandwichNum}(g_{5})  \
-j_{\relaxedNum}(g_{2})k_{\sandwichNum}(g_{6})  \
+j_{\relaxedNum}(g_{2})k_{\sandwichNum}(g_{9})  \\  \
-j_{\relaxedNum}(g_{4})k_{\sandwichNum}(g_{1})  \
+j_{\relaxedNum}(g_{4})k_{\sandwichNum}(g_{7})  \
+j_{\relaxedNum}(g_{4})k_{\sandwichNum}(g_{9})  \
+j_{\relaxedNum}(g_{5})k_{\sandwichNum}(g_{1})  \
-j_{\relaxedNum}(g_{5})k_{\sandwichNum}(g_{11})  \\  \
-j_{\relaxedNum}(g_{5})k_{\sandwichNum}(g_{2})  \
-j_{\relaxedNum}(g_{5})k_{\sandwichNum}(g_{3})  \
+j_{\relaxedNum}(g_{5})k_{\sandwichNum}(g_{6})  \
+j_{\relaxedNum}(g_{7})k_{\sandwichNum}(g_{1})  \
+j_{\relaxedNum}(g_{7})k_{\sandwichNum}(g_{10})  \\  \
-j_{\relaxedNum}(g_{7})k_{\sandwichNum}(g_{11})  \
-j_{\relaxedNum}(g_{7})k_{\sandwichNum}(g_{4})  \
+j_{\relaxedNum}(g_{7})k_{\sandwichNum}(g_{6})  \
+j_{\relaxedNum}(g_{8})k_{\sandwichNum}(g_{1})  \
+j_{\relaxedNum}(g_{8})k_{\sandwichNum}(g_{11})  \\  \
+j_{\relaxedNum}(g_{8})k_{\sandwichNum}(g_{6}) \Big)
\end{align*}

\begin{align*}
\compSandwich{j_{\sandwichNum}}{k_{\sandwichNum}} = a_1\Big( -2j_{\sandwichNum}(g_{1})k_{\sandwichNum}(g_{1})  \
+2j_{\sandwichNum}(g_{1})k_{\sandwichNum}(g_{8})  \
+2j_{\sandwichNum}(g_{8})k_{\sandwichNum}(g_{1})  \
+j_{\sandwichNum}(g_{1})k_{\sandwichNum}(g_{3})  \
-j_{\sandwichNum}(g_{1})k_{\sandwichNum}(g_{6})  \\  \
+j_{\sandwichNum}(g_{1})k_{\sandwichNum}(g_{9})  \
+j_{\sandwichNum}(g_{3})k_{\sandwichNum}(g_{1})  \
-j_{\sandwichNum}(g_{6})k_{\sandwichNum}(g_{1})  \
+j_{\sandwichNum}(g_{9})k_{\sandwichNum}(g_{1}) \Big) \\ + a_2\Big( \
-j_{\sandwichNum}(g_{1})k_{\sandwichNum}(g_{1})  \
+j_{\sandwichNum}(g_{1})k_{\sandwichNum}(g_{3})  \
-j_{\sandwichNum}(g_{1})k_{\sandwichNum}(g_{6})  \
+j_{\sandwichNum}(g_{1})k_{\sandwichNum}(g_{8})  \
+j_{\sandwichNum}(g_{1})k_{\sandwichNum}(g_{9})  \\  \
+j_{\sandwichNum}(g_{3})k_{\sandwichNum}(g_{8})  \
-j_{\sandwichNum}(g_{6})k_{\sandwichNum}(g_{8})  \
+j_{\sandwichNum}(g_{8})k_{\sandwichNum}(g_{1})  \
-j_{\sandwichNum}(g_{8})k_{\sandwichNum}(g_{3})  \
+j_{\sandwichNum}(g_{8})k_{\sandwichNum}(g_{6})  \\  \
-j_{\sandwichNum}(g_{8})k_{\sandwichNum}(g_{9})  \
+j_{\sandwichNum}(g_{9})k_{\sandwichNum}(g_{8}) \Big) \\ + a_3\Big( \
2j_{\sandwichNum}(g_{8})k_{\sandwichNum}(g_{3})  \
-2j_{\sandwichNum}(g_{8})k_{\sandwichNum}(g_{6})  \
+2j_{\sandwichNum}(g_{8})k_{\sandwichNum}(g_{9})  \
-j_{\sandwichNum}(g_{1})k_{\sandwichNum}(g_{3})  \
+j_{\sandwichNum}(g_{1})k_{\sandwichNum}(g_{6})  \\  \
-j_{\sandwichNum}(g_{1})k_{\sandwichNum}(g_{9})  \
+j_{\sandwichNum}(g_{3})k_{\sandwichNum}(g_{3})  \
-j_{\sandwichNum}(g_{3})k_{\sandwichNum}(g_{6})  \
+j_{\sandwichNum}(g_{3})k_{\sandwichNum}(g_{9})  \
-j_{\sandwichNum}(g_{6})k_{\sandwichNum}(g_{3})  \\  \
+j_{\sandwichNum}(g_{6})k_{\sandwichNum}(g_{6})  \
-j_{\sandwichNum}(g_{6})k_{\sandwichNum}(g_{9})  \
+j_{\sandwichNum}(g_{9})k_{\sandwichNum}(g_{3})  \
-j_{\sandwichNum}(g_{9})k_{\sandwichNum}(g_{6})  \
+j_{\sandwichNum}(g_{9})k_{\sandwichNum}(g_{9}) \Big) \\ + a_4\Big( \
-2j_{\sandwichNum}(g_{1})k_{\sandwichNum}(g_{11})  \
+2j_{\sandwichNum}(g_{1})k_{\sandwichNum}(g_{2})  \
+2j_{\sandwichNum}(g_{1})k_{\sandwichNum}(g_{4})  \
-2j_{\sandwichNum}(g_{1})k_{\sandwichNum}(g_{5})  \
-2j_{\sandwichNum}(g_{1})k_{\sandwichNum}(g_{6})  \\  \
-2j_{\sandwichNum}(g_{1})k_{\sandwichNum}(g_{7})  \
+2j_{\sandwichNum}(g_{8})k_{\sandwichNum}(g_{10})  \
-2j_{\sandwichNum}(g_{8})k_{\sandwichNum}(g_{2})  \
+2j_{\sandwichNum}(g_{8})k_{\sandwichNum}(g_{3})  \
-2j_{\sandwichNum}(g_{8})k_{\sandwichNum}(g_{4})  \\  \
+j_{\sandwichNum}(g_{3})k_{\sandwichNum}(g_{10})  \
-j_{\sandwichNum}(g_{3})k_{\sandwichNum}(g_{2})  \
+j_{\sandwichNum}(g_{3})k_{\sandwichNum}(g_{3})  \
-j_{\sandwichNum}(g_{3})k_{\sandwichNum}(g_{4})  \
-j_{\sandwichNum}(g_{6})k_{\sandwichNum}(g_{10})  \\  \
+j_{\sandwichNum}(g_{6})k_{\sandwichNum}(g_{2})  \
-j_{\sandwichNum}(g_{6})k_{\sandwichNum}(g_{3})  \
+j_{\sandwichNum}(g_{6})k_{\sandwichNum}(g_{4})  \
+j_{\sandwichNum}(g_{9})k_{\sandwichNum}(g_{10})  \
-j_{\sandwichNum}(g_{9})k_{\sandwichNum}(g_{2})  \\  \
+j_{\sandwichNum}(g_{9})k_{\sandwichNum}(g_{3})  \
-j_{\sandwichNum}(g_{9})k_{\sandwichNum}(g_{4}) \Big) \\ + a_5\Big( \
-j_{\sandwichNum}(g_{1})k_{\sandwichNum}(g_{11})  \
+j_{\sandwichNum}(g_{1})k_{\sandwichNum}(g_{2})  \
+j_{\sandwichNum}(g_{1})k_{\sandwichNum}(g_{4})  \
-j_{\sandwichNum}(g_{1})k_{\sandwichNum}(g_{5})  \
-j_{\sandwichNum}(g_{1})k_{\sandwichNum}(g_{6})  \\  \
-j_{\sandwichNum}(g_{1})k_{\sandwichNum}(g_{7})  \
+j_{\sandwichNum}(g_{3})k_{\sandwichNum}(g_{11})  \
-j_{\sandwichNum}(g_{3})k_{\sandwichNum}(g_{2})  \
-j_{\sandwichNum}(g_{3})k_{\sandwichNum}(g_{4})  \
+j_{\sandwichNum}(g_{3})k_{\sandwichNum}(g_{5})  \\  \
+j_{\sandwichNum}(g_{3})k_{\sandwichNum}(g_{6})  \
+j_{\sandwichNum}(g_{3})k_{\sandwichNum}(g_{7})  \
-j_{\sandwichNum}(g_{6})k_{\sandwichNum}(g_{11})  \
+j_{\sandwichNum}(g_{6})k_{\sandwichNum}(g_{2})  \
+j_{\sandwichNum}(g_{6})k_{\sandwichNum}(g_{4})  \\  \
-j_{\sandwichNum}(g_{6})k_{\sandwichNum}(g_{5})  \
-j_{\sandwichNum}(g_{6})k_{\sandwichNum}(g_{6})  \
-j_{\sandwichNum}(g_{6})k_{\sandwichNum}(g_{7})  \
+j_{\sandwichNum}(g_{8})k_{\sandwichNum}(g_{10})  \
-j_{\sandwichNum}(g_{8})k_{\sandwichNum}(g_{2})  \\  \
+j_{\sandwichNum}(g_{8})k_{\sandwichNum}(g_{3})  \
-j_{\sandwichNum}(g_{8})k_{\sandwichNum}(g_{4})  \
+j_{\sandwichNum}(g_{9})k_{\sandwichNum}(g_{11})  \
-j_{\sandwichNum}(g_{9})k_{\sandwichNum}(g_{2})  \
-j_{\sandwichNum}(g_{9})k_{\sandwichNum}(g_{4})  \\  \
+j_{\sandwichNum}(g_{9})k_{\sandwichNum}(g_{5})  \
+j_{\sandwichNum}(g_{9})k_{\sandwichNum}(g_{6})  \
+j_{\sandwichNum}(g_{9})k_{\sandwichNum}(g_{7}) \Big) \\ + a_6\Big( \
2j_{\sandwichNum}(g_{10})k_{\sandwichNum}(g_{7})  \
+2j_{\sandwichNum}(g_{10})k_{\sandwichNum}(g_{8})  \
+2j_{\sandwichNum}(g_{1})k_{\sandwichNum}(g_{11})  \
+2j_{\sandwichNum}(g_{1})k_{\sandwichNum}(g_{5})  \
+2j_{\sandwichNum}(g_{1})k_{\sandwichNum}(g_{6})  \\  \
+2j_{\sandwichNum}(g_{1})k_{\sandwichNum}(g_{7})  \
-2j_{\sandwichNum}(g_{2})k_{\sandwichNum}(g_{10})  \
-2j_{\sandwichNum}(g_{2})k_{\sandwichNum}(g_{4})  \
-2j_{\sandwichNum}(g_{2})k_{\sandwichNum}(g_{8})  \
-2j_{\sandwichNum}(g_{2})k_{\sandwichNum}(g_{9})  \\  \
+2j_{\sandwichNum}(g_{3})k_{\sandwichNum}(g_{5})  \
+2j_{\sandwichNum}(g_{3})k_{\sandwichNum}(g_{8})  \
-2j_{\sandwichNum}(g_{4})k_{\sandwichNum}(g_{11})  \
-2j_{\sandwichNum}(g_{4})k_{\sandwichNum}(g_{2})  \
-2j_{\sandwichNum}(g_{4})k_{\sandwichNum}(g_{8})  \\  \
-2j_{\sandwichNum}(g_{4})k_{\sandwichNum}(g_{9})  \
+3j_{\sandwichNum}(g_{4})k_{\sandwichNum}(g_{6})  \
-4j_{\sandwichNum}(g_{4})k_{\sandwichNum}(g_{3})  \
-j_{\sandwichNum}(g_{10})k_{\sandwichNum}(g_{1})  \
-j_{\sandwichNum}(g_{10})k_{\sandwichNum}(g_{2})  \\  \
-j_{\sandwichNum}(g_{10})k_{\sandwichNum}(g_{3})  \
-j_{\sandwichNum}(g_{10})k_{\sandwichNum}(g_{4})  \
-j_{\sandwichNum}(g_{10})k_{\sandwichNum}(g_{5})  \
-j_{\sandwichNum}(g_{10})k_{\sandwichNum}(g_{6})  \
-j_{\sandwichNum}(g_{1})k_{\sandwichNum}(g_{10})  \\  \
-j_{\sandwichNum}(g_{1})k_{\sandwichNum}(g_{2})  \
-j_{\sandwichNum}(g_{1})k_{\sandwichNum}(g_{3})  \
-j_{\sandwichNum}(g_{1})k_{\sandwichNum}(g_{4})  \
+j_{\sandwichNum}(g_{2})k_{\sandwichNum}(g_{1})  \
+j_{\sandwichNum}(g_{2})k_{\sandwichNum}(g_{11})  \\  \
-j_{\sandwichNum}(g_{2})k_{\sandwichNum}(g_{3})  \
+j_{\sandwichNum}(g_{2})k_{\sandwichNum}(g_{7})  \
-j_{\sandwichNum}(g_{3})k_{\sandwichNum}(g_{1})  \
-j_{\sandwichNum}(g_{3})k_{\sandwichNum}(g_{10})  \
-j_{\sandwichNum}(g_{3})k_{\sandwichNum}(g_{11})  \\  \
-j_{\sandwichNum}(g_{3})k_{\sandwichNum}(g_{2})  \
-j_{\sandwichNum}(g_{3})k_{\sandwichNum}(g_{4})  \
-j_{\sandwichNum}(g_{3})k_{\sandwichNum}(g_{7})  \
+j_{\sandwichNum}(g_{4})k_{\sandwichNum}(g_{1})  \
+j_{\sandwichNum}(g_{4})k_{\sandwichNum}(g_{10})  \\  \
+j_{\sandwichNum}(g_{4})k_{\sandwichNum}(g_{5}) \Big) \\ + a_7\Big( \
-2j_{\sandwichNum}(g_{10})k_{\sandwichNum}(g_{2})  \
-2j_{\sandwichNum}(g_{10})k_{\sandwichNum}(g_{7})  \
-2j_{\sandwichNum}(g_{3})k_{\sandwichNum}(g_{10})  \
-2j_{\sandwichNum}(g_{3})k_{\sandwichNum}(g_{4})  \
-2j_{\sandwichNum}(g_{3})k_{\sandwichNum}(g_{5})  \\  \
+2j_{\sandwichNum}(g_{4})k_{\sandwichNum}(g_{11})  \
-2j_{\sandwichNum}(g_{4})k_{\sandwichNum}(g_{6})  \
-j_{\sandwichNum}(g_{10})k_{\sandwichNum}(g_{3})  \
+j_{\sandwichNum}(g_{10})k_{\sandwichNum}(g_{4})  \
+j_{\sandwichNum}(g_{10})k_{\sandwichNum}(g_{5})  \\  \
+j_{\sandwichNum}(g_{10})k_{\sandwichNum}(g_{9})  \
-j_{\sandwichNum}(g_{2})k_{\sandwichNum}(g_{10})  \
-j_{\sandwichNum}(g_{2})k_{\sandwichNum}(g_{11})  \
-j_{\sandwichNum}(g_{2})k_{\sandwichNum}(g_{4})  \
+j_{\sandwichNum}(g_{2})k_{\sandwichNum}(g_{6})  \\  \
-j_{\sandwichNum}(g_{2})k_{\sandwichNum}(g_{7})  \
+j_{\sandwichNum}(g_{2})k_{\sandwichNum}(g_{9})  \
+j_{\sandwichNum}(g_{3})k_{\sandwichNum}(g_{11})  \
+j_{\sandwichNum}(g_{3})k_{\sandwichNum}(g_{2})  \
+j_{\sandwichNum}(g_{3})k_{\sandwichNum}(g_{3})  \\  \
-j_{\sandwichNum}(g_{3})k_{\sandwichNum}(g_{6})  \
+j_{\sandwichNum}(g_{3})k_{\sandwichNum}(g_{7})  \
+j_{\sandwichNum}(g_{3})k_{\sandwichNum}(g_{9})  \
-j_{\sandwichNum}(g_{4})k_{\sandwichNum}(g_{10})  \
-j_{\sandwichNum}(g_{4})k_{\sandwichNum}(g_{2})  \\  \
-j_{\sandwichNum}(g_{4})k_{\sandwichNum}(g_{5})  \
+j_{\sandwichNum}(g_{4})k_{\sandwichNum}(g_{9}) \Big) \\ + a_8\Big( \
2j_{\sandwichNum}(g_{10})k_{\sandwichNum}(g_{11})  \
+2j_{\sandwichNum}(g_{10})k_{\sandwichNum}(g_{2})  \
+2j_{\sandwichNum}(g_{10})k_{\sandwichNum}(g_{3})  \
+2j_{\sandwichNum}(g_{10})k_{\sandwichNum}(g_{7})  \
+2j_{\sandwichNum}(g_{1})k_{\sandwichNum}(g_{8})  \\  \
-2j_{\sandwichNum}(g_{2})k_{\sandwichNum}(g_{5})  \
-2j_{\sandwichNum}(g_{2})k_{\sandwichNum}(g_{6})  \
+2j_{\sandwichNum}(g_{3})k_{\sandwichNum}(g_{10})  \
+2j_{\sandwichNum}(g_{3})k_{\sandwichNum}(g_{4})  \
+2j_{\sandwichNum}(g_{3})k_{\sandwichNum}(g_{5})  \\  \
+2j_{\sandwichNum}(g_{3})k_{\sandwichNum}(g_{6})  \
-2j_{\sandwichNum}(g_{4})k_{\sandwichNum}(g_{11})  \
-2j_{\sandwichNum}(g_{4})k_{\sandwichNum}(g_{7})  \
-j_{\sandwichNum}(g_{10})k_{\sandwichNum}(g_{10})  \
-j_{\sandwichNum}(g_{10})k_{\sandwichNum}(g_{4})  \\  \
-j_{\sandwichNum}(g_{10})k_{\sandwichNum}(g_{5})  \
-j_{\sandwichNum}(g_{10})k_{\sandwichNum}(g_{6})  \
-j_{\sandwichNum}(g_{1})k_{\sandwichNum}(g_{1})  \
+j_{\sandwichNum}(g_{1})k_{\sandwichNum}(g_{3})  \
-j_{\sandwichNum}(g_{1})k_{\sandwichNum}(g_{6})  \\  \
+j_{\sandwichNum}(g_{1})k_{\sandwichNum}(g_{9})  \
+j_{\sandwichNum}(g_{2})k_{\sandwichNum}(g_{10})  \
+j_{\sandwichNum}(g_{2})k_{\sandwichNum}(g_{11})  \
+j_{\sandwichNum}(g_{2})k_{\sandwichNum}(g_{2})  \
+j_{\sandwichNum}(g_{2})k_{\sandwichNum}(g_{3})  \\  \
+j_{\sandwichNum}(g_{2})k_{\sandwichNum}(g_{4})  \
+j_{\sandwichNum}(g_{2})k_{\sandwichNum}(g_{7})  \
-j_{\sandwichNum}(g_{3})k_{\sandwichNum}(g_{11})  \
-j_{\sandwichNum}(g_{3})k_{\sandwichNum}(g_{2})  \
-j_{\sandwichNum}(g_{3})k_{\sandwichNum}(g_{3})  \\  \
-j_{\sandwichNum}(g_{3})k_{\sandwichNum}(g_{7})  \
+j_{\sandwichNum}(g_{4})k_{\sandwichNum}(g_{10})  \
+j_{\sandwichNum}(g_{4})k_{\sandwichNum}(g_{2})  \
+j_{\sandwichNum}(g_{4})k_{\sandwichNum}(g_{3})  \
+j_{\sandwichNum}(g_{4})k_{\sandwichNum}(g_{4})  \\  \
+j_{\sandwichNum}(g_{4})k_{\sandwichNum}(g_{5})  \
+j_{\sandwichNum}(g_{4})k_{\sandwichNum}(g_{6}) \Big) \\ + a_9\Big( \
2j_{\sandwichNum}(g_{11})k_{\sandwichNum}(g_{1})  \
+2j_{\sandwichNum}(g_{1})k_{\sandwichNum}(g_{11})  \
+2j_{\sandwichNum}(g_{1})k_{\sandwichNum}(g_{5})  \
+2j_{\sandwichNum}(g_{1})k_{\sandwichNum}(g_{6})  \
+2j_{\sandwichNum}(g_{1})k_{\sandwichNum}(g_{7})  \\  \
+2j_{\sandwichNum}(g_{5})k_{\sandwichNum}(g_{1})  \
+2j_{\sandwichNum}(g_{6})k_{\sandwichNum}(g_{1})  \
+2j_{\sandwichNum}(g_{7})k_{\sandwichNum}(g_{1})  \
-3j_{\sandwichNum}(g_{10})k_{\sandwichNum}(g_{2})  \
-3j_{\sandwichNum}(g_{10})k_{\sandwichNum}(g_{3})  \\  \
-3j_{\sandwichNum}(g_{2})k_{\sandwichNum}(g_{10})  \
-3j_{\sandwichNum}(g_{2})k_{\sandwichNum}(g_{4})  \
-3j_{\sandwichNum}(g_{3})k_{\sandwichNum}(g_{10})  \
-3j_{\sandwichNum}(g_{3})k_{\sandwichNum}(g_{4})  \
-3j_{\sandwichNum}(g_{4})k_{\sandwichNum}(g_{2})  \\  \
-3j_{\sandwichNum}(g_{4})k_{\sandwichNum}(g_{3})  \
-j_{\sandwichNum}(g_{10})k_{\sandwichNum}(g_{1})  \
-j_{\sandwichNum}(g_{1})k_{\sandwichNum}(g_{10})  \
-j_{\sandwichNum}(g_{1})k_{\sandwichNum}(g_{2})  \
-j_{\sandwichNum}(g_{1})k_{\sandwichNum}(g_{3})  \\  \
-j_{\sandwichNum}(g_{1})k_{\sandwichNum}(g_{4})  \
-j_{\sandwichNum}(g_{2})k_{\sandwichNum}(g_{1})  \
-j_{\sandwichNum}(g_{3})k_{\sandwichNum}(g_{1})  \
-j_{\sandwichNum}(g_{4})k_{\sandwichNum}(g_{1}) \Big) \\ + a_{10}\Big( \
-2j_{\sandwichNum}(g_{10})k_{\sandwichNum}(g_{3})  \
-2j_{\sandwichNum}(g_{3})k_{\sandwichNum}(g_{10})  \
-j_{\sandwichNum}(g_{10})k_{\sandwichNum}(g_{2})  \
+j_{\sandwichNum}(g_{10})k_{\sandwichNum}(g_{5})  \
+j_{\sandwichNum}(g_{10})k_{\sandwichNum}(g_{6})  \\  \
-j_{\sandwichNum}(g_{11})k_{\sandwichNum}(g_{2})  \
+j_{\sandwichNum}(g_{11})k_{\sandwichNum}(g_{3})  \
-j_{\sandwichNum}(g_{2})k_{\sandwichNum}(g_{10})  \
-j_{\sandwichNum}(g_{2})k_{\sandwichNum}(g_{11})  \
-j_{\sandwichNum}(g_{2})k_{\sandwichNum}(g_{7})  \\  \
+j_{\sandwichNum}(g_{3})k_{\sandwichNum}(g_{11})  \
-j_{\sandwichNum}(g_{3})k_{\sandwichNum}(g_{4})  \
+j_{\sandwichNum}(g_{3})k_{\sandwichNum}(g_{7})  \
-j_{\sandwichNum}(g_{4})k_{\sandwichNum}(g_{3})  \
-j_{\sandwichNum}(g_{4})k_{\sandwichNum}(g_{5})  \\  \
-j_{\sandwichNum}(g_{4})k_{\sandwichNum}(g_{6})  \
+j_{\sandwichNum}(g_{5})k_{\sandwichNum}(g_{10})  \
-j_{\sandwichNum}(g_{5})k_{\sandwichNum}(g_{4})  \
+j_{\sandwichNum}(g_{6})k_{\sandwichNum}(g_{10})  \
-j_{\sandwichNum}(g_{6})k_{\sandwichNum}(g_{4})  \\  \
-j_{\sandwichNum}(g_{7})k_{\sandwichNum}(g_{2})  \
+j_{\sandwichNum}(g_{7})k_{\sandwichNum}(g_{3}) \Big) \\ + a_{11}\Big( \
2j_{\sandwichNum}(g_{11})k_{\sandwichNum}(g_{4})  \
+2j_{\sandwichNum}(g_{1})k_{\sandwichNum}(g_{11})  \
-2j_{\sandwichNum}(g_{1})k_{\sandwichNum}(g_{2})  \
-2j_{\sandwichNum}(g_{1})k_{\sandwichNum}(g_{4})  \
+2j_{\sandwichNum}(g_{1})k_{\sandwichNum}(g_{5})  \\  \
+2j_{\sandwichNum}(g_{1})k_{\sandwichNum}(g_{6})  \
+2j_{\sandwichNum}(g_{1})k_{\sandwichNum}(g_{7})  \
+2j_{\sandwichNum}(g_{3})k_{\sandwichNum}(g_{4})  \
-2j_{\sandwichNum}(g_{5})k_{\sandwichNum}(g_{3})  \
-2j_{\sandwichNum}(g_{6})k_{\sandwichNum}(g_{4})  \\  \
-2j_{\sandwichNum}(g_{7})k_{\sandwichNum}(g_{10})  \
-2j_{\sandwichNum}(g_{8})k_{\sandwichNum}(g_{10})  \
+2j_{\sandwichNum}(g_{8})k_{\sandwichNum}(g_{2})  \
-2j_{\sandwichNum}(g_{8})k_{\sandwichNum}(g_{3})  \
+2j_{\sandwichNum}(g_{8})k_{\sandwichNum}(g_{4})  \\  \
+2j_{\sandwichNum}(g_{9})k_{\sandwichNum}(g_{2})  \
+2j_{\sandwichNum}(g_{9})k_{\sandwichNum}(g_{4})  \
-j_{\sandwichNum}(g_{10})k_{\sandwichNum}(g_{4})  \
-j_{\sandwichNum}(g_{10})k_{\sandwichNum}(g_{5})  \
-j_{\sandwichNum}(g_{10})k_{\sandwichNum}(g_{6})  \\  \
-j_{\sandwichNum}(g_{2})k_{\sandwichNum}(g_{10})  \
+j_{\sandwichNum}(g_{2})k_{\sandwichNum}(g_{11})  \
+j_{\sandwichNum}(g_{2})k_{\sandwichNum}(g_{3})  \
-j_{\sandwichNum}(g_{2})k_{\sandwichNum}(g_{4})  \
+j_{\sandwichNum}(g_{2})k_{\sandwichNum}(g_{7})  \\  \
-j_{\sandwichNum}(g_{3})k_{\sandwichNum}(g_{11})  \
+j_{\sandwichNum}(g_{3})k_{\sandwichNum}(g_{2})  \
-j_{\sandwichNum}(g_{3})k_{\sandwichNum}(g_{7})  \
+j_{\sandwichNum}(g_{4})k_{\sandwichNum}(g_{10})  \
-j_{\sandwichNum}(g_{4})k_{\sandwichNum}(g_{2})  \\  \
-j_{\sandwichNum}(g_{4})k_{\sandwichNum}(g_{3})  \
+j_{\sandwichNum}(g_{4})k_{\sandwichNum}(g_{5})  \
+j_{\sandwichNum}(g_{4})k_{\sandwichNum}(g_{6}) \Big) \\ + a_{12}\Big( \
2j_{\sandwichNum}(g_{1})k_{\sandwichNum}(g_{3})  \
-2j_{\sandwichNum}(g_{1})k_{\sandwichNum}(g_{6})  \
+2j_{\sandwichNum}(g_{1})k_{\sandwichNum}(g_{9})  \
-2j_{\sandwichNum}(g_{3})k_{\sandwichNum}(g_{5})  \
-2j_{\sandwichNum}(g_{5})k_{\sandwichNum}(g_{3})  \\  \
+2j_{\sandwichNum}(g_{5})k_{\sandwichNum}(g_{6})  \
-2j_{\sandwichNum}(g_{5})k_{\sandwichNum}(g_{9})  \
+2j_{\sandwichNum}(g_{6})k_{\sandwichNum}(g_{5})  \
-2j_{\sandwichNum}(g_{7})k_{\sandwichNum}(g_{9})  \
-2j_{\sandwichNum}(g_{9})k_{\sandwichNum}(g_{5})  \\  \
-2j_{\sandwichNum}(g_{9})k_{\sandwichNum}(g_{7})  \
-4j_{\sandwichNum}(g_{8})k_{\sandwichNum}(g_{3})  \
+4j_{\sandwichNum}(g_{8})k_{\sandwichNum}(g_{6})  \
-4j_{\sandwichNum}(g_{8})k_{\sandwichNum}(g_{9})  \
+j_{\sandwichNum}(g_{10})k_{\sandwichNum}(g_{5})  \\  \
-j_{\sandwichNum}(g_{11})k_{\sandwichNum}(g_{5})  \
+j_{\sandwichNum}(g_{11})k_{\sandwichNum}(g_{6})  \
+j_{\sandwichNum}(g_{2})k_{\sandwichNum}(g_{3})  \
+j_{\sandwichNum}(g_{2})k_{\sandwichNum}(g_{4})  \
-j_{\sandwichNum}(g_{2})k_{\sandwichNum}(g_{6})  \\  \
-j_{\sandwichNum}(g_{2})k_{\sandwichNum}(g_{7})  \
+j_{\sandwichNum}(g_{2})k_{\sandwichNum}(g_{9})  \
+j_{\sandwichNum}(g_{3})k_{\sandwichNum}(g_{2})  \
+j_{\sandwichNum}(g_{3})k_{\sandwichNum}(g_{4})  \
-j_{\sandwichNum}(g_{3})k_{\sandwichNum}(g_{7})  \\  \
+j_{\sandwichNum}(g_{4})k_{\sandwichNum}(g_{2})  \
+j_{\sandwichNum}(g_{4})k_{\sandwichNum}(g_{3})  \
-j_{\sandwichNum}(g_{4})k_{\sandwichNum}(g_{5})  \
-j_{\sandwichNum}(g_{4})k_{\sandwichNum}(g_{6})  \
+j_{\sandwichNum}(g_{4})k_{\sandwichNum}(g_{9})  \\  \
+j_{\sandwichNum}(g_{5})k_{\sandwichNum}(g_{10})  \
-j_{\sandwichNum}(g_{5})k_{\sandwichNum}(g_{11})  \
-j_{\sandwichNum}(g_{5})k_{\sandwichNum}(g_{4})  \
+j_{\sandwichNum}(g_{5})k_{\sandwichNum}(g_{7})  \
+j_{\sandwichNum}(g_{6})k_{\sandwichNum}(g_{11})  \\  \
-j_{\sandwichNum}(g_{6})k_{\sandwichNum}(g_{2})  \
-j_{\sandwichNum}(g_{6})k_{\sandwichNum}(g_{4})  \
+j_{\sandwichNum}(g_{6})k_{\sandwichNum}(g_{7})  \
-j_{\sandwichNum}(g_{7})k_{\sandwichNum}(g_{2})  \
-j_{\sandwichNum}(g_{7})k_{\sandwichNum}(g_{3})  \\  \
+j_{\sandwichNum}(g_{7})k_{\sandwichNum}(g_{5})  \
+j_{\sandwichNum}(g_{7})k_{\sandwichNum}(g_{6})  \
+j_{\sandwichNum}(g_{9})k_{\sandwichNum}(g_{2})  \
+j_{\sandwichNum}(g_{9})k_{\sandwichNum}(g_{4}) \Big) \\ + a_{13}\Big( \
2j_{\sandwichNum}(g_{10})k_{\sandwichNum}(g_{3})  \
-2j_{\sandwichNum}(g_{10})k_{\sandwichNum}(g_{7})  \
+2j_{\sandwichNum}(g_{11})k_{\sandwichNum}(g_{2})  \
-2j_{\sandwichNum}(g_{11})k_{\sandwichNum}(g_{5})  \
+2j_{\sandwichNum}(g_{11})k_{\sandwichNum}(g_{9})  \\  \
+2j_{\sandwichNum}(g_{2})k_{\sandwichNum}(g_{9})  \
+2j_{\sandwichNum}(g_{3})k_{\sandwichNum}(g_{10})  \
+2j_{\sandwichNum}(g_{3})k_{\sandwichNum}(g_{4})  \
+2j_{\sandwichNum}(g_{4})k_{\sandwichNum}(g_{11})  \
+2j_{\sandwichNum}(g_{4})k_{\sandwichNum}(g_{9})  \\  \
-2j_{\sandwichNum}(g_{5})k_{\sandwichNum}(g_{11})  \
-2j_{\sandwichNum}(g_{5})k_{\sandwichNum}(g_{3})  \
+2j_{\sandwichNum}(g_{5})k_{\sandwichNum}(g_{6})  \
+2j_{\sandwichNum}(g_{5})k_{\sandwichNum}(g_{7})  \
-2j_{\sandwichNum}(g_{5})k_{\sandwichNum}(g_{9})  \\  \
-2j_{\sandwichNum}(g_{6})k_{\sandwichNum}(g_{10})  \
+2j_{\sandwichNum}(g_{6})k_{\sandwichNum}(g_{11})  \
-2j_{\sandwichNum}(g_{6})k_{\sandwichNum}(g_{2})  \
+2j_{\sandwichNum}(g_{6})k_{\sandwichNum}(g_{3})  \
-2j_{\sandwichNum}(g_{6})k_{\sandwichNum}(g_{6})  \\  \
+2j_{\sandwichNum}(g_{6})k_{\sandwichNum}(g_{7})  \
+2j_{\sandwichNum}(g_{6})k_{\sandwichNum}(g_{9})  \
-2j_{\sandwichNum}(g_{7})k_{\sandwichNum}(g_{3})  \
+2j_{\sandwichNum}(g_{7})k_{\sandwichNum}(g_{5})  \
-2j_{\sandwichNum}(g_{7})k_{\sandwichNum}(g_{9})  \\  \
+2j_{\sandwichNum}(g_{9})k_{\sandwichNum}(g_{2})  \
+2j_{\sandwichNum}(g_{9})k_{\sandwichNum}(g_{4})  \
+3j_{\sandwichNum}(g_{3})k_{\sandwichNum}(g_{2})  \
-3j_{\sandwichNum}(g_{3})k_{\sandwichNum}(g_{7})  \
+3j_{\sandwichNum}(g_{4})k_{\sandwichNum}(g_{3})  \\  \
+4j_{\sandwichNum}(g_{1})k_{\sandwichNum}(g_{3})  \
-4j_{\sandwichNum}(g_{1})k_{\sandwichNum}(g_{6})  \
+4j_{\sandwichNum}(g_{1})k_{\sandwichNum}(g_{9})  \
+4j_{\sandwichNum}(g_{6})k_{\sandwichNum}(g_{5})  \
-4j_{\sandwichNum}(g_{9})k_{\sandwichNum}(g_{5})  \\  \
-4j_{\sandwichNum}(g_{9})k_{\sandwichNum}(g_{7})  \
-6j_{\sandwichNum}(g_{3})k_{\sandwichNum}(g_{5})  \
-8j_{\sandwichNum}(g_{8})k_{\sandwichNum}(g_{3})  \
+8j_{\sandwichNum}(g_{8})k_{\sandwichNum}(g_{6})  \
-8j_{\sandwichNum}(g_{8})k_{\sandwichNum}(g_{9})  \\  \
+j_{\sandwichNum}(g_{10})k_{\sandwichNum}(g_{4})  \
+j_{\sandwichNum}(g_{10})k_{\sandwichNum}(g_{5})  \
-j_{\sandwichNum}(g_{10})k_{\sandwichNum}(g_{6})  \
+j_{\sandwichNum}(g_{2})k_{\sandwichNum}(g_{10})  \
+j_{\sandwichNum}(g_{2})k_{\sandwichNum}(g_{11})  \\  \
+j_{\sandwichNum}(g_{2})k_{\sandwichNum}(g_{3})  \
+j_{\sandwichNum}(g_{2})k_{\sandwichNum}(g_{4})  \
-j_{\sandwichNum}(g_{2})k_{\sandwichNum}(g_{7})  \
-j_{\sandwichNum}(g_{3})k_{\sandwichNum}(g_{11})  \
-j_{\sandwichNum}(g_{4})k_{\sandwichNum}(g_{10})  \\  \
+j_{\sandwichNum}(g_{4})k_{\sandwichNum}(g_{2})  \
-j_{\sandwichNum}(g_{4})k_{\sandwichNum}(g_{5})  \
-j_{\sandwichNum}(g_{4})k_{\sandwichNum}(g_{6}) \Big) \\ + a_{14}\Big( \
2j_{\sandwichNum}(g_{11})k_{\sandwichNum}(g_{10})  \
+2j_{\sandwichNum}(g_{1})k_{\sandwichNum}(g_{11})  \
-2j_{\sandwichNum}(g_{1})k_{\sandwichNum}(g_{2})  \
-2j_{\sandwichNum}(g_{1})k_{\sandwichNum}(g_{4})  \
+2j_{\sandwichNum}(g_{1})k_{\sandwichNum}(g_{5})  \\  \
+2j_{\sandwichNum}(g_{1})k_{\sandwichNum}(g_{7})  \
-2j_{\sandwichNum}(g_{1})k_{\sandwichNum}(g_{8})  \
+2j_{\sandwichNum}(g_{3})k_{\sandwichNum}(g_{2})  \
+2j_{\sandwichNum}(g_{3})k_{\sandwichNum}(g_{4})  \
-2j_{\sandwichNum}(g_{5})k_{\sandwichNum}(g_{2})  \\  \
-2j_{\sandwichNum}(g_{6})k_{\sandwichNum}(g_{2})  \
+2j_{\sandwichNum}(g_{6})k_{\sandwichNum}(g_{3})  \
-2j_{\sandwichNum}(g_{6})k_{\sandwichNum}(g_{4})  \
-2j_{\sandwichNum}(g_{7})k_{\sandwichNum}(g_{4})  \
-2j_{\sandwichNum}(g_{8})k_{\sandwichNum}(g_{10})  \\  \
+2j_{\sandwichNum}(g_{8})k_{\sandwichNum}(g_{2})  \
-2j_{\sandwichNum}(g_{8})k_{\sandwichNum}(g_{3})  \
+2j_{\sandwichNum}(g_{8})k_{\sandwichNum}(g_{4})  \
+2j_{\sandwichNum}(g_{9})k_{\sandwichNum}(g_{2})  \
+2j_{\sandwichNum}(g_{9})k_{\sandwichNum}(g_{4})  \\  \
+3j_{\sandwichNum}(g_{1})k_{\sandwichNum}(g_{6})  \
-j_{\sandwichNum}(g_{10})k_{\sandwichNum}(g_{10})  \
+j_{\sandwichNum}(g_{1})k_{\sandwichNum}(g_{1})  \
-j_{\sandwichNum}(g_{1})k_{\sandwichNum}(g_{3})  \
-j_{\sandwichNum}(g_{1})k_{\sandwichNum}(g_{9})  \\  \
+j_{\sandwichNum}(g_{2})k_{\sandwichNum}(g_{2})  \
-j_{\sandwichNum}(g_{3})k_{\sandwichNum}(g_{3})  \
+j_{\sandwichNum}(g_{4})k_{\sandwichNum}(g_{4}) \Big)
\end{align*}

%% file: fivePoints.bbl
\providecommand{\href}[2]{#2}\begingroup\raggedright\begin{thebibliography}{10}

\bibitem{Elvang:2016qvq}
H.~Elvang, C.R.T.~Jones and S.G.~Naculich, \emph{{Soft photon and graviton
  theorems in effective field theory}},
  \href{https://doi.org/10.1103/PhysRevLett.118.231601}{\emph{Phys. Rev. Lett.}
  {\bfseries 118} (2017) 231601}
  [\href{https://arxiv.org/abs/1611.07534}{{\ttfamily 1611.07534}}].

\bibitem{Shadmi:2018xan}
Y.~Shadmi and Y.~Weiss, \emph{{Effective Field Theory Amplitudes the On-Shell
  Way: Scalar and Vector Couplings to Gluons}},
  \href{https://doi.org/10.1007/JHEP02(2019)165}{\emph{JHEP} {\bfseries 02}
  (2019) 165} [\href{https://arxiv.org/abs/1809.09644}{{\ttfamily
  1809.09644}}].

\bibitem{Cheung:2018oki}
C.~Cheung, K.~Kampf, J.~Novotny, C.-H.~Shen, J.~Trnka and C.~Wen, \emph{{Vector
  Effective Field Theories from Soft Limits}},
  \href{https://doi.org/10.1103/PhysRevLett.120.261602}{\emph{Phys. Rev. Lett.}
  {\bfseries 120} (2018) 261602}
  [\href{https://arxiv.org/abs/1801.01496}{{\ttfamily 1801.01496}}].

\bibitem{Kampf:2019mcd}
K.~Kampf, J.~Novotny, M.~Shifman and J.~Trnka, \emph{{New Soft Theorems for
  Goldstone Boson Amplitudes}},
  \href{https://doi.org/10.1103/PhysRevLett.124.111601}{\emph{Phys. Rev. Lett.}
  {\bfseries 124} (2020) 111601}
  [\href{https://arxiv.org/abs/1910.04766}{{\ttfamily 1910.04766}}].

\bibitem{Elvang:2019twd}
H.~Elvang, M.~Hadjiantonis, C.R.T.~Jones and S.~Paranjape,
  \emph{{All-Multiplicity One-Loop Amplitudes in Born-Infeld Electrodynamics
  from Generalized Unitarity}},
  \href{https://arxiv.org/abs/1906.05321}{{\ttfamily 1906.05321}}.

\bibitem{Ma:2019gtx}
T.~Ma, J.~Shu and M.-L.~Xiao, \emph{{Standard Model Effective Field Theory from
  On-shell Amplitudes}},  \href{https://arxiv.org/abs/1902.06752}{{\ttfamily
  1902.06752}}.

\bibitem{Durieux:2019eor}
G.~Durieux, T.~Kitahara, Y.~Shadmi and Y.~Weiss, \emph{{The electroweak
  effective field theory from on-shell amplitudes}},
  \href{https://doi.org/10.1007/JHEP01(2020)119}{\emph{JHEP} {\bfseries 01}
  (2020) 119} [\href{https://arxiv.org/abs/1909.10551}{{\ttfamily
  1909.10551}}].

\bibitem{Bern:2019wie}
Z.~Bern, J.~Parra-Martinez and E.~Sawyer, \emph{{Non-renormalization and
  operator mixing via on-shell methods}},
  \href{https://arxiv.org/abs/1910.05831}{{\ttfamily 1910.05831}}.

\bibitem{Bern:2020ikv}
Z.~Bern, J.~Parra-Martinez and E.~Sawyer, \emph{{Structure of two-loop SMEFT
  anomalous dimensions via on-shell methods}},
  \href{https://doi.org/10.1007/JHEP10(2020)211}{\emph{JHEP} {\bfseries 10}
  (2020) 211} [\href{https://arxiv.org/abs/2005.12917}{{\ttfamily
  2005.12917}}].

\bibitem{Durieux:2020gip}
G.~Durieux, T.~Kitahara, C.S.~Machado, Y.~Shadmi and Y.~Weiss,
  \emph{{Constructing massive on-shell contact terms}},
  \href{https://doi.org/10.1007/JHEP12(2020)175}{\emph{JHEP} {\bfseries 12}
  (2020) 175} [\href{https://arxiv.org/abs/2008.09652}{{\ttfamily
  2008.09652}}].

\bibitem{Rodina:2021isd}
L.~Rodina and Z.~Yin, \emph{{Exploring the Landscape for Soft Theorems of
  Nonlinear Sigma Models}},  \href{https://arxiv.org/abs/2102.08396}{{\ttfamily
  2102.08396}}.

\bibitem{Arkani-Hamed:2020blm}
N.~Arkani-Hamed, T.-C.~Huang and Y.-T.~Huang, \emph{{The EFT-Hedron}},
  \href{https://arxiv.org/abs/2012.15849}{{\ttfamily 2012.15849}}.

\bibitem{Bern:2021ppb}
Z.~Bern, D.~Kosmopoulos and A.~Zhiboedov, \emph{{Gravitational Effective Field
  Theory Islands, Low-Spin Dominance, and the Four-Graviton Amplitude}},
  \href{https://arxiv.org/abs/2103.12728}{{\ttfamily 2103.12728}}.

\bibitem{BCJ}
Z.~Bern, J.J.M.~Carrasco and H.~Johansson, \emph{{New relations for
  gauge-theory amplitudes}},
  \href{https://doi.org/10.1103/PhysRevD.78.085011}{\emph{Phys. Rev.}
  {\bfseries D78} (2008) 085011}
  [\href{https://arxiv.org/abs/0805.3993}{{\ttfamily 0805.3993}}].

\bibitem{BCJLoop}
Z.~Bern, J.J.M.~Carrasco and H.~Johansson, \emph{{Perturbative quantum gravity
  as a double copy of gauge theory}},
  \href{https://doi.org/10.1103/PhysRevLett.105.061602}{\emph{Phys. Rev. Lett.}
  {\bfseries 105} (2010) 061602}
  [\href{https://arxiv.org/abs/1004.0476}{{\ttfamily 1004.0476}}].

\bibitem{KLT}
H.~Kawai, D.C.~Lewellen and S.H.H.~Tye, \emph{{A relation between tree
  amplitudes of closed and open strings}},
  \href{https://doi.org/10.1016/0550-3213(86)90362-7}{\emph{Nucl. Phys.}
  {\bfseries B269} (1986) 1}.

\bibitem{BjerrumBohr:2009rd}
N.E.J.~Bjerrum-Bohr, P.H.~Damgaard and P.~Vanhove, \emph{{Minimal Basis for
  Gauge Theory Amplitudes}},
  \href{https://doi.org/10.1103/PhysRevLett.103.161602}{\emph{Phys. Rev. Lett.}
  {\bfseries 103} (2009) 161602}
  [\href{https://arxiv.org/abs/0907.1425}{{\ttfamily 0907.1425}}].

\bibitem{Stieberger:2009hq}
S.~Stieberger, \emph{{Open and closed vs. pure open string Disk amplitudes}},
  \href{https://arxiv.org/abs/0907.2211}{{\ttfamily 0907.2211}}.

\bibitem{Carrasco:2019yyn}
J.J.M.~Carrasco, L.~Rodina, Z.~Yin and S.~Zekioglu, \emph{{Simple encoding of
  higher derivative gauge and gravity counterterms}},
  \href{https://arxiv.org/abs/1910.12850}{{\ttfamily 1910.12850}}.

\bibitem{Broedel2013tta}
J.~Broedel, O.~Schlotterer and S.~Stieberger, \emph{{Polylogarithms, multiple
  zeta values and superstring amplitudes}},
  \href{https://doi.org/10.1002/prop.201300019}{\emph{Fortsch. Phys.}
  {\bfseries 61} (2013) 812} [\href{https://arxiv.org/abs/1304.7267}{{\ttfamily
  1304.7267}}].

\bibitem{Carrasco2016ldy}
J.J.M.~Carrasco, C.R.~Mafra and O.~Schlotterer, \emph{{Abelian Z-theory: NLSM
  amplitudes and $\alpha$'-corrections from the open string}},
  \href{https://doi.org/10.1007/JHEP06(2017)093}{\emph{JHEP} {\bfseries 06}
  (2017) 093} [\href{https://arxiv.org/abs/1608.02569}{{\ttfamily
  1608.02569}}].

\bibitem{Carrasco2016ygv}
J.J.M.~Carrasco, C.R.~Mafra and O.~Schlotterer, \emph{{Semi-abelian Z-theory:
  NLSM$+\phi^{3}$ from the open string}},
  \href{https://doi.org/10.1007/JHEP08(2017)135}{\emph{JHEP} {\bfseries 08}
  (2017) 135} [\href{https://arxiv.org/abs/1612.06446}{{\ttfamily
  1612.06446}}].

\bibitem{Mafra2016mcc}
C.R.~Mafra and O.~Schlotterer, \emph{{Non-abelian $Z$-theory: Berends-Giele
  recursion for the $\alpha'$-expansion of disk integrals}},
  \href{https://doi.org/10.1007/JHEP01(2017)031}{\emph{JHEP} {\bfseries 01}
  (2017) 031} [\href{https://arxiv.org/abs/1609.07078}{{\ttfamily
  1609.07078}}].

\bibitem{Bern:2012uf}
Z.~Bern, J.J.M.~Carrasco, L.J.~Dixon, H.~Johansson and R.~Roiban,
  \emph{{Simplifying Multiloop Integrands and Ultraviolet Divergences of Gauge
  Theory and Gravity Amplitudes}},
  \href{https://doi.org/10.1103/PhysRevD.85.105014}{\emph{Phys. Rev. D}
  {\bfseries 85} (2012) 105014}
  [\href{https://arxiv.org/abs/1201.5366}{{\ttfamily 1201.5366}}].

\bibitem{ancFiles}
  \emph{fivePointEFT auxiliary Mathematica files},  [\href{https://github.com/sunazekioglu/fivePointEFT}{https://github.com/sunazekioglu/fivePointEFT}].
	

\bibitem{BCJreview}
Z.~Bern, J.J.~Carrasco, M.~Chiodaroli, H.~Johansson and R.~Roiban, \emph{{The
  Duality Between Color and Kinematics and its Applications}},
  \href{https://arxiv.org/abs/1909.01358}{{\ttfamily 1909.01358}}.

\bibitem{KiermaierTalk}
M.~Kiermaier, \emph{{Gravity as the square of gauge theory}},  in
  \emph{Amplitudes 2010}, Queen Mary, University of London, [\href{https://strings.ph.qmul.ac.uk/~theory/Amplitudes2010/Talks/MK2010.pdf}{https://strings.ph.qmul.ac.uk/$\sim$theory/Amplitudes2010/Talks/MK2010.pdf}].


\bibitem{Bern:2017tuc}
Z.~Bern, A.~Edison, D.~Kosower and J.~Parra-Martinez, \emph{{Curvature-squared
  multiplets, evanescent effects, and the U(1) anomaly in ${\cal N}=4$
  supergravity}}, \href{https://doi.org/10.1103/PhysRevD.96.066004}{\emph{Phys.
  Rev.} {\bfseries D96} (2017) 066004}
  [\href{https://arxiv.org/abs/1706.01486}{{\ttfamily 1706.01486}}].

\bibitem{Barreiro:2013dpa}
L.A.~Barreiro and R.~Medina, \emph{{RNS derivation of $N$-point disk amplitudes
  from the revisited S-matrix approach}},
  \href{https://doi.org/10.1016/j.nuclphysb.2014.07.015}{\emph{Nucl. Phys.}
  {\bfseries B886} (2014) 870}
  [\href{https://arxiv.org/abs/1310.5942}{{\ttfamily 1310.5942}}].

\bibitem{Boels:2016xhc}
R.H.~Boels and R.~Medina, \emph{{Graviton and gluon scattering from first
  principles}},
  \href{https://doi.org/10.1103/PhysRevLett.118.061602}{\emph{Phys. Rev. Lett.}
  {\bfseries 118} (2017) 061602}
  [\href{https://arxiv.org/abs/1607.08246}{{\ttfamily 1607.08246}}].

\bibitem{Bargheer2012gv}
T.~Bargheer, S.~He and T.~McLoughlin, \emph{{New relations for
  three-dimensional supersymmetric scattering amplitudes}},
  \href{https://doi.org/10.1103/PhysRevLett.108.231601}{\emph{Phys. Rev. Lett.}
  {\bfseries 108} (2012) 231601}
  [\href{https://arxiv.org/abs/1203.0562}{{\ttfamily 1203.0562}}].

\bibitem{Huang2012wr}
Y.-t.~Huang and H.~Johansson, \emph{{Equivalent $D=3$ supergravity amplitudes
  from double copies of three-algebra and two-algebra gauge theories}},
  \href{https://doi.org/10.1103/PhysRevLett.110.171601}{\emph{Phys. Rev. Lett.}
  {\bfseries 110} (2013) 171601}
  [\href{https://arxiv.org/abs/1210.2255}{{\ttfamily 1210.2255}}].

\bibitem{Huang:2013kca}
Y.-t.~Huang, H.~Johansson and S.~Lee, \emph{{On three-algebra and
  bi-fundamental matter amplitudes and integrability of supergravity}},
  \href{https://doi.org/10.1007/JHEP11(2013)050}{\emph{JHEP} {\bfseries 11}
  (2013) 050} [\href{https://arxiv.org/abs/1307.2222}{{\ttfamily 1307.2222}}].

\bibitem{Chandia:2003sh}
O.~Chandia and R.~Medina, \emph{{Four point effective actions in open and
  closed superstring theory}},
  \href{https://doi.org/10.1088/1126-6708/2003/11/003}{\emph{JHEP} {\bfseries
  11} (2003) 003} [\href{https://arxiv.org/abs/hep-th/0310015}{{\ttfamily
  hep-th/0310015}}].

\bibitem{Bandiera:2020aqn}
R.~Bandiera and C.R.~Mafra, \emph{{A closed-formula solution to the color-trace
  decomposition problem}},  \href{https://arxiv.org/abs/2009.02534}{{\ttfamily
  2009.02534}}.

\bibitem{Boels:2013jua}
R.H.~Boels, \emph{{On the field theory expansion of superstring five point
  amplitudes}},
  \href{https://doi.org/10.1016/j.nuclphysb.2013.08.009}{\emph{Nucl. Phys. B}
  {\bfseries 876} (2013) 215}
  [\href{https://arxiv.org/abs/1304.7918}{{\ttfamily 1304.7918}}].

\bibitem{callumTalk}
H.~Chi, H.~Elvang, A.~Herderschee, C.~Jones and S.~Paranjape,
  \emph{Generalizations of the KLT formula},  in \emph{QCD Meets Gravity VI},
  Northwestern University, 
  [\href{https://indico.desy.de/event/27454/contributions/93682/}{https://indico.desy.de/event/27454/contributions/93682/}].

\bibitem{Carrasco:2019qwr}
J.J.M.~Carrasco and L.~Rodina, \emph{{UV considerations on scattering
  amplitudes in a web of theories}},
  \href{https://arxiv.org/abs/1908.08033}{{\ttfamily 1908.08033}}.

\bibitem{Huang:2016tag}
Y.-t.~Huang, O.~Schlotterer and C.~Wen, \emph{{Universality in string
  interactions}}, \href{https://doi.org/10.1007/JHEP09(2016)155}{\emph{JHEP}
  {\bfseries 09} (2016) 155}
  [\href{https://arxiv.org/abs/1602.01674}{{\ttfamily 1602.01674}}].

\bibitem{Johansson2017srf}
H.~Johansson and J.~Nohle, \emph{{Conformal gravity from gauge theory}},
  \href{https://arxiv.org/abs/1707.02965}{{\ttfamily 1707.02965}}.

\bibitem{Mafra2011nv}
C.R.~Mafra, O.~Schlotterer and S.~Stieberger, \emph{{Complete N-Point
  Superstring Disk Amplitude I. Pure Spinor Computation}},
  \href{https://doi.org/10.1016/j.nuclphysb.2013.04.023}{\emph{Nucl. Phys. B}
  {\bfseries 873} (2013) 419}
  [\href{https://arxiv.org/abs/1106.2645}{{\ttfamily 1106.2645}}].

\bibitem{Schlotterer2012ny}
O.~Schlotterer and S.~Stieberger, \emph{{Motivic multiple zeta values and
  superstring amplitudes}},
  \href{https://doi.org/10.1088/1751-8113/46/47/475401}{\emph{J. Phys.}
  {\bfseries A46} (2013) 475401}
  [\href{https://arxiv.org/abs/1205.1516}{{\ttfamily 1205.1516}}].

\bibitem{Mafra2011nw}
C.R.~Mafra, O.~Schlotterer and S.~Stieberger, \emph{{Complete N-Point
  Superstring Disk Amplitude II. Amplitude and Hypergeometric Function
  Structure}},
  \href{https://doi.org/10.1016/j.nuclphysb.2013.04.022}{\emph{Nucl. Phys. B}
  {\bfseries 873} (2013) 461}
  [\href{https://arxiv.org/abs/1106.2646}{{\ttfamily 1106.2646}}].

\bibitem{Broedel:2013tta}
J.~Broedel, O.~Schlotterer and S.~Stieberger, \emph{{Polylogarithms, Multiple
  Zeta Values and Superstring Amplitudes}},
  \href{https://doi.org/10.1002/prop.201300019}{\emph{Fortsch. Phys.}
  {\bfseries 61} (2013) 812} [\href{https://arxiv.org/abs/1304.7267}{{\ttfamily
  1304.7267}}].

\bibitem{Green2013bza}
M.B.~Green, C.R.~Mafra and O.~Schlotterer, \emph{{Multiparticle one-loop
  amplitudes and S-duality in closed superstring theory}},
  \href{https://doi.org/10.1007/JHEP10(2013)188}{\emph{JHEP} {\bfseries 10}
  (2013) 188} [\href{https://arxiv.org/abs/1307.3534}{{\ttfamily 1307.3534}}].

\bibitem{Henning:2015alf}
B.~Henning, X.~Lu, T.~Melia and H.~Murayama, \emph{{2, 84, 30, 993, 560, 15456,
  11962, 261485, ...: Higher dimension operators in the SM EFT}},
  \href{https://doi.org/10.1007/JHEP08(2017)016}{\emph{JHEP} {\bfseries 08}
  (2017) 016} [\href{https://arxiv.org/abs/1512.03433}{{\ttfamily
  1512.03433}}].

\bibitem{Lehman:2015via}
L.~Lehman and A.~Martin, \emph{{Hilbert Series for Constructing Lagrangians:
  expanding the phenomenologist's toolbox}},
  \href{https://doi.org/10.1103/PhysRevD.91.105014}{\emph{Phys. Rev. D}
  {\bfseries 91} (2015) 105014}
  [\href{https://arxiv.org/abs/1503.07537}{{\ttfamily 1503.07537}}].

\bibitem{Lehman:2015coa}
L.~Lehman and A.~Martin, \emph{{Low-derivative operators of the Standard Model
  effective field theory via Hilbert series methods}},
  \href{https://doi.org/10.1007/JHEP02(2016)081}{\emph{JHEP} {\bfseries 02}
  (2016) 081} [\href{https://arxiv.org/abs/1510.00372}{{\ttfamily
  1510.00372}}].

\bibitem{Henning:2015daa}
B.~Henning, X.~Lu, T.~Melia and H.~Murayama, \emph{{Hilbert series and operator
  bases with derivatives in effective field theories}},
  \href{https://doi.org/10.1007/s00220-015-2518-2}{\emph{Commun. Math. Phys.}
  {\bfseries 347} (2016) 363}
  [\href{https://arxiv.org/abs/1507.07240}{{\ttfamily 1507.07240}}].

\bibitem{Henning:2017fpj}
B.~Henning, X.~Lu, T.~Melia and H.~Murayama, \emph{{Operator bases,
  $S$-matrices, and their partition functions}},
  \href{https://doi.org/10.1007/JHEP10(2017)199}{\emph{JHEP} {\bfseries 10}
  (2017) 199} [\href{https://arxiv.org/abs/1706.08520}{{\ttfamily
  1706.08520}}].

\bibitem{SimplifyingBCJ}
Z.~Bern, J.J.M.~Carrasco, L.J.~Dixon, H.~Johansson and R.~Roiban,
  \emph{{Simplifying multiloop integrands and ultraviolet divergences of gauge
  theory and gravity amplitudes}},
  \href{https://doi.org/10.1103/PhysRevD.85.105014}{\emph{Phys. Rev.}
  {\bfseries D85} (2012) 105014}
  [\href{https://arxiv.org/abs/1201.5366}{{\ttfamily 1201.5366}}].

\bibitem{Low:2019wuv}
I.~Low and Z.~Yin, \emph{{New Flavor-Kinematics Dualities and Extensions of
  Nonlinear Sigma Models}},
  \href{https://doi.org/10.1016/j.physletb.2020.135544}{\emph{Phys. Lett. B}
  {\bfseries 807} (2020) 135544}
  [\href{https://arxiv.org/abs/1911.08490}{{\ttfamily 1911.08490}}].

\bibitem{Low:2020ubn}
I.~Low, L.~Rodina and Z.~Yin, \emph{{Double Copy in Higher Derivative Operators
  of Nambu-Goldstone Bosons}},
  \href{https://doi.org/10.1103/PhysRevD.103.025004}{\emph{Phys. Rev. D}
  {\bfseries 103} (2021) 025004}
  [\href{https://arxiv.org/abs/2009.00008}{{\ttfamily 2009.00008}}].

\end{thebibliography}\endgroup
